\documentclass[twocolumn,tighten,times]{aastex631}

\usepackage[T1]{fontenc}
\usepackage{times}
\usepackage{booktabs}
\usepackage{amsmath} 
\usepackage{mathrsfs} 
\usepackage{ulem}

\usepackage{soul}
\usepackage[section]{placeins}

\newcommand{\OIII}{\hbox{[O\,{\sc iii}]}}
\newcommand{\NII}{\hbox{[N\,{\sc ii}]}}
\newcommand{\Feii}{\hbox{Fe\,{\sc ii}}}

\newcommand{\sigmaline}{\hbox{$\sigma_{\rm line}$}}
\newcommand{\fwhmsigma}{\hbox{FWHM/$\sigma_{\rm line}$}}
\newcommand{\deltav}{\hbox{$\Delta v$}}

\defcitealias{du_monitoring_2018}{I}
\defcitealias{brotherton2020}{II}
\defcitealias{bao2022}{III}
\defcitealias{Zastrocky2024}{IV}

\def\authoremaila{\href{mailto:dupu@ihep.ac.cn}{dupu@ihep.ac.cn}}
\def\authoremailb{\href{mailto:mbrother@uwyo.edu}{mbrother@uwyo.edu}}

\shorttitle{Long-term Variation and Evolution of the Broad H$\beta$ Emission-Line Profiles}
\shortauthors{F.-N. Fang et al.}

\begin{document}
\title{Monitoring AGNs with H$\beta$ Asymmetry. V. \\
Long-term Variation and Evolution of the Broad H$\beta$ Emission-Line Profiles}

\author[0009-0009-9945-673X]{Feng-Na Fang}
\affiliation{Key Laboratory for Particle Astrophysics, Institute of High Energy Physics, Chinese Academy of Sciences, 19B Yuquan Road, Beijing 100049, China; \authoremaila}
\affiliation{School of Physical Science, University of Chinese Academy of Sciences, 19A Yuquan Road, Beijing 100049, China}

\author[0000-0002-5830-3544]{Pu Du}
\affiliation{Key Laboratory for Particle Astrophysics, Institute of High Energy Physics, Chinese Academy of Sciences, 19B Yuquan Road, Beijing 100049, China; \authoremaila}

\author[0000-0002-1207-0909]{Michael S. Brotherton}
\affiliation{Department of Physics and Astronomy, University of Wyoming, Laramie, WY 82071, USA; \authoremailb}

\author[0000-0003-1081-2929]{Jacob N. McLane}
\affiliation{Department of Physics and Astronomy, University of Wyoming, Laramie, WY 82071, USA; \authoremailb}

\author[0009-0003-5173-0548]{T. E. Zastrocky}
\affiliation{Department of Physics and Astronomy, University of Wyoming, Laramie, WY 82071, USA; \authoremailb}
\affiliation{Physics and Astronomy Department, Regis University, Denver, CO 80212, USA}

\author{Kianna A. Olson}
\affiliation{Department of Physics and Astronomy, University of Wyoming, Laramie, WY 82071, USA; \authoremailb}

\author[0000-0003-2024-1648]{Dong-Wei Bao}
\affiliation{National Astronomical Observatories, Chinese Academy of Sciences, 20A Datun Road, Chaoyang District, Beijing 100101, China}

\author[0009-0005-4152-2088]{Shuo Zhai}
\affiliation{National Astronomical Observatories, Chinese Academy of Sciences, 20A Datun Road, Chaoyang District, Beijing 100101, China}

\author{Hua-Rui Bai}
\affiliation{Key Laboratory for Particle Astrophysics, Institute of High Energy Physics, Chinese Academy of Sciences, 19B Yuquan Road, Beijing 100049, China; \authoremaila}
\affiliation{School of Physical Science, University of Chinese Academy of Sciences, 19A Yuquan Road, Beijing 100049, China}

\author[0009-0004-5259-0900]{Yi-Xin Fu}
\affiliation{Key Laboratory for Particle Astrophysics, Institute of High Energy Physics, Chinese Academy of Sciences, 19B Yuquan Road, Beijing 100049, China; \authoremaila}
\affiliation{School of Physical Science, University of Chinese Academy of Sciences, 19A Yuquan Road, Beijing 100049, China}

\author{Bi-Xuan Zhao}
\affiliation{Shanghai Observatory, Chinese Academy of Sciences, 80 Nandan Road, Shanghai 200030, China}

\author[0000-0003-4280-7673]{Yong-Jie Chen}
\affiliation{Key Laboratory for Particle Astrophysics, Institute of High Energy Physics, Chinese Academy of Sciences, 19B Yuquan Road, Beijing 100049, China; \authoremaila}
\affiliation{Dongguan Neutron Science Center, 1 Zhongziyuan Road, Dongguan 523808, China}

\author[0009-0006-2592-7699]{Yue-Chang Peng}
\affiliation{Key Laboratory for Particle Astrophysics, Institute of High Energy Physics, Chinese Academy of Sciences, 19B Yuquan Road, Beijing 100049, China; \authoremaila}
\affiliation{School of Physical Science, University of Chinese Academy of Sciences, 19A Yuquan Road, Beijing 100049, China}

\author[0000-0003-4042-7191]{Yu-Yang Songsheng}
\affiliation{Key Laboratory for Particle Astrophysics, Institute of High Energy Physics, Chinese Academy of Sciences, 19B Yuquan Road, Beijing 100049, China; \authoremaila}

\author[0000-0001-5841-9179]{Yan-Rong Li}
\affiliation{Key Laboratory for Particle Astrophysics, Institute of High Energy Physics, Chinese Academy of Sciences, 19B Yuquan Road, Beijing 100049, China; \authoremaila}

\author{Chen Hu}
\affiliation{Key Laboratory for Particle Astrophysics, Institute of High Energy Physics, Chinese Academy of Sciences, 19B Yuquan Road, Beijing 100049, China; \authoremaila}

\author[0000-0001-5981-6440]{Ming Xiao}
\affiliation{Key Laboratory for Particle Astrophysics, Institute of High Energy Physics, Chinese Academy of Sciences, 19B Yuquan Road, Beijing 100049, China; \authoremaila}

\author[0000-0003-3825-0710]{Bo-Wei Jiang}
\affiliation{Key Laboratory for Particle Astrophysics, Institute of High Energy Physics, Chinese Academy of Sciences, 19B Yuquan Road, Beijing 100049, China; \authoremaila}
\affiliation{School of Physical Science, University of Chinese Academy of Sciences, 19A Yuquan Road, Beijing 100049, China}

\author{Yi-Lin Wang}
\affiliation{Key Laboratory for Particle Astrophysics, Institute of High Energy Physics, Chinese Academy of Sciences, 19B Yuquan Road, Beijing 100049, China; \authoremaila}
\affiliation{School of Physical Science, University of Chinese Academy of Sciences, 19A Yuquan Road, Beijing 100049, China}

\author{Hao Zhang}
\affiliation{Key Laboratory for Particle Astrophysics, Institute of High Energy Physics, Chinese Academy of Sciences, 19B Yuquan Road, Beijing 100049, China; \authoremaila}
\affiliation{School of Physical Science, University of Chinese Academy of Sciences, 19A Yuquan Road, Beijing 100049, China}

\author{Yu Zhao}
\affiliation{Key Laboratory for Particle Astrophysics, Institute of High Energy Physics, Chinese Academy of Sciences, 19B Yuquan Road, Beijing 100049, China; \authoremaila}
\affiliation{School of Physical Science, University of Chinese Academy of Sciences, 19A Yuquan Road, Beijing 100049, China}

\author{Jia-Qi Feng}
\affiliation{Key Laboratory for Particle Astrophysics, Institute of High Energy Physics, Chinese Academy of Sciences, 19B Yuquan Road, Beijing 100049, China; \authoremaila}
\affiliation{School of Physical Science, University of Chinese Academy of Sciences, 19A Yuquan Road, Beijing 100049, China}

\author{Yi-Peng Zhao}
\affiliation{Key Laboratory for Particle Astrophysics, Institute of High Energy Physics, Chinese Academy of Sciences, 19B Yuquan Road, Beijing 100049, China; \authoremaila}
\affiliation{School of Physical Science, University of Chinese Academy of Sciences, 19A Yuquan Road, Beijing 100049, China}

\author[0000-0003-0534-6388]{David H. Kasper}
\affiliation{Department of Physics and Astronomy, University of Wyoming, Laramie, WY 82071, USA; \authoremailb}

\author[0000-0002-8818-6780]{William T. Chick}
\affiliation{Department of Physics and Astronomy, University of Wyoming, Laramie, WY 82071, USA; \authoremailb}

\author{My L. Nguyen}
\affiliation{Department of Physics and Astronomy, University of Wyoming, Laramie, WY 82071, USA; \authoremailb}

\author[0000-0002-4423-4584]{Jaya Maithil}
\affiliation{Department of Physics and Astronomy, University of Wyoming, Laramie, WY 82071, USA; \authoremailb}

\author[0000-0002-4475-4176]{H. A. Kobulnicky}
\affiliation{Department of Physics and Astronomy, University of Wyoming, Laramie, WY 82071, USA; \authoremailb}

\author[0000-0002-5782-9093]{D. A. Dale}
\affiliation{Department of Physics and Astronomy, University of Wyoming, Laramie, WY 82071, USA; \authoremailb}

\author{Derek Hand}
\affiliation{Department of Physics and Astronomy, University of Wyoming, Laramie, WY 82071, USA; \authoremailb}

\author{C. Adelman}
\affiliation{Department of Physics and Astronomy, University of Wyoming, Laramie, WY 82071, USA; \authoremailb}
\affiliation{Department of Physics \& Astronomy, Cal Poly Pomona, Pomona, CA 91768, USA}

\author[0009-0006-5373-6515]{Z. Carter}
\affiliation{Department of Physics and Astronomy, University of Wyoming, Laramie, WY 82071, USA; \authoremailb}
\affiliation{Department of Physics and Astronomy, Trinity University, San Antonio, TX 78212, USA}

\author{A. M. Murphree}
\affiliation{Department of Physics and Astronomy, University of Wyoming, Laramie, WY 82071, USA; \authoremailb}
\affiliation{Department of Physics, Rhodes College, Memphis, TN 38112, USA}

\author[0000-0001-5636-3108]{M. Oeur}
\affiliation{Department of Physics and Astronomy, University of Wyoming, Laramie, WY 82071, USA; \authoremailb}
\affiliation{Department of Physics and Astronomy, State Long Beach, Long Beach, CA 90840, USA}

\author{S. Schonsberg}
\affiliation{Department of Physics and Astronomy, University of Wyoming, Laramie, WY 82071, USA; \authoremailb}
\affiliation{Department of Physics and Astronomy, University of Montana, Missoula, MT 59812, USA}

\author{T. Roth} 
\affiliation{Department of Physics and Astronomy, University of Wyoming, Laramie, WY 82071, USA; \authoremailb}
\affiliation{Department of Physics \& Astronomy, California State University, Sacramento, CA 95747, USA}

\author[0000-0003-2662-0526]{Hartmut Winkler}
\affiliation{Department of Physics, University of Johannesburg, P.O. Box 524, 2006 Auckland Park, South Africa}

\author[0000-0002-6058-4912]{Paola Marziani}
\affiliation{Istituto Nazionale di Astrofisica (INAF), Osservatorio Astronomico di Padova, I-35122 Padova, Italy}

\author[0000-0001-6441-9044]{Mauro D'Onofrio} 
\affiliation{Dipartimento di Fisica \& Astronomia ``Galileo Galilei'', Universit\`{a} di Padova, Padova, Italy}

\author[0000-0003-1728-0304]{Keith Horne}
\affiliation{SUPA School of Physics and Astronomy, North Haugh, St. Andrews, KY16 9SS, Scotland, UK}

\author[0000-0001-6947-5846]{Luis C. Ho}
\affiliation{Kavli Institute for Astronomy and Astrophysics, Peking University, Beijing 100871, China}
\affiliation{Department of Astronomy, School of Physics, Peking University, Beijing 100871, China}

\author{Jin-Ming Bai}
\affiliation{Yunnan Observatories, Chinese Academy of Sciences, Kunming 650011, China}

\author[0000-0001-9449-9268]{Jian-Min Wang}
\altaffiliation{PI of the MAHA project.}
\affiliation{Key Laboratory for Particle Astrophysics, Institute of High Energy Physics, Chinese Academy of Sciences, 19B Yuquan Road, Beijing 100049, China; \authoremaila}
\affiliation{National Astronomical Observatories, Chinese Academy of Sciences, 20A Datun Road, Chaoyang District, Beijing 100101, China}
\affiliation{School of Astronomy and Space Science, University of Chinese Academy of Sciences, 19A Yuquan Road, Beijing 100049, China}

\collaboration{99}{(MAHA collaboration)}

\begin{abstract}
The physical origins of the diverse emission-line asymmetries observed in the spectra of active galactic nuclei (AGNs) remain incompletely understood. Monitoring the temporal variations of line profiles offers a promising approach to investigating the underlying physics. In this study, we present an analysis of the broad H$\beta$ emission line profiles of eight AGNs observed from the end of 2016 to May 2023 as part of the reverberation mapping campaign titled ``Monitoring AGNs with H$\beta$ Asymmetry'' (MAHA), utilizing data obtained from the Wyoming Infrared Observatory (WIRO) 2.3-meter telescope. We measure the temporal variations of line asymmetry, width, and central velocity shift for the eight objects. Our findings reveal that the variation in asymmetry is positively correlated with H$\beta$ flux in five of the eight objects, while the remaining objects exhibit negative or complex correlations. Furthermore, we observe anti-correlations between line width and H$\beta$ flux for most objects, indicating the presence of the ``breathing'' phenomenon in their H$\beta$ emission lines. In contrast, two objects demonstrate an ``anti-breathing'' phenomenon or complex behavior. We discuss the physical origins of the temporal variations in line profiles and propose the possibility of decomposing the variations in H$\beta$ asymmetry and width into components: one that corresponds to short-term variations in H$\beta$ flux and another that reflects long-term variations in continuum light curves, perhaps driven by radiation pressure.
\end{abstract}

\keywords{galaxies: active – galaxies: nuclei – quasars: emission lines – quasars: supermassive black holes - techniques: spectroscopic - line: profiles}

\section{Introduction} \label{sec:intro}
Broad emission lines (BELs) with velocity widths of thousands of km\,s$^{-1}$ are prominent diagnostic features in the spectra of active galactic nuclei (AGNs). It is long established that a significant fraction of AGNs show apparently asymmetric profiles in the BELs, especially the broad Balmer lines (e.g., H$\beta$, H$\alpha$), displaying blue or red asymmetry, double peaks, or even complex sub-features within the line profile \citep[e.g.,][]{robertis1985, boroson1992, Sulentic2000, Baskin2005}. However, the physical origin of the asymmetry in the BEL profiles is complex and remains far from fully understood. 

Observationally, the census of BEL asymmetries in large AGN samples or the variability behavior of the BELs are two ways to reveal the origin of the asymmetries in those BEL profiles. For example, the H$\beta$ line profiles were revealed to exhibit more redward asymmetry in those AGNs with weaker \Feii\ and stronger \OIII\ emission lines, or vice versa, based on principal component analysis of the properties of low-redshift quasars \citep{boroson1992, brotherton1996}. Subsequently, it was found that the asymmetry of the H$\beta$ profile systematically changes along the Eigenvector 1 sequence \citep[e.g.,][]{sulentic2002, Zamfir2010}. The physical drivers of the Eigenvector 1 are probably the accretion rates and inclinations of AGNs \citep[e.g.,][]{boroson1992, Sulentic2000, Shen2014}, however, how these mechanisms control the broad-line regions (BLRs) and the line asymmetries are not clear. Furthermore, \cite{hu_systematic_2008} proposed that the profiles of H$\beta$ can be decomposed into two components, with the narrower ones exhibiting the same velocity shift and width as the \Feii\ lines. The variations of the asymmetric and double-peaked profiles of the BELs in some individual objects across timescales similar or longer than the kinematic timescales were investigated, revealing possible explanations for some specific cases \citep[e.g.,][]{storchi-begmann2003, Eracleous2009, Shapovalova2010, Bon2012, Li2016_NGC5548, Fries2023}. Recently, \cite{Oknyansky2021, Oknyansky2023} attributed the variation in BEL profiles and light curves of changing-look objects NGC~3516 and NGC~2617 to Compton-thick clouds that partially obscure the line of sight, based on long-term multiwavelength monitoring.

Numerous phenomenological and physical models have been proposed to explain the asymmetric profiles of BELs. For instance, the BEL asymmetries may arise as a natural consequence of optically thick clouds with radial motion (inflow, outflow) \citep{Capriotti1979}, and/or with self-absorption \citep{Ferland1979}. The double-peak component of Balmer lines could be attributed to a relativistic eccentric disk \citep{Eracleous1994, Eracleous1995}, or more recently, supermassive binary black holes \citep[e.g.,][]{Shen2010, Bon2012, Li2016_NGC5548}. Moreover, \cite{Wang2017_tidally_disrupted} proposed a model of the BLRs in which the gas originates from tidally disrupted clumps from the torus. Alternative models concerning the line profile asymmetry include partial dust obscuration of the clouds within BLRs \citep[e.g.,][]{Gaskell2018, Kara2021_AGN_STORM}, warped disks \citep[e.g.,][]{Pringle1996}, hot spots \citep[e.g.,][]{Newman1997, Jovanovic2010}, and spiral arms \citep[e.g.,][]{storchi-begmann2003, Du2023}. These models are applied on a case-by-case basis, emphasizing the complexity of the origin of emission line asymmetry on one hand, while also suggesting that there may be underlying fundamental physical processes at play in BLRs that have yet to be unveiled on the other hand.

Reverberation mapping (RM, see e.g., \citealt{blandford_reverberation_1982, peterson_reverberation_1993, peterson_central_2004}) is a powerful tool for investigating the geometry and kinematics of the BLR gas by monitoring the delayed responses of the BELs to the varying continuum. In RM, model-independent velocity-resolved lags \citep[e.g.,][]{bentz_lick_2010, denney2010, du_supermassive_2016, De_Rosa2018, chen_broad-line_2023, Li2024, Yao2024} and transfer functions (also known as ``velocity-delay maps'', e.g., \citealt{Grier2013, Xiao2018, Horne2021}) are commonly utilized to reveal the general BLR geometry and kinematics. It is thus intuitive that RM can play a key role in understanding the BEL asymmetries. However, the velocity-resolved lags and transfer functions only correspond to a portion of the BLR gas that responds to the varying continuum. The non-responsive gas in BLRs cannot be probed using these methods. Furthermore, most RM campaigns are shorter than the kinematic timescales of the BLRs, thus the temporal evolution of the BLRs (and the corresponding changes of the BEL asymmetries) is not captured. Therefore, only the RM campaigns with both long time spans and high cadence can reveal the origin of BEL asymmetries. 

Since the end of 2016, we started a long-term RM campaign dedicated to the AGNs with asymmetric H$\beta$ line profiles at the 2.3-meter telescope of the Wyoming Infrared Observatory (WIRO). The name of the campaign is ``Monitoring AGNs with H$\beta$ Asymmetry'' (MAHA, see \citealt{du_monitoring_2018, brotherton2020}). A primary objective of this campaign is to understand the physical mechanisms behind the line asymmetries observed in AGNs. The light curves and the velocity-resolved lags for a total of 34 objects have been presented by \cite{du_monitoring_2018}, \cite{brotherton2020}, \cite{bao2022}, and \cite{Zastrocky2024} (hereafter Paper \citetalias{du_monitoring_2018}, \citetalias{brotherton2020}, \citetalias{bao2022}, \citetalias{Zastrocky2024}, respectively). In Papers \citetalias{bao2022} and \citetalias{Zastrocky2024}, we discovered that the velocity-resolved lags constructed in RM observations do not directly correlate with the red or blue asymmetries of the BEL profiles (e.g., the AGNs with red-asymmetric line profiles in single-epoch spectra can be associated with either inflow, outflow, or Keplerian/virialized-like velocity-resolved lags). This motivates us to contemplate the necessity of revisiting the temporal variations of the BEL asymmetries based on the long-term RM data obtained in MAHA. 

Furthermore, the ``breathing'' behaviors exhibited by BELs are a common phenomenon in RM studies, serving as a valuable tool for probing the gas distributions within the BLR. This phenomenon is characterized by changes in line width during variations in the continuum, indicating shifts in the radius of the optimal emitting regions \citep{Korista_Goad_2004, Cackett2006, Goad2014}. Specifically, in the case of Balmer lines, their widths typically increase when the continuum decreases, as observed in studies by \cite{Wilhite2006}, \cite{park2012}, \cite{Barth2015}, \cite{Dexter2019}, \cite{Homan2020}, \cite{Wangshu2020}, and \cite{Fries2023}. Therefore, it is crucial to investigate whether AGNs exhibiting different BEL asymmetries also demonstrate distinct ``breathing'' behaviors.

In this paper, we focus on studying the asymmetry variability and ``breathing'' behavior of the H$\beta$ emission line, as well as their correlations with the continuum and emission-line light curves. The data reduction is described in Section \ref{sec:observations}. Section \ref{sec:measurements} describes the measurements, including asymmetry, line width, velocity shift, light curves, and time lags. In Section \ref{sec:results}, we present the data analyses and the results. Some discussions regarding the physical explanations are provided in Section \ref{sec:discussion}. Finally, we summarize our results in Section \ref{sec:Summary}.

\section{Observations and Data Reduction} \label{sec:observations}

\subsection{Sample Selection}
We selected 8 objects from Papers \citetalias{du_monitoring_2018}-\citetalias{Zastrocky2024} to investigate their asymmetry variations. To quantify the BEL asymmetry, we adopt the dimensionless parameter \citep{robertis1985} defined as
\begin{equation}
    A=\frac{\lambda_c(3/4)-\lambda_c(1/4)}{\Delta \lambda (1/2)},
    \label{equation: A_1985}
\end{equation}
where $\lambda_c(3/4)$ and $\lambda_c(1/4)$ are the central wavelengths at 3/4 and 1/4 of the peak of the line profile, respectively, and $\Delta \lambda (1/2)$ denotes the full width at half maximum (FWHM) of the line. A positive value of $A$ indicates blueward asymmetries, whereas a negative value represents redward asymmetries (refer to the schematic diagram in Figure 1 of Paper \citetalias{du_monitoring_2018}). Note that the asymmetry $A$ is only sensitive to the shape of the line profile, instead of the velocity shift of the line center (see Section \ref{sec:FWHM} for details). 

To identify targets exhibiting substantial asymmetry variability, we utilize ${\rm var}(A) ={(A_{\rm{max}} - A_{\rm{min}})}/{\sigma_{\rm{sys}}}$ to assess the significance, where $A_{\rm{max}}$ and $A_{\rm{min}}$ represent the maximum and minimum values of the asymmetry, respectively. Here, $\sigma_{\rm{sys}}$ denotes the scatter of $A$ obtained by removing the median-smoothed curve with a median filter of 5 points \citep{du_supermassive_2014}. 
The goal of this paper is not to present a complete sample but to analyze the temporal evolution of BEL asymmetry in a subset of AGNs where asymmetry variations can be robustly measured. Accordingly, we selected the AGNs from Papers \citetalias{du_monitoring_2018} to \citetalias{Zastrocky2024} that exhibited the highest values of ${\rm var}(A)$ (with the criterion of ${\rm var}(A) > 15$). The coordinates and redshifts of the eight selected objects are listed in Table \ref{tab:objects}. Notably, all eight objects display red asymmetries, making them a relatively homogeneous sample for asymmetry analysis. A more comprehensive analysis of larger and more diverse samples will be conducted in future studies.

\subsection{Spectroscopy and Photometry} 
\label{sec:Spectroscopy and Photometry}
The 8 objects were continuously monitored up to May 2023. Some of the data (such as light curves and velocity-resolved lags) have been previously reported in Papers \citetalias{du_monitoring_2018}-\citetalias{Zastrocky2024}. The objective of this study is to comprehensively analyze all available data to investigate the asymmetry variations of these objects, hence we have included data beyond that presented in Papers \citetalias{du_monitoring_2018}-\citetalias{Zastrocky2024}. The time spans covered by the new data for the objects discussed in this paper are listed in Table \ref{tab:objects}. The observational procedures and data reduction techniques for the spectroscopic observations have been detailed in Papers \citetalias{du_monitoring_2018}-\citetalias{Zastrocky2024}. Here, we offer a brief overview for completeness.

The WIRO 2.3\,m telescope was operated remotely, as described by \cite{Findlay_wiro_2016}. For the observations of the 8 objects, a 900 l/m grating was utilized, providing a spectral dispersion of 1.49\,\AA\,pixel$^{-1}$ over the wavelength range of 4000$\sim$7000\,\AA. To mitigate the impact of light loss, a 5$\arcsec$ wide long-slit was employed, wider than the typical seeing conditions of $2\,\arcsec \sim 3\,\arcsec$ in Wyoming. 

The data reduction was carried out using standard procedures via IRAF v2.16. Flux calibration was first conducted using one or more standard stars (BD+$28^{\circ}$4211, G191B2B, Feige 34, and Hz 44) taken on the same nights as the AGN spectra. However, the accuracy of the traditional standard-star-based calibration was found to be insufficient. To improve our flux accuracy, we further calibrated the spectra based on narrow $\OIII\lambda5007$ emission line, which originates from the narrow-line region and remains constant throughout the campaign. As the line spread function varied across different nights, we broadened the spectra to ensure that the \OIII\ line exhibited a consistent profile in all spectra before proceeding with the \OIII-based calibration. This calibration approach effectively minimized fluctuations in the light curves caused by weather conditions, particularly for consecutive nights. Further details regarding the \OIII-based calibration process can be found in Paper \citetalias{du_monitoring_2018}. 

For Mrk~841, additional 28 spectra were collected at the South African Astronomical Observatory (SAAO) 1.9\,m telescope with a 4$\arcsec$.04 slit. The dispersion of the spectrum is 1.31\,\AA\,pixel$^{-1}$. We also obtained 5 spectra based on the Asiago Astrophysical Observatory with a 1.82\,m telescope, using the Asiago Faint Object Spectrograph and Camera (AFOSC) in a 4$\arcsec$.2 slit. More details on the SAAO and Asiago observations can be found in Paper \citetalias{bao2022}.

As a supplement to the 5100\,\AA\ continuum from the spectroscopic observations, we also incorporated the publicly-available photometric data from time-domain surveys including the Zwicky Transient Facility (ZTF; \citealt{Masci2019_ZTF}) and the All-Sky Automated Survey for SuperNovae (ASAS-SN; \citealt{Shappee2014_ASASSN, Kochanek2017_ASASSN}). The ZTF survey of transients and variable stars commenced in 2018 utilizing the Palomar 48-inch Schmidt telescope. ASAS-SN, which began in 2013, obtains reliable photometry for objects with magnitudes down to approximately 17\,mag across the entire visible sky and currently operates 24 telescopes. In the following Section \ref{sec:Long-term trend}, we utilized data from the Catalina Real-time Transient Survey (CRTS; \citealt{Drake2009_CRTS}) to explore potential correlations between asymmetry variations and long-term continuum changes. CRTS started in 2007 and covers 33,000$\,\rm deg^2$ using three dedicated telescopes.

\section{Measurements} \label{sec:measurements}
\subsection{\texorpdfstring{H$\beta$ Asymmetry}{H beta Asymmetry}}
\label{sec:Asymmetry}
To assess the asymmetry of the broad H$\beta$ emission line in each spectrum, it is essential to remove the underlying continuum and the nearby narrow lines (H$\beta$ and $\OIII\lambda\lambda4959,5007$). The continuum background was determined through linear interpolation between two continuum windows surrounding H$\beta$, as provided in Table \ref{tab:measurement windows}.

The flux ratios between the narrow H$\beta$ and $\OIII\lambda\lambda4959,5007$ were deduced from the mean spectra (see Section \ref{sec:Spectroscopy and Photometry}) of the objects. Subsequently, assuming uniform profiles for all narrow lines, we obtained the contributions of the three narrow lines based on the ratios and the $\OIII\lambda5007$ profile (see more details in Paper \citetalias{du_monitoring_2018}). After removing the continuum and the narrow lines, we can isolate the pure broad H$\beta$ lines to evaluate their asymmetries. 

\begin{figure*}[ht]
  \centering
    \includegraphics[width=0.23\textwidth]{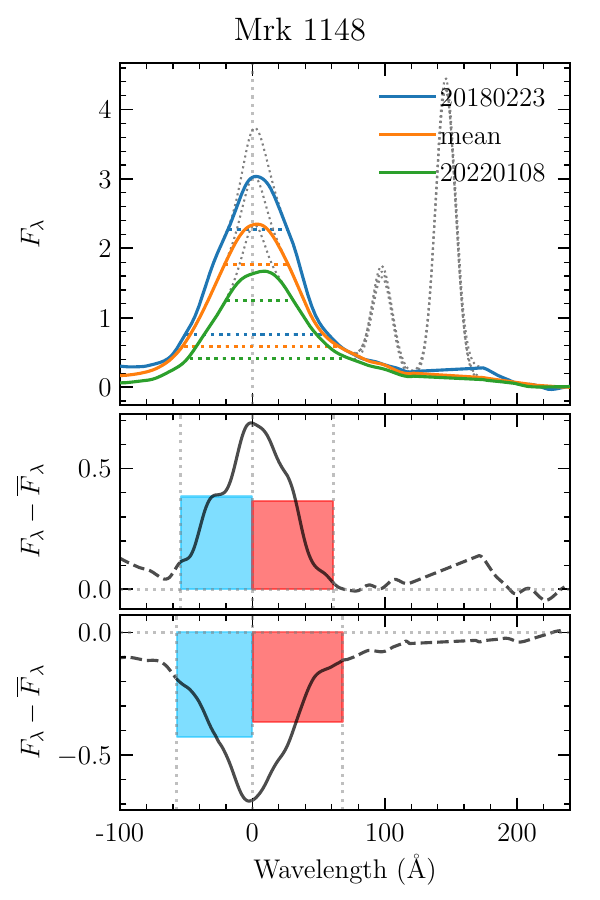}
    \includegraphics[width=0.23\textwidth]{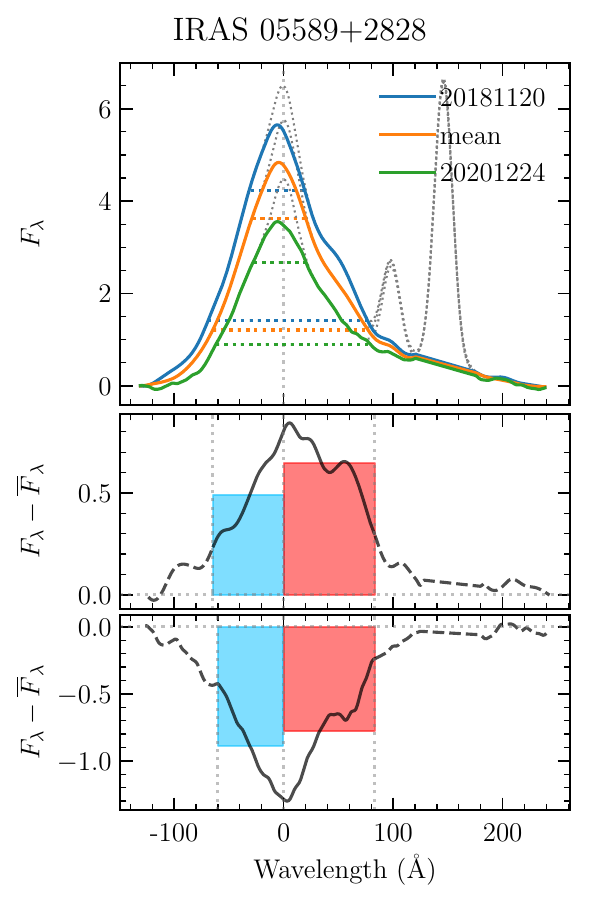}
    \includegraphics[width=0.23\textwidth]{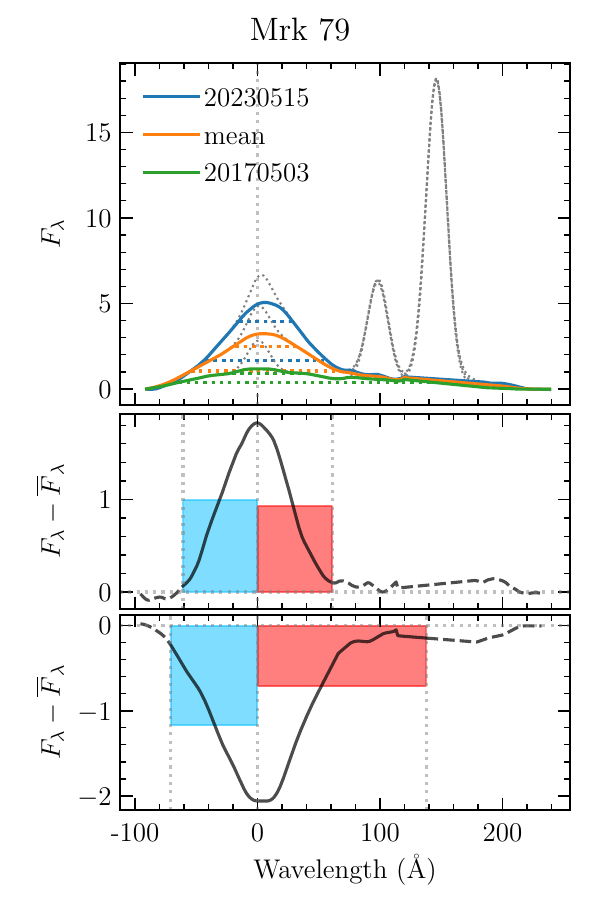}
    \includegraphics[width=0.23\textwidth]{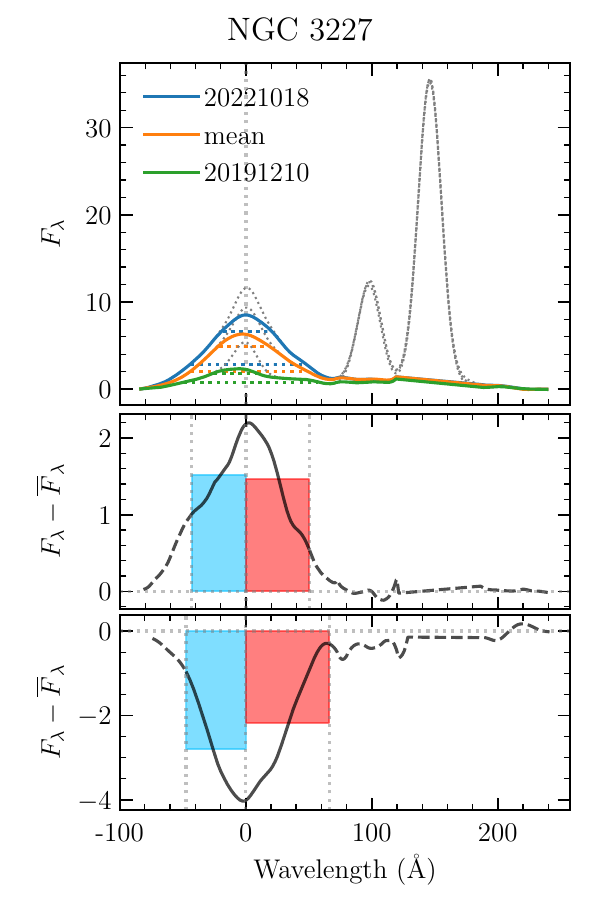}
    \includegraphics[width=0.23\textwidth]{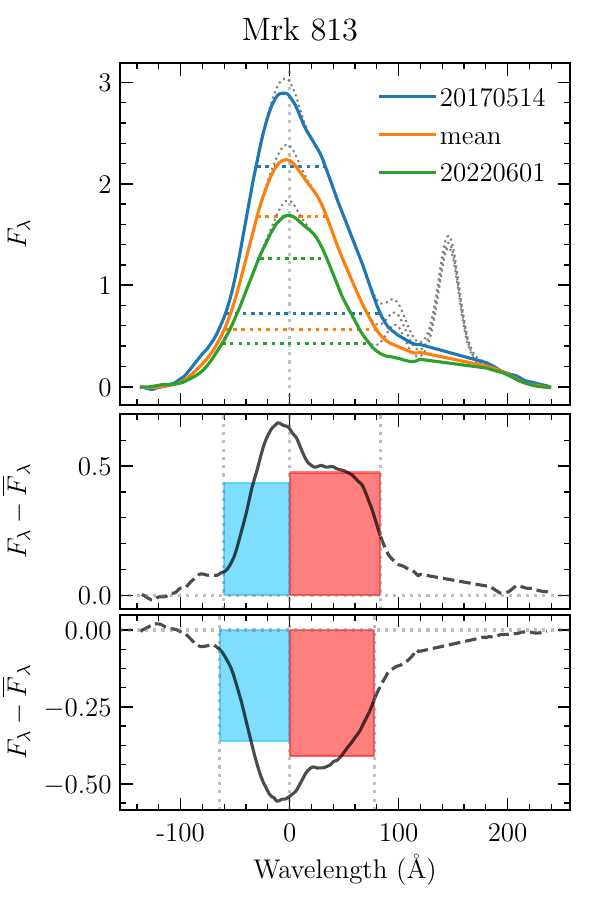}
    \includegraphics[width=0.23\textwidth]{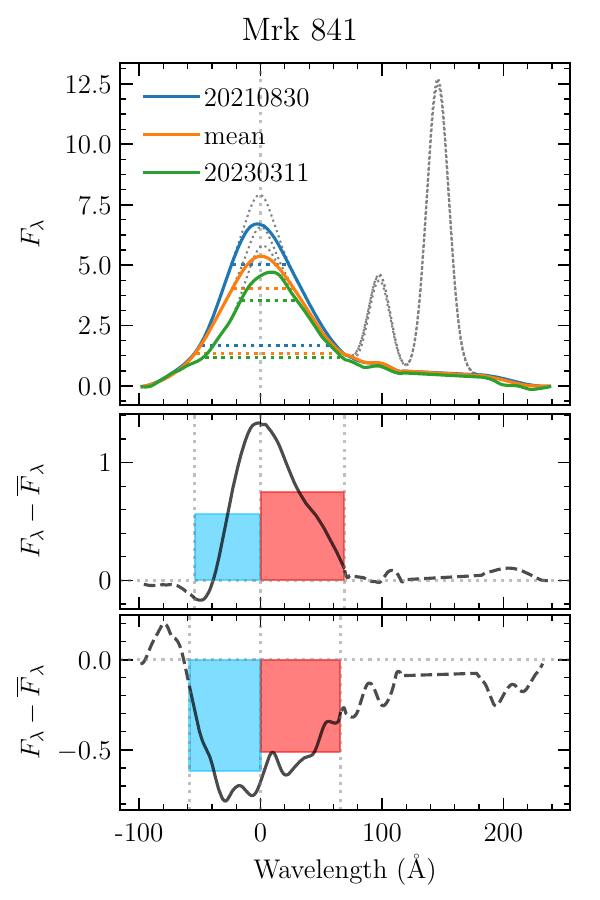}
    \includegraphics[width=0.23\textwidth]{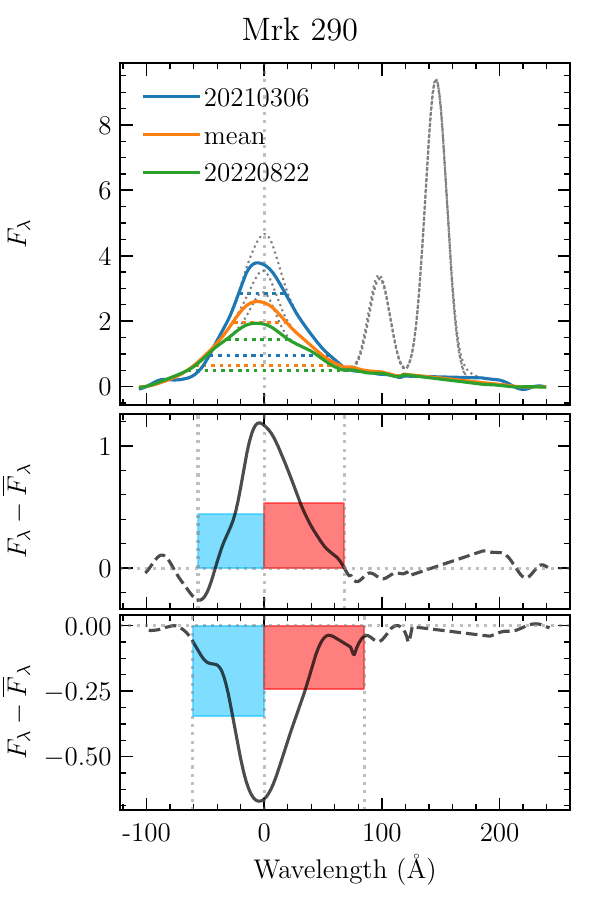}
    \includegraphics[width=0.23\textwidth]{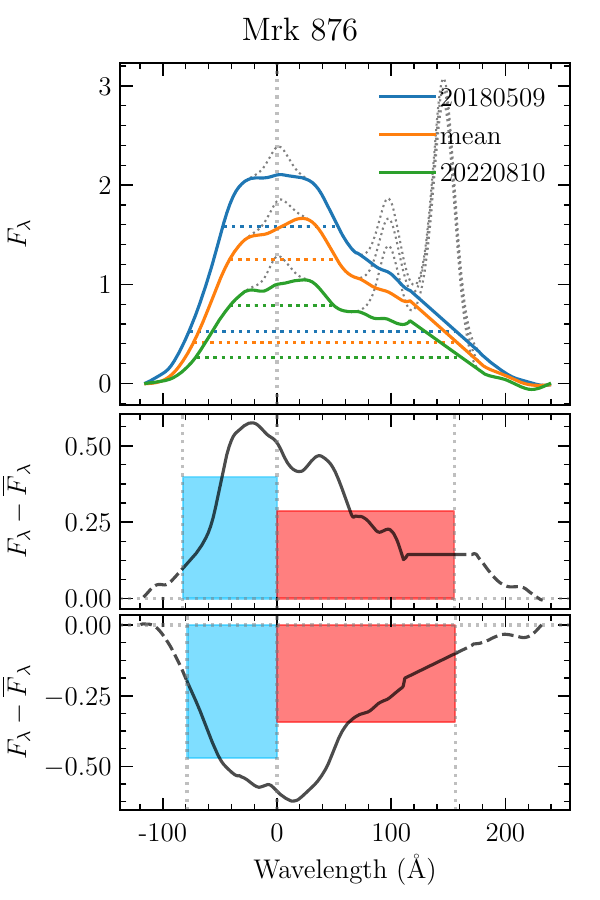}
  \caption{Examples of asymmetry measurement and comparisons between the line profiles in bright (blue solid lines), faint (green solid lines), and mean spectra (orange solid lines). The units are $\rm 10^{-15}\,erg\,s^{-1}\,cm^{-2}\,$\AA$^{-1}$. The horizontal colored dotted lines mark the 3/4 and 1/4 levels of the peak intensity. The black dotted lines represent the spectra before the narrow-line subtraction. The middle panels are the residuals and subtracting the mean spectra from the bright spectra, while the bottom panels are those from the faint spectra. The blue and red bars indicate the average flux in the corresponding ranges. The residuals were calculated after aligning the profiles according to their velocity shift $\Delta v$. }
  \label{fig:spectra}
\end{figure*}

As examples, we present the broad H$\beta$ profiles in the brightest, faintest, and the mean spectra of each object in Figure \ref{fig:spectra}. The mean spectrum is defined by 
\begin{equation}
    {\bar{F}}_{\lambda} =\frac{1}{N}\sum_{i=1}^NF_{\lambda}^i,
\end{equation}
where $N$ represents the number of spectra for an individual object, and $F_{\lambda}^i$ denotes the spectrum on the $i$-th day. In the lower panels of each object in Figure \ref{fig:spectra}, the residual after subtracting the mean spectra from the brightest and faintest spectra are shown. The blue and red bars mark the average intensities within their corresponding wavelength ranges. If the blue (or red) bar is higher than the red (or blue) one, it indicates the brightest/faintest profile is bluer (or redder).

The Galactic reddening was corrected for each spectrum prior to measuring line profiles and light curves (see Section \ref{sec:light curves}), by using the extinction law in \cite{Cardelli1989} and $A_{\mathrm{v}}$ values sourced from \cite{Schlafly2011}. In the cases of NGC~3227 and Mrk~79, the determination of $\lambda_c(1/4)$ is consistently affected by the uncertainty in the subtraction of the \OIII\ line. Consequently, we elected to disregard the lowest 10\% of the profiles and calculated the $A$ parameter based on the intensity levels at 32.5\%, 55.0\%, and 77.5\%, rather than the conventional 1/4, 1/2, and 3/4. 

The uncertainties were estimated through a bootstrap method as follows: (1) resample $N$ points (with repetition) from each of the spectra before the \OIII-based calibration to construct resampled spectra; (2) perform the \OIII-based calibration and combine the spectra at the same night; (3) subtract the continuum and narrow lines, and measure $A$ of the broad H$\beta$. This process was repeated 500 times. Finally, we determined the measured asymmetries as the median values of the distributions, with 1$\sigma$ corresponding to the uncertainties, defined by the 15.9\% and 84.1\% quantiles. The variations of the H$\beta$ asymmetry are illustrated in Figure \ref{fig:jd_Mrk1148}.

\subsection{Width and Shift of H\texorpdfstring{$\beta$}{beta} Line}\label{sec:FWHM}
Besides the asymmetry $A$, we also measure the FWHM and dispersion \sigmaline\ of H$\beta$ to quantify the width variation of its profile. The line dispersion is defined as
\begin{equation}
\sigma_{\rm line}^{2}(\lambda)= \left\langle \lambda^2 \right\rangle -\lambda_0^2=\frac{ \int \lambda^2 F_{\lambda} d\lambda }{ \int F_{\lambda} d\lambda} - \lambda^2_0,
\label{equ:sigmaline}
\end{equation}
where
\begin{equation}
\lambda_0 = \frac{ \int \lambda F_{\lambda} d\lambda }{ \int F_{\lambda} d\lambda}
\end{equation}
is the centroid wavelength, and $F_{\lambda}$ is the flux density \citep{peterson_central_2004}. The integrated window of \sigmaline\ is from 4760 to 4980\,\AA\ in the rest frame. We removed the contributions of the line spread function, which arise from the instrument, seeing, and artificial broadening during the \OIII-calibration, in a quadratic manner. This line spread function (approximately $1200{\rm~ km~s^{-1}}$) was estimated by comparing the observed widths of the $\OIII\lambda5007$ lines in the objects with their intrinsic values as reported in \cite{Whittle1992}. The uncertainties were estimated via a similar bootstrap method in Section \ref{sec:Asymmetry}. The variations of the FWHM and \sigmaline\ for all the objects are presented in Figure \ref{fig:jd_Mrk1148}. 

It is important to note that the parameter $A$ is independent of systematic shifts in the emission line. To investigate the velocity shifts of the emission lines during the campaign, we calculated the velocity shift \deltav, defined as the difference between $\lambda_c(3/4)$ and 4861\,\AA, for all spectra of the 8 objects. The velocity shifts \deltav\ are also shown in Figure \ref{fig:jd_Mrk1148}.

The $A$ parameters, line widths, and line shifts obtained from the Asiago data are not included in Figures \ref{fig:jd_Mrk1148} to \ref{fig:ratio1} due to their relatively wider line spread functions compared to the WIRO and SAAO data. Furthermore, several significant outliers resulting from poor weather conditions and low signal-to-noise ratios have been excluded from the dataset ($\sim$2\%--5\% spectra for different objects from WIRO and 7\% of the spectra of Mrk~841 from SAAO).

\subsection{Light Curves}\label{sec:light curves}
To evaluate the asymmetry variations in relation to the variability of the continuum and H$\beta$ emission lines, we measure the continuum and emission-line light curves from the spectra. While some portions of the light curves for the objects were previously reported in Papers \citetalias{du_monitoring_2018}-\citetalias{Zastrocky2024}, we extend our analysis to include additional light curves not covered in those publications. The methodology for measuring the light curves has been outlined in Papers \citetalias{du_monitoring_2018}-\citetalias{Zastrocky2024}, but is briefly summarized here for clarity and completeness.

The continuum flux is calculated by averaging the flux between 5075 and 5125\,\AA\ in the rest frame. H$\beta$ fluxes were determined through integration after subtracting a linear continuum background. The ranges for the local continuum and H$\beta$ flux windows were defined based on the signals in the root-mean-square (rms) spectra. These windows are exactly the same as those utilized in Papers \citetalias{du_monitoring_2018}-\citetalias{Zastrocky2024}, except for NGC~3227 (a broader window ranging from 4788 to 4928\,\AA\ is used for the H$\beta$ flux measurements). Similar to the approach in Papers \citetalias{du_monitoring_2018}-\citetalias{Zastrocky2024}, the differences between consecutive nights were assessed using a 5-point median filter to account for systematic uncertainties arising from varying weather conditions and the tracking inaccuracy of the telescope, and were incorporated into the error bars through quadratic summation. The continuum and H$\beta$ light curves are provided in Figure \ref{fig:jd_Mrk1148}. 

Given the different apertures in the photometric and spectroscopic observations conducted by various telescopes, it is necessary to perform intercalibration prior to merging the photometric light curves with the spectroscopic 5100\,\AA\ light curves. We combined the ZTF data (in g and r bands), ASAS-SN data (in g and V bands), and CRTS data (without filter) with the spectroscopic 5100\,\AA\ light curves using the PyCALI code\footnote{https://github.com/LiyrAstroph/PyCALI} \citep{Li2014_pyclai}. PyCALI employs a damped random walk process to model the variations and utilizes a Bayesian framework to determine the optimal parameters (scale and shift factors) by exploring the posterior probability distribution with the diffusive nested sampling algorithm \citep{Brewer2011_DNS}. The combined light curves after the intercalibration are shown in Figure \ref{fig:jd_Mrk1148}.

In addition to H$\beta$, we can also measure the flux of H$\alpha$ lines for five objects (IRAS~05589+2828, Mrk~79, NGC~3227, Mrk~841, and Mrk~290). The flux of their broad H$\alpha$ was derived by integrating the flux between 6430 and 6680\,\AA\ (in the rest frame) after subtracting the local linear continuum decided by two windows, 6075-6125\,\AA\ and 6760-6770\,\AA. Subsequently, we removed the contribution from the narrow H$\alpha$ and the $\NII\lambda\lambda6548,6583$ emission lines, whose flux ratios were fixed to the values obtained by fitting higher resolution spectra in \cite{Ho1995}, \cite{York2000_SDSS}, and \cite{Gavazzi2013} through the spectral decomposition code DASpec\footnote{https://github.com/PuDu-Astro/DASpec} 
\citep{Du2024}. We did not find a high-resolution spectrum for IRAS~05589+2828, thus the narrow lines were obtained from its mean spectrum. The narrow-line-subtracted H$\alpha$ light curves are shown in Figure \ref{fig:jd_ha_hb}.

\subsection{Cross-correlation Function} \label{sec:ccf}
We employ the interpolated cross-correlation function (ICCF; \citealt{gaskell_line_1986, gaskell_accuracy_1987, White_Peterson_1994}) to estimate the time delays between the variabilities of the profiles ($A$, FWHM, line dispersion \sigmaline, \fwhmsigma, \deltav) and the continuum or emission-line light curves. The CCFs are shown in the right panels of Figure \ref{fig:jd_Mrk1148}. The blue and orange lines represent the time delays of the corresponding curves to the 5100\,\AA\ continuum, and H$\beta$ light curves, respectively. We adopt the centroids of the CCFs above 80\% of the peak correlation coefficients as the time-lag measurements. The uncertainties of the time lags are obtained through the cross-correlation centroid distributions (CCCDs) from the ``flux randomization/random subset sampling'' method \citep{peterson_uncertainties_1998, peterson_central_2004}.
We get the lags from the first significant peaks or troughs (with $r_{\rm max}>0.5$ or $r_{\rm max}<-0.5$) of the CCFs in the range of $t \gtrsim 0$, and provide their values in Table \ref{tab:ccf}. The CCCDs are shown in Figure \ref{fig:jd_Mrk1148}.

\section{Results} \label{sec:results}
\subsection{Variabilities of Line Asymmetry} \label{sec:profile}
From Figure \ref{fig:jd_Mrk1148}, it can be seen that for Mrk~1148, Mrk~79, NGC~3227, Mrk~841, Mrk~290, their emission line asymmetries exhibit clearly similar or identical structures and features to the H$\beta$ light curves on long-term or short-term (single-year) time scales, while for other targets, some opposite or weaker correlations are displayed. In order to explore and quantify the relationships between the asymmetry variabilities and the H$\beta$ light curves, we initially present their correlations in Figure \ref{fig:ratio_A_all}. The Pearson correlation coefficient and the null probabilities are marked in the upper-left corner of the panel for each object. 

As can be seen, four objects (Mrk~1148, Mrk~79, NGC~3227, Mrk~290) show strong positive correlations between parameter $A$ and H$\beta$ flux with Pearson coefficients larger than 0.5, i.e., the larger flux, the bluer profile, and the larger value of parameter $A$. Mrk~841 demonstrates a moderately positive correlation, with a Pearson coefficient of 0.34, yet this correlation is highly statistically significant with a null probability of $1.20 \times 10^{-8}$. IRAS~05589+2828 displays a complex behavior; $A$ is inversely correlated with the flux when $\log F_{\rm H\beta}$ exceeds approximately 2.65, and proportional when the H$\beta$ flux is below 2.65. In contrast, Mrk~813 shows a strong anti-correlation, while Mrk~876 does not exhibit a significant correlation.

Furthermore, we conduct the BCES regression for each object, assuming that 
\begin{equation}
A=\alpha_1 + \beta_1 \log {F}_{\mathrm{H}\beta},
\label{equ:A-Hb}
\end{equation}
where $\alpha_1$ and $\beta_1$ are two coefficients. The slope $\beta_1$ is provided in the upper-right corner of each panel in Figure \ref{fig:ratio_A_all}. Mrk~1148, Mrk~79, NGC~3227, and Mrk~841 have the largest $|\beta_1|$. Although Mrk~876 shows statistically no correlation, it seems that $|\beta_1|$ would be inversely correlated at the higher state when $\log F_{\rm H\beta}$ is roughly larger than 2.5.

\subsection{Variabilities of Line Width and Breathing Phenomenon} \label{sec:line breathing}
The correlation between the BEL time lag and the continuum flux (or the BEL flux) is commonly referred to as the ``line breathing'' phenomenon \citep[e.g.,][]{peterson2002, Korista_Goad_2004, Cackett2006, Netzer2010, Goad2014}, where the underlying photoionization physics suggests that the responsivity-weighted radius of the BLR increases with the continuum flux. If the BLR gas surrounding the SMBH is primarily governed by its gravitational potential, meaning that the kinematics of the BLR follow Keplerian or virialized motion, the velocity of the gas clouds should be higher at the inner radius and lower at the outer radius. Consequently, the velocity width and flux of the BELs exhibit an anti-correlation \citep{Wilhite2006, park2012, Barth2015, zhang2019, Wangshu2020, Fries2023}. Upon visual inspection of Figure \ref{fig:jd_Mrk1148}, the curves of the H$\beta$ FWHM and line dispersion for Mrk~1148, Mrk~79, NGC~3227, Mrk~290, Mrk~876 show a significant negative correlation with the H$\beta$ light curves, with nearly zero time delays (or slightly delayed compared to the H$\beta$ flux, but the delay is smaller than that with the continuum light curves). Therefore, utilizing H$\beta$ flux instead of the continuum flux is likely more suitable for investigating the ``line breathing'' phenomenon as it can circumvent the need to account for time lags. Similarly to the analysis of H$\beta$ asymmetry, we depict the correlations between the line widths and H$\beta$ flux for each object in Figure \ref{fig:ratio1}.

Assuming that the BLR velocity is $v \propto R^{-0.5}$ and the typical size of the BLR is $R_{\rm{BLR}} \propto L_{\rm H\beta}^{0.5}$ \citep{Kaspi2005, du_supermassive_2015}, the correlation between the line width and the flux of H$\beta$ is expected to be $\Delta \log W = -0.25 \Delta\log L_{\rm H\beta}$, where $W$ could represent FWHM or line dispersion. We conduct linear regression on the correlation between the line width and the H$\beta$ flux, assuming
\begin{equation}
\log W = \alpha_2 + \beta_2 \log {F}_{\mathrm{H}\beta}
\label{equ:FWHM-Hb}
\end{equation}
and report the slope $\beta_2$ in the upper-right corner of the panels of Figure \ref{fig:ratio1} where the null probability is higher than $10^{-3}$ (indicating significant correlations). Figure \ref{fig:ratio1} illustrates that 5 objects (Mrk~1148, Mrk~79, NGC~3227, Mrk~290, and Mrk~876) exhibit typical ``breathing'' behaviors. Conversely, Mrk~813 displays an ``anti-breathing'' pattern where its line widths are directly proportional to the line flux (and concurrently, its parameter $A$ behavior differs from that of other sources). The FWHM correlations of IRAS~05589+2828 are complex, demonstrating a positive relationship in the low state and a negative relationship in the high state. For Mrk~876, the FWHM shows a sharp increase in the high state, while remaining relatively saturated in the low state. The ``breathing'' phenomenon is still evident, although the slope $\beta_2$ is different in high and low states. This possibly indicates that its radius of H$\beta$ response expands rapidly during the high state and tends to stabilize during the low state. 

The \fwhmsigma\ ratio describes the profile of the emission line, with $\fwhmsigma=2.355$ (equivalent to 0.372 on a logarithmic scale) for a Gaussian profile, a higher value for a boxy profile, and a lower value for a peaky profile. In Figure \ref{fig:ratio1}, we also provide the correlations between \fwhmsigma\ and the H$\beta$ flux. The \fwhmsigma\ ratios of Mrk~79, Mrk~290, and Mrk~876 display a significant anti-correlation with their H$\beta$ flux, indicating that a brighter BLR/continuum leads to a peakier line profile. In contrast, Mrk~841 shows a positive correlation with the flux, suggesting that a brighter BLR/continuum results in a boxier profile. The behavior of the \fwhmsigma\ ratio of IRAS~05589+2828 mirrors that of its FWHM, while those of Mrk~1148, NGC~3227, and Mrk~813 do not exhibit a correlation with H$\beta$ flux. In addition, we present the velocity shift \deltav\ of the emission line, which is defined by the velocity at $\lambda_c(3/4)$, in Figure \ref{fig:ratio1}. For most of the objects, \deltav\ is not correlated with the H$\beta$ flux. Only Mrk~813 exhibits an inverse correlation between \deltav\ and the flux.

\section{Discussion} \label{sec:discussion} 
Based on Figures \ref{fig:jd_Mrk1148} and \ref{fig:ratio_A_all}, the majority of the objects (Mrk~1148, Mrk~79, NGC~3227, Mrk~841, and Mrk~290) exhibit a ``bluer-profile-when-brighter'' (BPWB) phenomenon. Two objects (Mrk~813 and Mrk~876) display a ``redder-profile-when-brighter'' (RPWB) trend, while one object (IRAS~05589+2828) demonstrates a hybrid behavior, showing a BPWB pattern in the lower state and an RPWB trend in the higher state. In the following qualitative discussion, we dive into the physical origin of these line-profile variations.

\subsection{Anisotropic Radiation of BLR Clouds}\label{secc:anisotropy}
The first natural idea for the physical driver of the different behaviors of BPWB and RPWB is the different BLR kinematics of these objects. Papers \citetalias{bao2022} and \citetalias{Zastrocky2024} demonstrated that the line asymmetry and the BLR kinematics derived from the velocity-resolved lags, although based on a limited sample, are not strongly correlated. This is because the AGNs with the same line asymmetry in this sample can be associated with either inflow, outflow, or Keplerian/virialized BLR kinematics. The correlation between the variations of the emission-line asymmetry and flux shows a similar behavior. Specifically, we list the BLR kinematics derived from Papers \citetalias{brotherton2020}, \citetalias{bao2022}, and \citetalias{Zastrocky2024} in Table \ref{tab:kinematics}. The BLRs of almost all the objects show inflowing kinematics or the features of Keplerian disk/virialized motion with some inflowing velocities. Mrk~1148 shows a combination of Keplerian disk/virialized motion with weak outflowing velocity. Mrk~290 presents velocity-resolved lags corresponding to complex BLR kinematics. Notably, there is no straightforward correspondence between the positive or negative value of $\beta_1$ and the presence of inflow or outflow within the BLR kinematics. Therefore, we speculate that the differing BLR kinematics in these objects alone do not suffice as the primary physical driver for the BPWB or RPWB phenomena.

The H$\beta$ lines of the objects in the present paper exhibit a tendency towards red asymmetry ($A$ is negative). A physical explanation for this line asymmetry could be the anisotropic radiation of the BLR clouds. One possible scenario for the red asymmetry is an expanding BLR, where the line emission from the clouds on the near side is weaker than that from the clouds on the far side. This could be a result of the photoionization physics, where the clouds are optically thick and most of the radiation is emitted from the irradiated face of the clouds \citep[e.g.,][]{Capriotti1979, Ferland1992}. Another possible scenario for the red asymmetry is an inflowing BLR, where there is weaker line emission from the clouds on the far side, possibly due to shielding effects from the clouds on the near side (also mentioned by \citealt{Capriotti1979}), from the disk (the extension of the accretion disk), or from the dust inside the BLR \citep[e.g.,][]{Reynolds1997_BDs, Osterbrock1989, Dong2008_BDs, Gaskell2017}. The BLR kinematics of the objects tend to be inflow dominated or have contributions from inflow velocities (see Table \ref{tab:kinematics}). Therefore, the second scenario could be more plausible. 

If the degree of radiation anisotropy varies in response to changes in the continuum and H$\beta$ fluxes, the asymmetry will adjust accordingly. For example, if the radiation anisotropy weakens as the continuum or H$\beta$ flux increases, we would expect the red asymmetry of the line profile to diminish, potentially revealing the BPWB phenomenon. The weakening of radiation anisotropy in response to rising continuum or H$\beta$ flux may occur under two conditions: (1) the shielding effect could diminish as the ionizing flux increases, due to a greater number of clouds becoming ionized or an increase in the ionization degree of the clouds, and (2) the dust within the BLR may sublimate as the ionizing flux intensifies. If either or both of these conditions are met, we can observe the BPWB phenomenon.

\subsection{Dust Extinction in BLR}\label{sec:hahb}
As discussed in Section \ref{secc:anisotropy}, if the radiation anisotropy of BLRs diminishes during the rise in continuum or H$\beta$ fluxes, we would observe the BPWB phenomenon. The sublimation of dust within the BLRs is likely a physical mechanism contributing to this effect. The Balmer decrement (BD) of broad hydrogen lines can serve as an indicator of dust extinction in BLRs \citep[e.g.,][]{Osterbrock1989, Reynolds1997_BDs, Dong2008_BDs, Gaskell2017}. The typical flux ratio of H$\alpha$ to H$\beta$ for BLRs without dust extinction is approximately 3, and possibly as low as 2.7, as determined from the broad emission lines in extremely blue AGNs \citep[e.g.,][]{Dong2008_BDs, Gaskell2017}. We plot the light curves of the broad H$\alpha$ emission lines within the observed wavelength ranges (IRAS~05589+2828, Mrk~79, NGC~3227, Mrk~841, and Mrk~290) and their correlations with H$\beta$ fluxes in Figure \ref{fig:jd_ha_hb}. We perform linear regression to the correlation between the H$\alpha$ and H$\beta$ fluxes as 
\begin{equation}
F_{\mathrm{H}\alpha}=\alpha_3 + \beta_3 F_{\mathrm{H}\beta},
\label{equ:Ha-Hb}
\end{equation}
and provide the values of $\beta_3$ in Figure \ref{fig:jd_ha_hb}. For the five objects from which we can measure the H$\alpha$ flux, all exhibit clear linear correlations between the H$\alpha$ and H$\beta$ fluxes. Among these, the slope $\beta_3$ for three objects (IRAS~05589+2828, NGC~3227, and Mrk~841) is significantly greater than 2.7, while it is close to or less than 2.7 for the remaining two objects (Mrk~79 and Mrk~290). 
Compared to the direct ratio of $\mathrm{H\alpha}$ flux to $\mathrm{H\beta}$ flux, the slope $\beta_3$ eliminates potential uncertainties arising from narrow line subtraction. 

The flux ratios of the narrow H$\alpha$ and H$\beta$ lines are 4.90, 3.16, 3.72, 3.14, and 3.05 for IRAS~05589+2828, Mrk~79, NGC~3227, Mrk~841, and Mrk~290, respectively. Here, the flux ratios of the narrow H$\alpha$ and H$\beta$ of the latter four objects are measured from the spectra with higher resolution in Sloan Digital Sky Survey \citep{York2000_SDSS}, or from \cite{Gavazzi2013} and \cite{Ho1995} (see also Section \ref{sec:light curves}). The narrow line ratio of IRAS~05589+2828 is measured from its mean spectrum. The higher value of the narrow $\rm H\alpha/H\beta$ in IRAS~05589+2828 is probably caused by extinction in its host galaxy. \cite{Mehdipour2021} suggested that NGC~3227 displays significant internal reddening, as indicated by both the shape of the optical-UV continuum and the flux ratios of its AGN emission lines. 

Only the higher Balmer decrement in the broad lines in Mrk~841 suggests the presence of dust in its BLR. Furthermore, the observed linear correlations between the broad H$\alpha$ and H$\beta$ (with $\beta_3$ remaining unchanged) imply that extinction does not vary with the increase in H$\beta$ flux, which does not strongly support the sublimation of dust in these objects. Therefore, the mechanism discussed in the last paragraph of Section \ref{secc:anisotropy} may be more plausible.

\subsection{Abnormal Behavior in ``Breathing'' Phenomenon}
In the present sample, most objects exhibit a normal ``breathing'' behavior, characterized by a narrowing of the H$\beta$ width as their flux increases. This normal phenomenon of ``breathing'' can be easily understood. If the BLR geometry and kinematics resemble a Keplerian disk or involve virialized motion, or if there is inflow with higher velocities at smaller radii and lower velocities at larger radii (the velocity-resolved lags indicate that their BLR geometry and kinematics show seldom outflow, see Table \ref{tab:kinematics}), the H$\beta$ width would decrease with an increase in continuum/H$\beta$ flux. 

However, there are two notable exceptions: IRAS~05589+2828 and Mrk~813. In these objects, the H$\beta$ FWHM first increases and then decreases with rising H$\beta$ flux, rendering their behavior relatively unusual compared to the other objects in the sample. A possible explanation for this anomaly is that their BLR inflow velocity has a more complex relationship to the distance to the SMBH. If the inflow velocity first increases and then decreases from the outer to the inner region of the BLR -- potentially due to stronger radiation pressure in the very inner region -- we would observe such anomalous ``breathing'' behavior. 

Additionally, these two objects also exhibit the relatively rare RPWB phenomenon in their H$\beta$ asymmetry variations, suggesting a possible connection between the unique behaviors observed in their asymmetry variation and the ``breathing'' effect. In the future, more detailed modeling is needed to simultaneously explain the H$\beta$ asymmetry and ``breathing'' phenomena.

\subsection{Potential Evidence for Radiation Pressure}
\label{sec:Long-term trend}
From Figure \ref{fig:jd_Mrk1148}, it is evident that the short-term variability features of the line-profile variations ($A$ and line widths) are correlated (Mrk~1148, Mrk~79, NGC~3227, Mrk~841, and Mrk~290) or anti-correlated (Mrk~813) with those of the H$\beta$ light curves (with no clear time delays). However, the behaviors of the long-term line-profile variations on timescales of years are distinctly different from the corresponding H$\beta$ curves. The origin of the different long-term trends between the line-profile variations and the H$\beta$ light curves is of significant interest.

Recent reverberation mapping (RM) data have shown that radiation pressure may play a crucial role in the geometry and kinematics of the BLR. The variations in time lags measured from individual-year light curves in NGC~5548 and NGC~4151 are both correlated with their long-term continuum light curves, with the time lags closely matching the dynamical timescales of their BLRs \citep{Lu2016, Lu2022, chen_broad-line_2023}. This provides potential evidence for the influence of radiation pressure on their BLRs, suggesting that radiation pressure may gradually alter the BLR's geometry or kinematics over extended timescales. Theoretical calculations also indicate that radiation pressure may influence the line profiles \citep[][]{Netzer2010}. Motivated by these findings, we investigate whether the different trends of the line-profile variations and the H$\beta$ light curves are a result of radiation pressure.

To explore this, we construct a model in the following form:
\begin{equation}
    M = c_1 F_{\mathrm{H}\beta} + c_2 F_{\mathrm{cont}}*G(\mu, \sigma) +c_3,
    \label{equ:model for radiation pressure}
\end{equation}
where $M$ is $A$ or $\rm FWHM$, $F_{\mathrm{H}\beta}$ and $F_{\mathrm{cont}}$ are the H$\beta$ and continuum flux, $G$ denotes a Gaussian function characterized by parameters $\mu$ and $\sigma$, $*$ indicates a convolution, and $c_1$, $c_2$ and $c_3$ are three coefficients. The first term corresponds to the short-term component that is correlated with the H$\beta$ light curve. The second term represents the influence of radiation pressure, which is delayed and broadened with respect to the long-term continuum variation (the broadening can ensure that this term primarily represents the long-term large-scale variation). We collect the continuum light curves as long as we can (including ASAS-SN, CRTS, and ZTF data) and tentatively obtain the optimal values of the parameters using the \texttt{emcee} algorithm in the Python package \texttt{lmfit}. The fitting results, along with the corresponding probability distributions of the parameters, are presented in Figure \ref{fig:longterm light curves}.

After including the contribution from the long-term variation of the continuum, the fitting results are generally satisfactory. The time shift of the Gaussian function $G$ can be compared with the typical timescales of their BLRs to assess the self-consistency of radiation pressure as the physical driver behind the long-term trends observed in the $A$ and line width variations. The dynamical timescale for a BLR can be estimated by \citep{peterson_reverberation_1993}:
\begin{equation}
    \tau_{\mathrm{dyn}}=\frac{c\tau_{\mathrm{H}\beta}}{V_{\mathrm{FWHM}}} = 1.4 \tau_{10}V^{-1}_{6000}\,\mathrm{yr},
    \label{equ:dynamical timescale}
\end{equation}
where $\tau_{10}=\tau_{\mathrm{H}\beta}/10\,$days and $V_{6000}=V_{\mathrm{FWHM}}/6000$\,km\,s$^{-1}$. On this timescale, the BLR geometry and kinematics may be stable. The thermal timescale, defined as the ratio of heating rate to dissipation rate, is expressed as:
\begin{equation}
    \tau_{\mathrm{th}}=\tau_{\mathrm{dyn}} / \alpha,
    \label{equ:thermal timescale}
\end{equation}
which is evidently longer than $\tau_{\mathrm{dyn}}$ with a viscosity parameter $\alpha\approx0.1$. The thermal timescale represents the time required for the BLR to reach thermal equilibrium \citep{Frank2002}. The recombination timescale, associated with local ionization equilibrium, is given by \citep{Osterbrock1989}:
\begin{equation}
    \tau_{\mathrm{rec}}=(n_{\rm e} \alpha_{\mathrm{B}})^{-1}=0.1n_{10}^{-1}\,\mathrm{hr},
    \label{equ:recombination timescale}
\end{equation}
where $n_{10} = n_{\rm e} / 10^{10}\,{\rm cm^{-3}}$ is the electron density and $\alpha_{\rm B}$ is the hydrogen case B recombination coefficient. For the BLR, the recombination process occurs so rapidly that it can be ignored when analyzing the long-term behavior of the light curves.

We compare the time shift of $G$ (denoted as $\tau_{A}$ for the $A$ variation and as $\tau_{\rm FWHM}$ for the FWHM variation) with the different timescales in Figure \ref{fig:timescale}. It is evident that $\tau_{A}$ and $\tau_{\rm FWHM}$ for the objects are generally consistent with the dynamical timescales. This consistency suggests that radiation pressure may contribute to the differing long-term trends observed in the $A$ and line width variations, in comparison to the H$\beta$ light curves. However, it is important to note that the total duration of the continuum light curves collected for the objects in this study is insufficient to adequately capture variations associated with the thermal timescales. Consequently, it would be difficult to entirely exclude alternative explanations, such as those related to thermal mechanisms, which warrant further discussion in future studies.

Therefore, the variations in $A$ and line widths likely originate from two distinct processes: (1) a component that is correlated with the H$\beta$ light curve (or delayed relative to the continuum light curve by a time lag corresponding to the light-travel time), which may result from changes in the illuminated region of the BLR or the process mentioned in Section \ref{secc:anisotropy}, and (2) a component that reflects variations in BLR geometry or kinematics induced by radiation pressure, occurring on dynamical timescales. We present the correlations between the asymmetry (and FWHM) after subtracting the term corresponding to $c_2$ (i.e., $M - c_2 F_{\mathrm{cont}}*G$) and the H$\beta$ flux in Figures \ref{fig:ratio_A_model} and \ref{fig:ratio_FWHM_model}. It is evident that the correlations between H$\beta$ flux and asymmetry $A$ for Mrk~841, Mrk~290, and Mrk~876 are significantly stronger than those depicted in Figure \ref{fig:ratio_A_all}, and also for every object between H$\beta$ flux and FWHM depicted in Figure \ref{fig:ratio1} (e.g., Mrk~813 changes from an ``anti-breathing'' pattern to a normal ``breathing'' pattern), strongly suggesting the potential validity of decomposing these two processes.

\subsection{Peculiar BLR Structures} \label{sec:Special BLR models}
Anisotropy in the BLR may arise from irradiation by a warped accretion disk \citep{Pringle1996, Li2016ApJ_NGC5548, Jiangbowei2021}, hot spots within the accretion disk \citep[e.g.,][]{Newman1997, Jovanovic2010}, or peculiar structures in the BLR, such as spiral arms \citep[e.g.,][]{Gilbert1999, Storchi-Bergmann2003, Schimoia2012, Wang2022_spiral, Du2023}, as well as hot spots in the BLR \citep{Fries2023}. These factors may contribute to the abnormal behaviors observed in the ``breathing'' phenomenon and the asymmetry variations in IRAS~05589+2828 and Mrk~813. While detailed modeling is beyond the scope of the present paper, such analyses will be essential in the future to explore the implications of these models.

\subsection{Notes for Individual Objects} \label{sec:notes}
\subsubsection{Mrk~1148}
In Figure \ref{fig:jd_Mrk1148}, the profile of the broad H$\beta$ in Mrk~1148 is nearly symmetric ($A\sim0$) during the first year ($A$ is similar to $-0.024$ in \citealt{robertis1985} and $-0.0047$ in \citealt{boroson1992}). Subsequently, the profile becomes progressively redder, but exhibiting a peak similar to that observed in the continuum or H$\beta$ light curve around 2020-2021. The time lag measured between $A$ and the continuum light curve is larger than that between H$\beta$ and the continuum light curves, which may be attributed to a component driven by the radiation pressure (see Section \ref{sec:Long-term trend}). Additionally, the FWHM variation of the H$\beta$ closely follows the continuum light curve, with a time lag of approximately 30$\,$days, approaching the measured time lag of $22.6^{+2.7}_{-5.2}$ days between the H$\beta$ and the continuum in Paper \citetalias{bao2022}. 

In the third year, the FWHM dropped sharply from 5,000 to 4,000$\rm \, km\,s^{-1}$, while the \sigmaline\ experienced a more modest decline. Consequently, the ratio \fwhmsigma\ also exhibited a clear decrease during this year. Additionally, the velocity shift \deltav\ remains positive but shows fluctuations. 

\subsubsection{IRAS~05589+2828}
The most prominent feature of the continuum and H$\beta$ light curves is a valley-like shape, with a notable drop occurring in the third year. The variation in $A$ for IRAS~05589+2828 does not closely follow its light curves, although it exhibits some similar features. The long-term trend of the $A$ curve is more comparable to that of the \fwhmsigma\ variation. The FWHM decreases from nearly 6,000\,km\,s$^{-1}$ to 5,000\,km\,s$^{-1}$ and then moderately increases to 6,000\,km\,s$^{-1}$ again over the last three years, while \sigmaline\ varies in a sinusoidal-like manner over the five years. Additionally, the blue shift of H$\beta$ gradually decreases during this period.

In Figure \ref{fig:ratio_A_all}, the correlation between H$\beta$ asymmetry and flux is positive in the low state but negative in the high state. We conduct separate linear regression analyses for these two segments, presenting the results in Figure \ref{fig:ratio_A_all}. The slope $\beta_1$ is 0.72 in the low state and $-0.93$ in the high state, suggesting that there may be additional mechanisms at work beyond the anisotropic radiation processes discussed in Section \ref{secc:anisotropy}. 

\subsubsection{Mrk~79}
This object has been previously monitored by \cite{peterson_uncertainties_1998}, Paper \citetalias{brotherton2020}, and \cite{Lukaixing2019b}. In Paper \citetalias{brotherton2020}, the velocity-delay map reveals that the geometry and kinematics of its BLR are complex, featuring both a disk with inflowing velocities and a disk with outflowing velocities, based on observations from 2016-2017 (corresponding to the first year in the present study). The long-term trend of $A$ generally aligns with the rising trends of both the continuum and H$\beta$ light curves. The correlations between $A$ and H$\beta$ flux, as well as between line widths and flux, are consistent with BPWB and the normal ``breathing'' phenomenon.

\subsubsection{NGC~3227}
In the first year, the line profile is generally symmetric or slightly blue-asymmetric, which aligns with the $A$ value reported by \cite{robertis1985} ($A = 0.046$). \cite{denney2010} presented a mildly red asymmetric profile in the mean spectrum and a double-peaked profile in the rms spectrum for this object in 2007, with its velocity-resolved time lags exhibiting an outflow signature. In contrast, \cite{De_Rosa2018} observed a blue asymmetric H$\beta$ profile and an inflow signature in 2012 and 2014, characterized by slightly higher luminosity and shorter time lags compared to those recorded in 2007. Consequently, the profiles from these two campaigns also exhibit the BPWB phenomenon, consistent with the behavior of $A$ shown in Figure \ref{fig:ratio_A_all}.

The \fwhmsigma\ values remain below 2.355 throughout the campaign. The light curve of \sigmaline\ is nearly parallel to the FWHM, with the exception of 2021, when \sigmaline\ is elevated relative to the trend of the H$\beta$ light curve. The variation in $A$ for this object generally follows the H$\beta$ light curve up until 2023; however, a divergence occurs post-2023, contributing scatter to the correlation shown in Figure \ref{fig:ratio_A_all}.

\subsubsection{Mrk~813}
Mrk~813 exhibits RPWB and anti-breathing behaviors, which distinguish it from most other objects studied in this work. The profiles show red asymmetries and remain largely unchanged during the first two years, not following the variations in the light curves. The long-term rising trend of its $A$ variation correlates with the continuum variation, with a time lag of 6.9 to 7.7 years (see Figure \ref{fig:longterm light curve8}). After incorporating the long-term contribution (attributed to radiation pressure) discussed in Section \ref{sec:Long-term trend}, the variation of FWHM can be well fitted. When the contribution from the long-term component is removed, there is almost no correlation between the residuals and the H$\beta$ flux (see Figure \ref{fig:ratio_FWHM_model}), indicating that the variation in the line profile is likely dominated by the influence of radiation pressure. 

\subsubsection{Mrk~841}
The asymmetry $A$ for this object is consistent with the value reported by \cite{robertis1985} ($A=-0.054$). It exhibits the BPWB phenomenon, particularly in the short-term variation features (see Figure \ref{fig:jd_Mrk841}). However, the H$\beta$ flux shows a gradual decrease over the last three years, which is distinctly different from the short-term BPWB behavior observed in $A$.

The FWHM curve displays a clear structure: it remains constant during the first two years (before 2019), then gradually narrows, and subsequently broadens again in 2023. The long-term trend of the velocity shift $\Delta v$ also exhibits significant changes; it decreases prior to 2019 and then begins to rise. From the long-term continuum light curves presented in Figure \ref{fig:longterm light curve6}, we can easily identify a similar long-term trend corresponding to the variation in $A$ and H$\beta$ line width.

\subsubsection{Mrk~290}
The H$\beta$ profile of this object exhibited blue asymmetry in \cite{boroson1992} ($A=0.044$) but displayed red asymmetries in our campaign, alongside the BPWB phenomenon. The dispersions in the correlations between the asymmetry $A$ and H$\beta$ flux, as well as between FWHM and H$\beta$ flux, shown in Figures \ref{fig:ratio_A_all} and \ref{fig:ratio1}, can be significantly reduced by removing the contributions from the long-term component driven by radiation pressure. This suggests that radiation pressure plays a crucial role in shaping the H$\beta$ line profile.

\subsubsection{Mrk~876}
The overall line asymmetry of Mrk~876 during our campaign is red, consistent with previous studies \citep{robertis1985, boroson1992, erkens1995monitoring, kaspi_reverberation_2000, Kaspi2005}. The asymmetry gradually becomes bluer before 2020, after which it fluctuates, exhibiting both redder and bluer features starting in 2022. Notably, the H$\beta$ asymmetry and flux are clearly anti-correlated.

The FWHM of Mrk~876 ranges from 8,000 to 12,000$\rm \, km\,s^{-1}$ and shows an anti-correlation with the H$\beta$ flux, with a time lag of approximately 56 days. Considering the time lag between the continuum and H$\beta$ light curves is $48.3^{+5.0}_{-3.8}$ days (as discussed in Paper \citetalias{bao2022}), this suggests that the response time of the FWHM of broad H$\beta$ to its flux is twice the lag between H$\beta$ FWHM and the continuum flux. Additionally, the lag of \sigmaline\ appears to be longer, delayed with respect to the H$\beta$ flux (see Figure \ref{fig:jd_Mrk876}).

\section{Summary} \label{sec:Summary}
In this fifth paper of the series, we present a comprehensive analysis of the temporal variations in the asymmetry and width of the H$\beta$ line profiles in eight objects, utilizing high-cadence and long-duration RM data obtained from the MAHA project. The key findings of this study are summarized as follows:

\begin{enumerate}
\item We report that five objects (Mrk~1148, Mrk~79, NGC~3227, Mrk~841, and Mrk~290) exhibit a strong correlation between H$\beta$ asymmetry and H$\beta$ flux, indicating their BPWB behavior. The asymmetry of Mrk~813 shows a strong anti-correlation with H$\beta$ flux. IRAS~05589+2828 demonstrates a complex relationship between H$\beta$ asymmetry and flux, exhibiting a positive correlation in the low state and a negative correlation in the high state. Mrk~876 does not show a significant correlation (or weak correlation). 

\item The typical ``breathing'' phenomenon is observed for the broad H$\beta$ emission line in most objects, while a more complex or ``anti-breathing'' behavior is exhibited in IRAS~05589+2828 and Mrk~813. In the case of Mrk~841, the ``breathing'' phenomenon is relatively weak. The temporal evolution of H$\beta$ velocity shifts is also provided for the objects.

\item We provide some discussion regarding the physical origins of the observed BPWB and RPWB phenomena. We speculate that variations in the anisotropy of the BLRs could be a contributing factor. Additionally, we investigate the potential role of radiation pressure in the temporal variations of line asymmetry and width.
\end{enumerate}

\section*{ACKNOWLEDGEMENTS}
We thank WIRO engineers James Weger, Conrad Vogel, and Andrew Hudson for their indispensable and invaluable assistance. We thank the staff of the SAAO 1.9\,m telescope and the Asiago 1.82\,m telescope for their invaluable support. We thank the South African Astronomical Observatory for the allocation of telescope time, and Francois van Wyk for obtaining some of the spectra. We acknowledge the Copernico telescope (Asiago, Italy) of the INAF-Osservatorio Astronomico di Padova, for part of the observations in this research. 
This research is supported by National Key R\&D Program of China (2023YFA1607903 and 2021YFA1600404); by NSFC-11991051, -11991054, -12273041, -12022301, -12333003; by the Key Research Program of Frontier Sciences, CAS, grant QYZDJ-SSW-SLH007
and by the China Manned Space Project with no. CMS-CSST-2021-A06. LCH was supported by the National Key R\&D Program of China (2022YFF0503401), the National Science Foundation of China (11991052, 12233001), and the China Manned Space Project (CMS-CSST-2021-A04, CMS-CSST-2021-A06). T.E.Z. was funded by Wyoming NASA Space Grant Consortium, NASA Grant \#80NSSC20M0113. 

This research also makes use of the public data from ZTF, ASAS-SN, and CRTS. Part of the Zwicky Transient Facility project is observed from the Samuel Oschin 48-inch Telescope at the Palomar Observatory. ZTF is supported by the National Science Foundation under grant No. AST-1440341 and involves a collaboration with Caltech, IPAC, the Weizmann Institute for Science, the Oskar Klein Center at Stockholm University, the University of Maryland, the University of Washington, Deutsches Elektronen-Synchrotron and Humboldt University, Los Alamos National Laboratories, the TANGO Consortium of Taiwan, the University of Wisconsin at Milwaukee, and Lawrence Berkeley National Laboratories. Operations are conducted by COO, IPAC, and UW.
ASAS-SN is supported by the Gordon and Betty Moore Foundation through grant GBMF5490 to the Ohio State University, and NSF grants AST-1515927 and AST-1908570. Development of ASAS-SN has been supported by NSF grant AST-0908816, the Mt. Cuba Astronomical Foundation, the Center for Cosmology and AstroParticle Physics at the Ohio State University, the Chinese Academy of Sciences South America Center for Astronomy (CASSACA), the Villum Foundation, and George Skestos. This work makes use of the public data from the ASAS-SN project. 
The CSS survey is funded by the National Aeronautics and Space Administration under Grant No. NNG05GF22G issued through the Science Mission Directorate Near-Earth Objects Observations Program. The CRTS survey is supported by the U.S. National Science Foundation under grants AST-0909182 and AST-1313422.

\software{PyCALI \citep{Li2014_pyclai}, DASpec \citep{Du2024}. }

\begin{deluxetable*}{lcccccccc}[ht]
\tablecaption{Basic Information and Durations of the Objects \label{tab:objects}}
\tablewidth{0pt}
\tablehead{ \colhead{Name} & \colhead{Other Names} & \colhead{RA} & \colhead{DEC} & \colhead{Redshift} & \colhead{$N_{\rm{spec}}$} & \colhead{Previous} & \colhead{New} & \colhead{Telescope} }
\startdata
Mrk~1148 & PG 0049+171 & 00:51:54.7 & $+$17:25:59 & 0.0645 & 217   &  2017.10–2021.02 & 2021.08-2023.02 & WIRO \\ 
IRAS~05589+2828 &   & 06:02:10.7 & $+$28:28:22 & 0.0294 & 307   &  2018.09-2023.05 &  & WIRO \\ 
Mrk~79 &   & 07:42:32.8 & $+$49:48:35 & 0.0222 & 432   & 2016.12-2017.05 & 2019.09-2023.05 & WIRO \\ 
NGC~3227 &   & 10:23:30.6 & $+$19:51:54 & 0.0039 & 297 &  2016.12-2017.05 & 2019.11-2023.05 & WIRO \\ 
Mrk~813 &   & 14:27:25.1 & $+$19:49:52 & 0.1099 & 97 & 2016.12-2023.05 &  & WIRO \\ 
Mrk~841 & PG 1501+106 & 15:04:01.2 & $+$10:26:16 & 0.0364 & 239  & 2016.12–2020.06 & 2020.12-2023.05 & WIRO \\ 
 &  &  &  &  & 28  & 2019.06-2020.08 & 2021.07-2023.08 & SAAO \\ 
 &  &  &  &  & 5  & 2019.01-2020.03 &  & Asiago \\ 
Mrk~290 & PG 1534+580 & 15:35:52.3 & $+$57:54:09 & 0.0302 & 250   & 2020.02–2021.05 & 2021.08-2023.05 & WIRO \\ 
Mrk~876 & PG 1613+658 & 16:13:57.1 & $+$65:43:10 & 0.1211 & 300   & 2016.12–2021.04 & 2021.05-2023.05 & WIRO \\ 
\enddata
\tablecomments{$N_{\rm{spec}}$ is the number of spectroscopic epochs. RA and DEC are for the J2000 epoch.}
\end{deluxetable*} 
%%%%%%%%%%%%%%%%%%%%%%%%%%%
\begin{deluxetable*}{lcc}[ht]\label{tab:measurement windows}
\tablecaption{Measurement Windows for Asymmetry $A$ in the Rest-frame }
\tablewidth{0pt}
\tablehead{ \colhead{Name} & continuum (blue) (\AA) & continuum (red) (\AA)}
\startdata
Mrk~1148 & 4580-4595 & 5075-5125 \\
IRAS~05589+2828 & 4727-4733 & 5075-5125 \\
Mrk~79 & 4760-4780 & 5075-5125 \\
NGC~3227 & 4765-4788 & 5075-5125 \\
Mrk~813 & 4710-4740 & 5075-5125 \\
Mrk~841 & 4750-4775 & 5075-5125 \\
Mrk~290 & 4740-4770 & 5075-5125 \\
Mrk~876 & 4730-4760 & 5075-5125
\enddata
\end{deluxetable*}

%%%%%%%%%%%%%%%%%%%%%%%%%%%
\begin{deluxetable*}{llll}[ht]
\tablecaption{BLR Kinematics vs. H$\beta$ Profile Variability  \label{tab:kinematics}}\tablewidth{0pt}
\tablehead{\colhead{Name} & \colhead{BLR kinematics} & \colhead{$\beta_{1}$} & \colhead{Ref.}} 
%\decimalcolnumbers
\startdata
Mrk~1148        &  Keplerian Disk/Virialized Motion + Weak Outflow   & $0.78$  & Paper \citetalias{bao2022}        \\ 
IRAS~05589+2828 &  Keplerian Disk/Virialized Motion + Weak Inflow    & $0.72,-0.93$  & Paper \citetalias{Zastrocky2024}  \\ 
Mrk~79          &  Disk (+ outflow velocity) + Inflow                & $0.79$  & Paper \citetalias{brotherton2020} \\ 
NGC~3227        &  Disk + Inflow                                     & $0.69$  & Paper \citetalias{brotherton2020} \\ 
Mrk~813         &  Inflow                                            & $-0.60$ & Paper \citetalias{Zastrocky2024}  \\ 
Mrk~841         &  Inflow                                            & $0.72$  & Paper \citetalias{bao2022}        \\ 
Mrk~290         &  Complicated                                       & $0.51$  & Paper \citetalias{bao2022}        \\ 
Mrk~876         &  Inflow                                            & $-0.56$ & Paper \citetalias{bao2022}        \\ 
\enddata
\tablecomments{For IRAS~05589+2828, the two numbers of $\beta_1$ are the slopes for $\log F_{\rm H\beta}<2.65$ and $\log F_{\rm H\beta}>2.65$, respectively.}
\end{deluxetable*}
%%%%%%%%%%%%%%%%%%%%%%%%%%%
\begin{deluxetable*}{llccccc}[ht]
\tablecaption{Rest-frame Time Lags \label{tab:ccf}}\tablewidth{0pt}
\tablehead{ \colhead{Name} & & $A$ & FWHM & $\sigmaline$ & $\fwhmsigma$ & $\deltav$}
\startdata
Mrk\ 1148 & $F_{\mathrm{cont}}$& $...$& $71.1_{-13.0}^{+27.0}$& $...$& $66.1_{-6.1}^{+6.4}$& $...$\\
 & $F_{\mathrm{H}\beta}$  & $64.7_{-36.3}^{+44.1}$& $29.6_{-7.0}^{+15.3}$& $...$& $...$& $...$\\
IRAS\ 05589+2828 & $F_{\mathrm{cont}}$& $602.8_{-12.9}^{+16.1}$& $586.0_{-31.0}^{+31.5}$& $199.4_{-20.0}^{+12.6}$& $593.4_{-20.0}^{+19.3}$& $...$\\
 & $F_{\mathrm{H}\beta}$  & $716.4_{-15.2}^{+23.6}$& $588.0_{-88.6}^{+61.1}$& $200.2_{-35.5}^{+12.7}$& $587.3_{-50.4}^{+61.4}$& $736.4_{-9.3}^{+14.2}$\\
Mrk\ 79 & $F_{\mathrm{cont}}$& $147.2_{-20.7}^{+26.5}$& $41.9_{-17.2}^{+14.7}$& $18.8_{-2.7}^{+5.2}$& $...$& $...$\\
 & $F_{\mathrm{H}\beta}$  & $-4.4_{-5.2}^{+11.4}$& $11.2_{-5.4}^{+7.7}$& $-1.5_{-3.4}^{+5.2}$& $...$& $...$\\
NGC\ 3227 & $F_{\mathrm{cont}}$& $-0.7_{-5.4}^{+7.6}$& $...$& $5.9_{-1.4}^{+1.3}$& $...$& $7.4_{-2.9}^{+2.2}$\\
 & $F_{\mathrm{H}\beta}$  & $4.7_{-6.7}^{+4.3}$& $...$& $1.1_{-1.1}^{+1.0}$& $...$& $...$\\
Mrk\ 813 & $F_{\mathrm{cont}}$& $1204.5_{-26.1}^{+22.6}$& $1026.9_{-148.3}^{+74.2}$& $1551.8_{-425.1}^{+122.3}$& $...$& $1340.7_{-118.0}^{+63.8}$\\
 & $F_{\mathrm{H}\beta}$  & $266.7_{-47.2}^{+72.3}$& $...$& $315.0_{-31.9}^{+31.3}$& $...$& $205.5_{-16.0}^{+17.7}$\\
Mrk\ 841 & $F_{\mathrm{cont}}$& $...$& $...$& $...$& $2.9_{-5.6}^{+6.4}$& $...$\\
 & $F_{\mathrm{H}\beta}$  & $...$& $...$& $...$& $...$& $...$\\
Mrk\ 290 & $F_{\mathrm{cont}}$& $26.1_{-3.3}^{+3.2}$& $40.2_{-7.4}^{+5.3}$& $22.1_{-3.4}^{+3.3}$& $57.2_{-23.2}^{+10.4}$& $233.9_{-26.1}^{+28.2}$\\
 & $F_{\mathrm{H}\beta}$  & $3.1_{-4.3}^{+4.8}$& $8.0_{-4.5}^{+6.9}$& $-5.8_{-4.4}^{+4.6}$& $43.3_{-16.1}^{+10.3}$& $241.1_{-41.1}^{+26.5}$\\
Mrk\ 876 & $F_{\mathrm{cont}}$& $...$& $144.2_{-7.6}^{+6.8}$& $167.2_{-6.9}^{+6.4}$& $134.9_{-10.6}^{+9.4}$& $...$\\
 & $F_{\mathrm{H}\beta}$  & $...$& $55.9_{-7.8}^{+7.0}$& $88.2_{-8.1}^{+9.0}$& $41.9_{-7.8}^{+8.6}$& $...$\\
\enddata
\tablecomments{The time lags are in units of days.}
\end{deluxetable*}
%%%%%%%%%%%%%%%%%%%%%%%%%%%
%%%%%%%%%%%%%%%%%%%%%%%%%%%
\begin{figure*}[ht]
  \centering
  \includegraphics[width=1\textwidth]{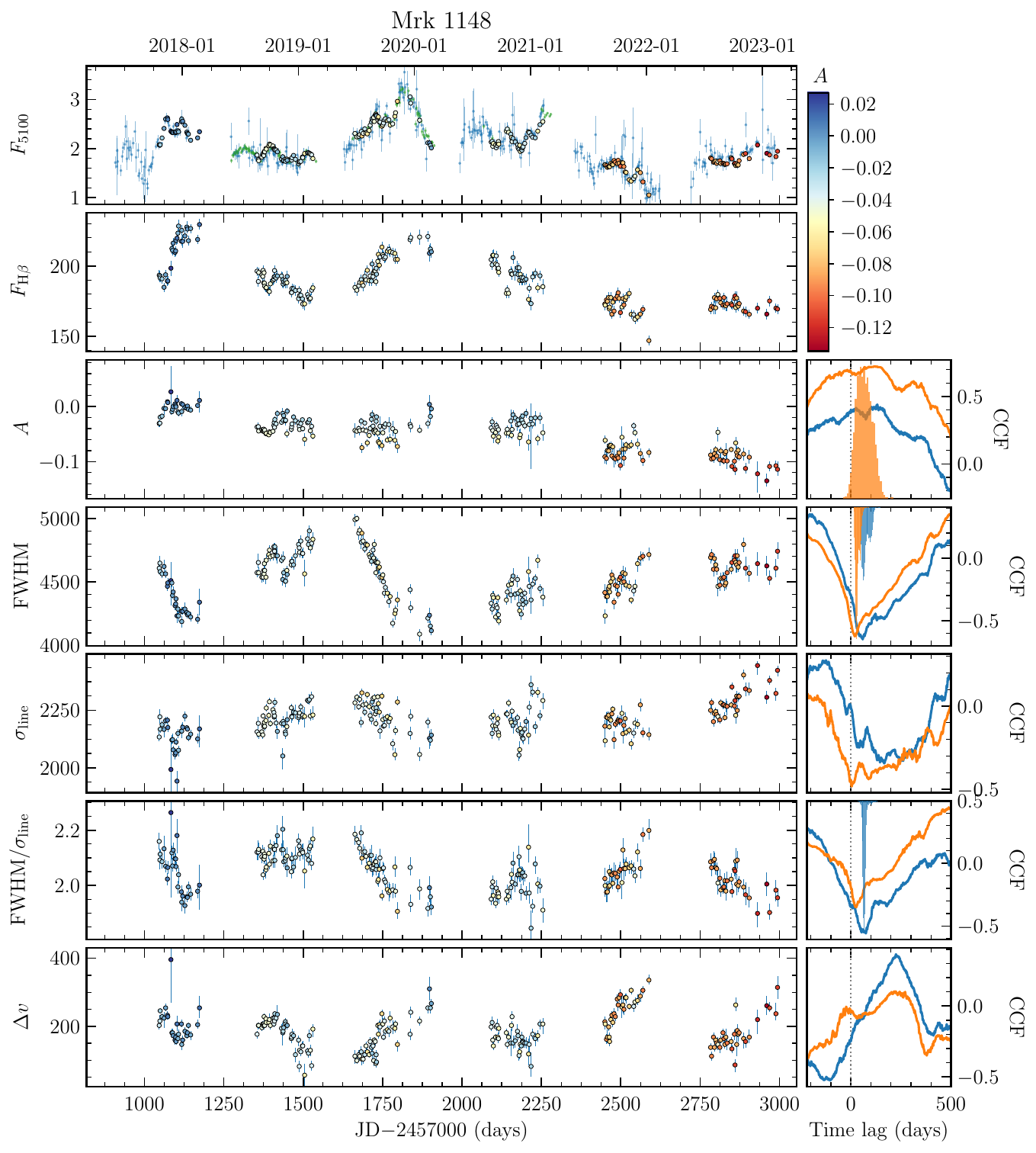}
  \caption{Continuum and H$\beta$ light curves, temporal variabilities of the H$\beta$ profiles, and CCF analyses. The left panels from top to bottom are the 5100\,\AA\ continuum, H$\beta$, asymmetry $A$, FWHM, line dispersion \sigmaline, \fwhmsigma, and \deltav, respectively. In the right panels, the colored lines represent the CCFs between the observables and the continuum (in blue) and the H$\beta$ (in orange) light curves, while the histograms are the corresponding time lag distributions. The 5100\,\AA\ continuum and H$\beta$ light curves are in units of $\rm 10^{-15}\,erg\,s^{-1}\,cm^{-2}\,$\AA$^{-1}$ and $\rm 10^{-15}\,erg\,s^{-1}\,cm^{-2}$, respectively. The FWHM, \sigmaline, and \deltav\ are in units of $\rm{km\,s^{-1}}$. }
  \label{fig:jd_Mrk1148}
\end{figure*}

\begin{figure*}
    \centering
    \includegraphics[width=1\textwidth]{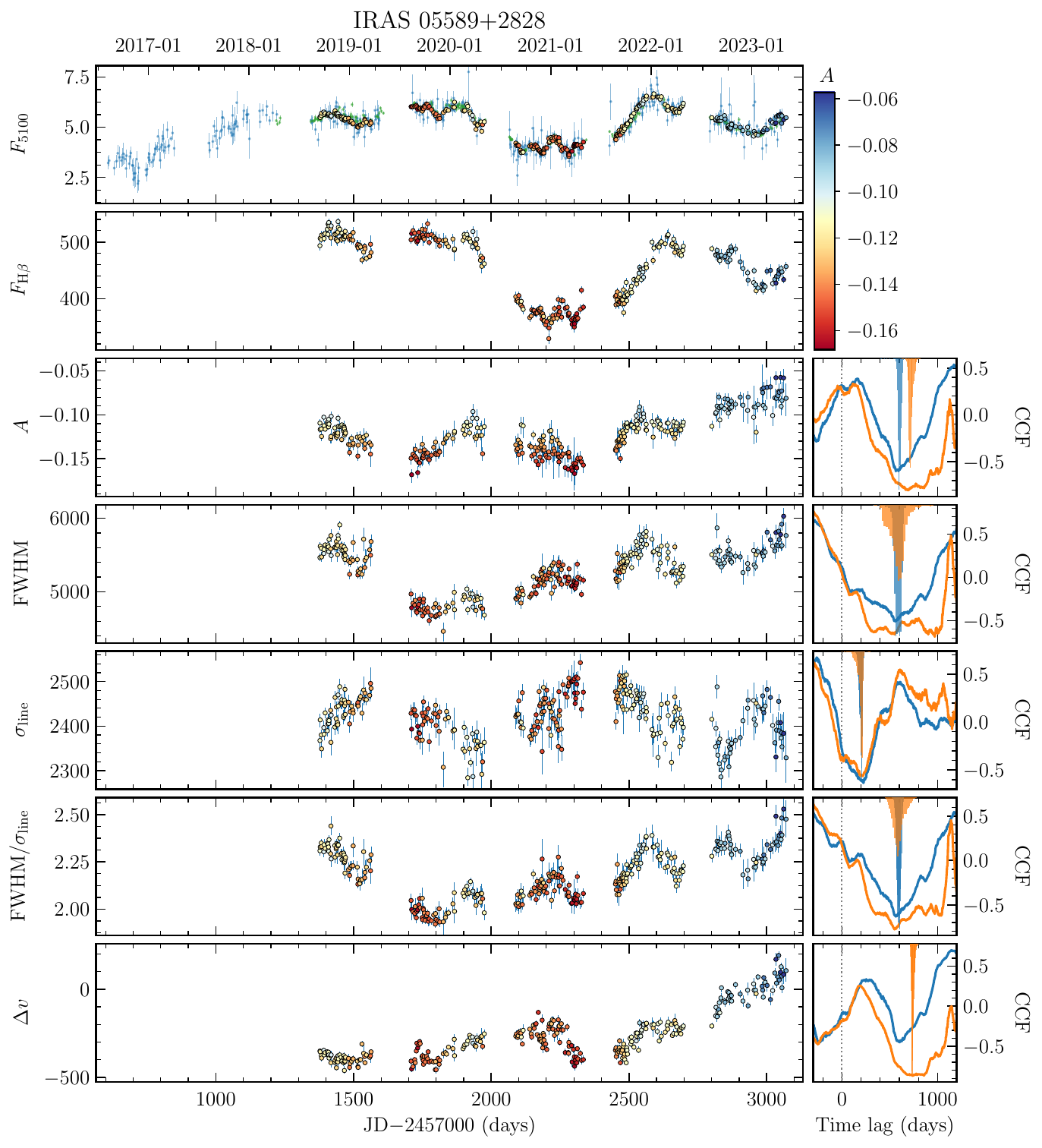}
    \addtocounter{figure}{-1}
    \caption{Continued.}
    \label{fig:jd_IRAS05589+2828}
\end{figure*}
\begin{figure*}
    \centering
    \includegraphics[width=1\textwidth]{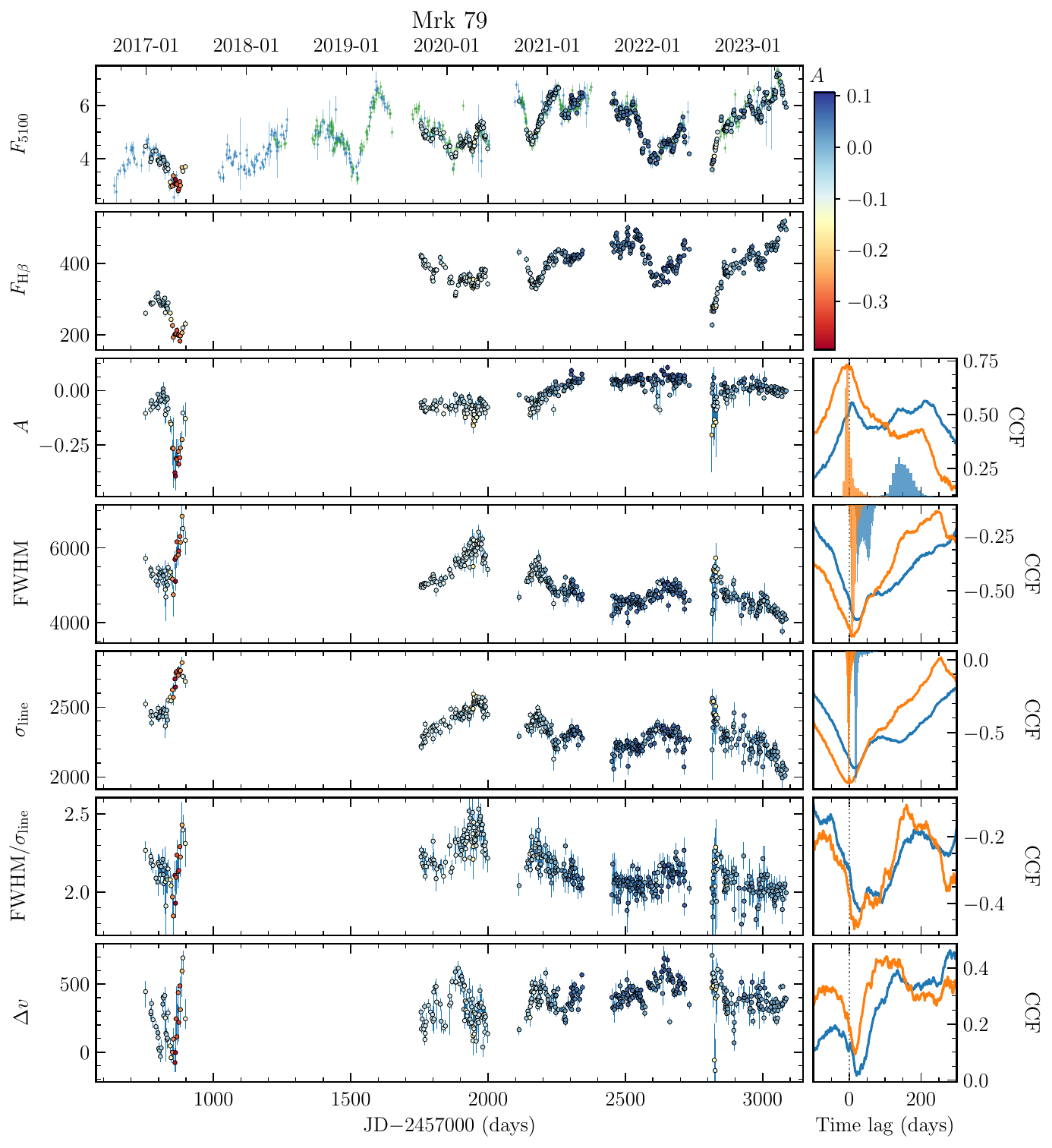}
    \addtocounter{figure}{-1}
    \caption{Continued.}
    \label{fig:jd_Mrk79}
\end{figure*}
\begin{figure*}
    \centering
    \includegraphics[width=1\textwidth]{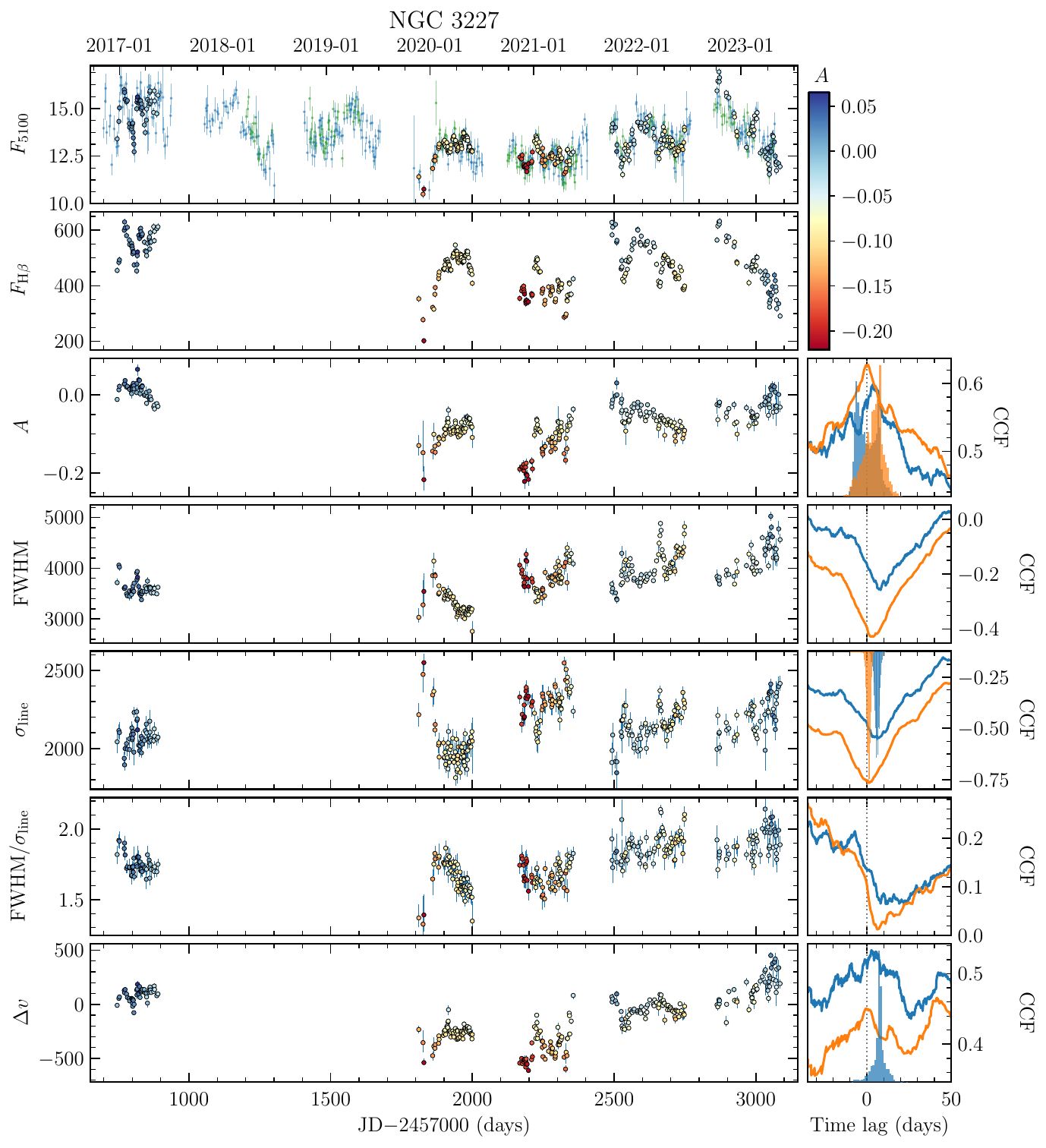}
    \addtocounter{figure}{-1}
    \caption{Continued.}
    \label{fig:jd_NGC3227}
\end{figure*}
\begin{figure*}
    \centering
    \includegraphics[width=1\textwidth]{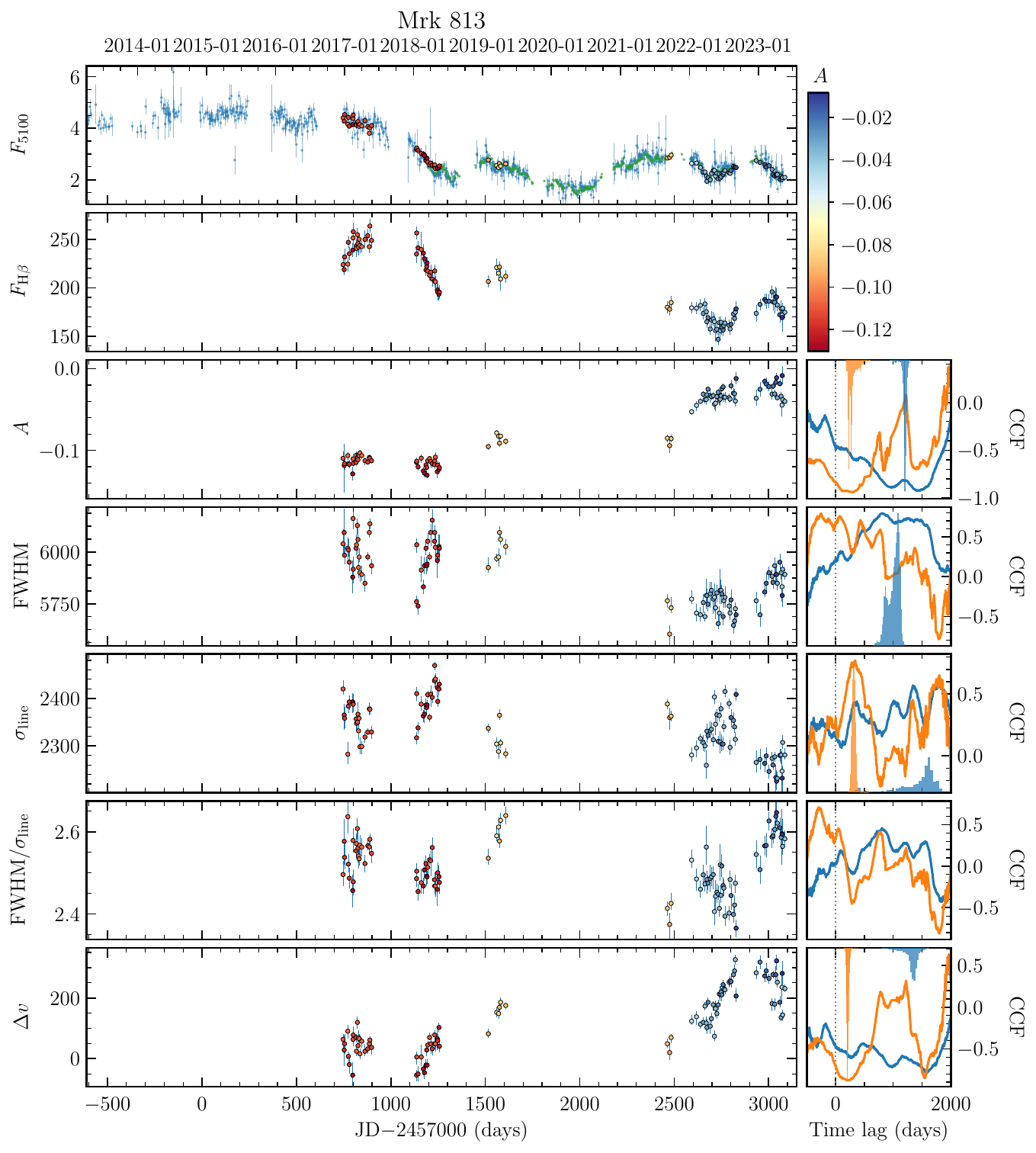}
    \addtocounter{figure}{-1}
    \caption{Continued.}
    \label{fig:jd_Mrk813}
\end{figure*}
\begin{figure*}
    \centering
    \includegraphics[width=1\textwidth]{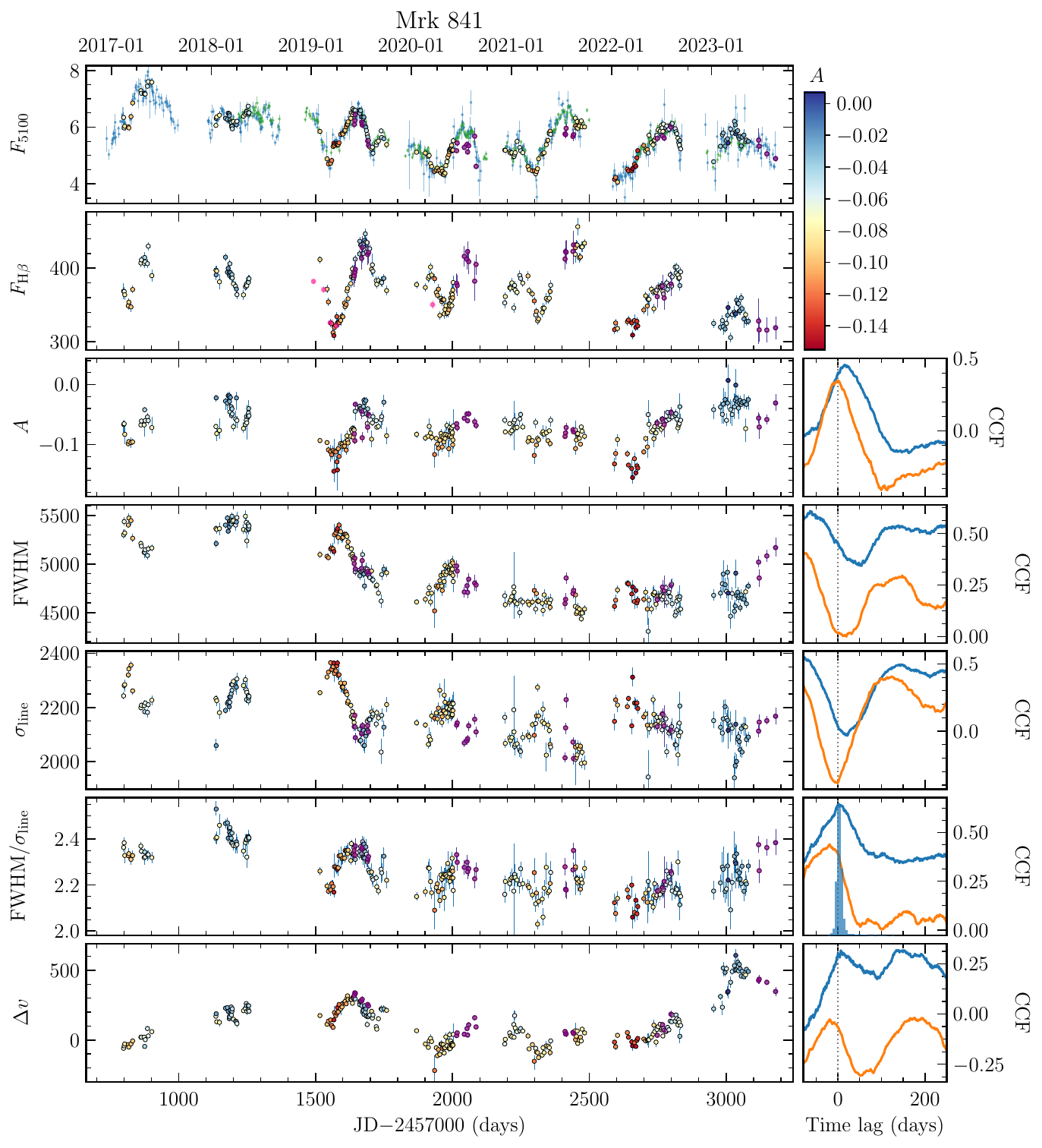}
    \addtocounter{figure}{-1}
    \caption{Continued.}
    \label{fig:jd_Mrk841}
\end{figure*}
\begin{figure*}
    \centering
    \includegraphics[width=1\textwidth]{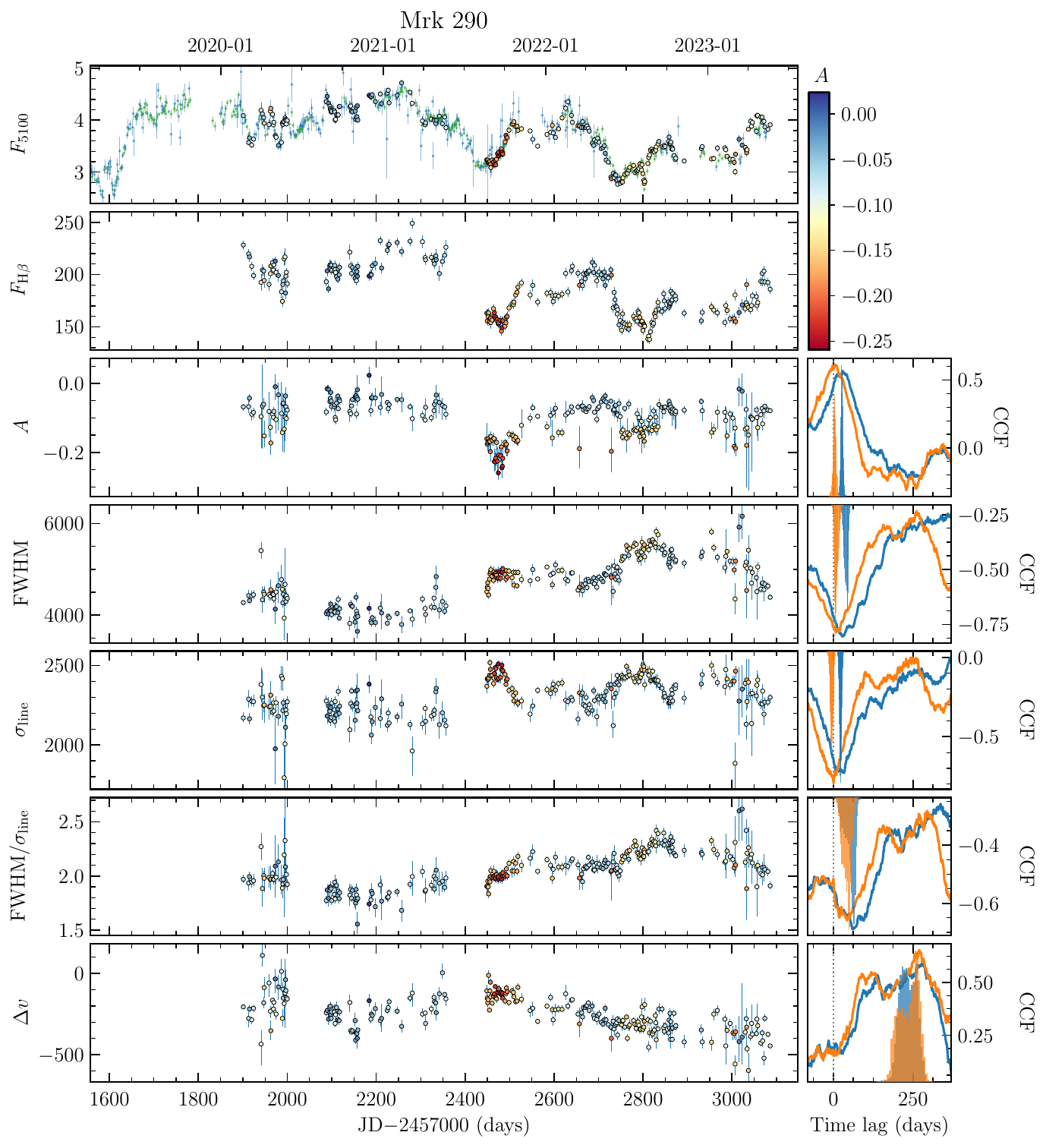}
    \addtocounter{figure}{-1}
    \caption{Continued.}
    \label{fig:jd_PG1534+580}
\end{figure*}
\begin{figure*}
    \centering
    \includegraphics[width=1\textwidth]{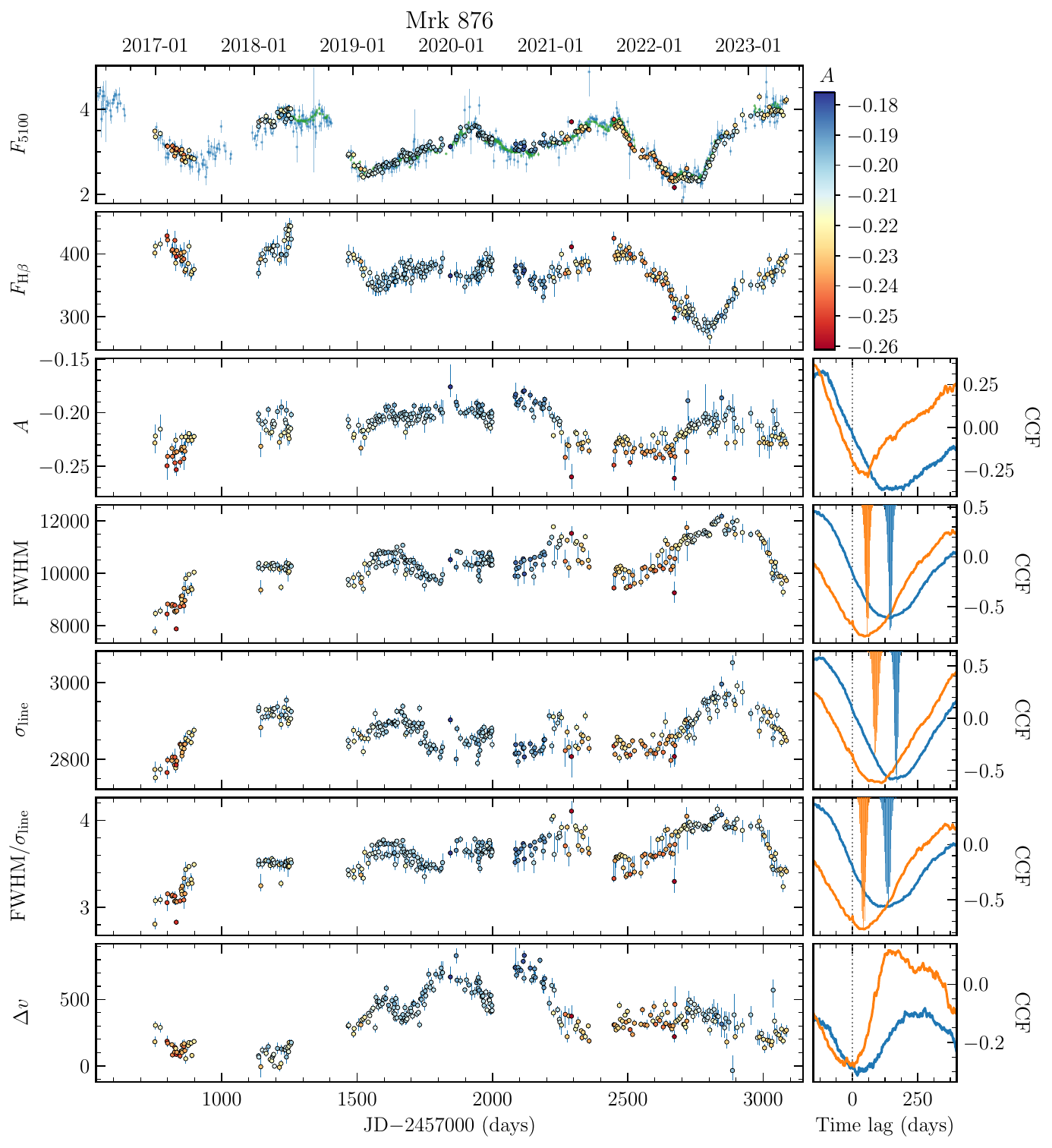}
    \addtocounter{figure}{-1}
    \caption{Continued.}
    \label{fig:jd_Mrk876}
\end{figure*}
%%%%%%%%%%%%%%%%%%%%%%%%%%%
\begin{figure*}
    \centering
    \includegraphics[width=1\textwidth]{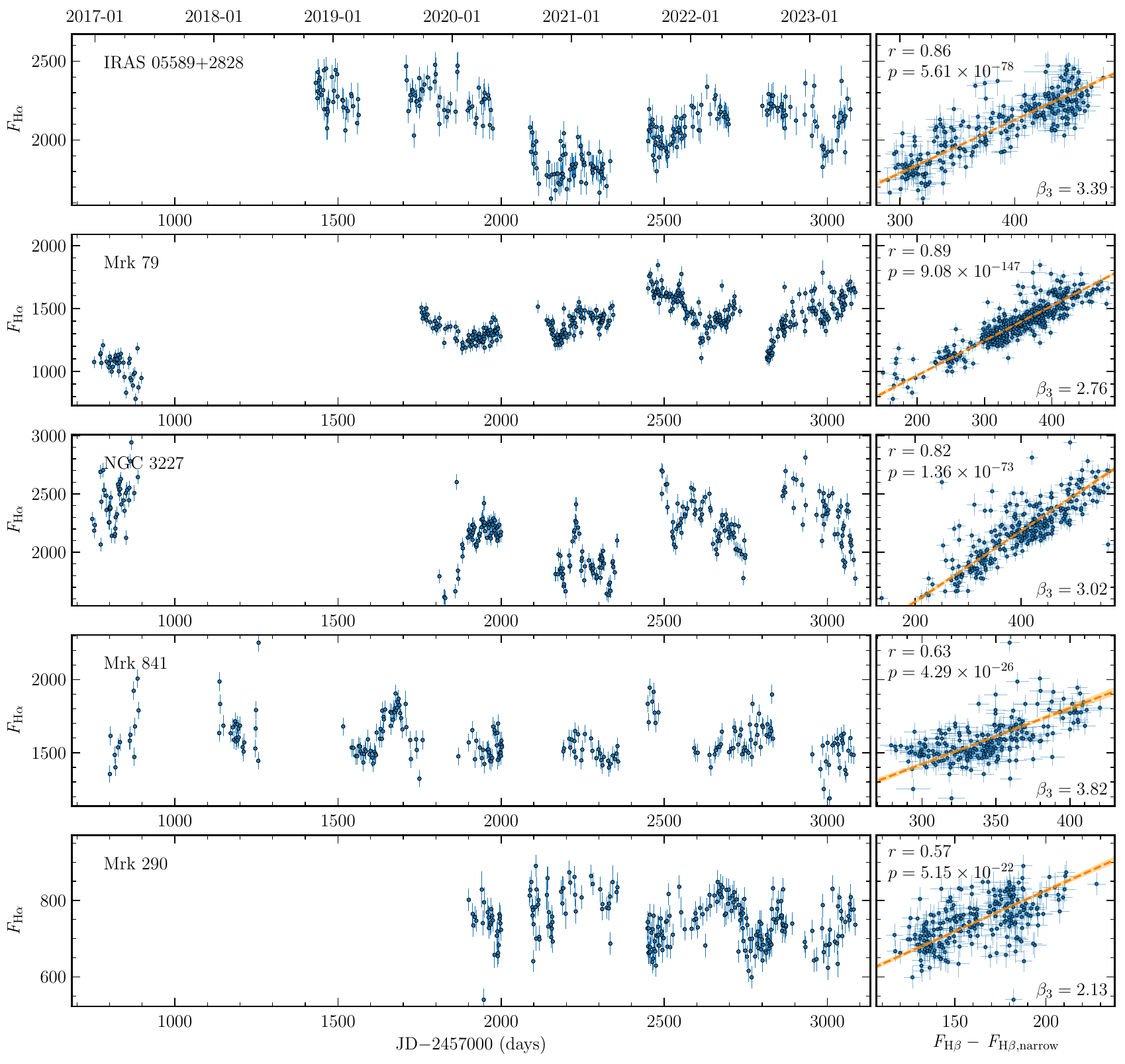}
    \caption{The H$\alpha$ light curves and their correlations with H$\beta$ flux. The units of H$\alpha$ and H$\beta$ are both $\rm 10^{-15}\,erg\,s^{-1}\,cm^{-2}$. $r$ and $p$ are the Pearson's correlation coefficients and the null probabilities, and $\beta_3$ is the slope. The contribution from the narrow line in the $F_{\rm H\beta}$ here has been removed.}
    \label{fig:jd_ha_hb}
\end{figure*}
%%%%%%%%%%%%%%%%%%%%%%%%%%%
\begin{figure*}[ht]
  \centering
  \includegraphics[width=1\textwidth]{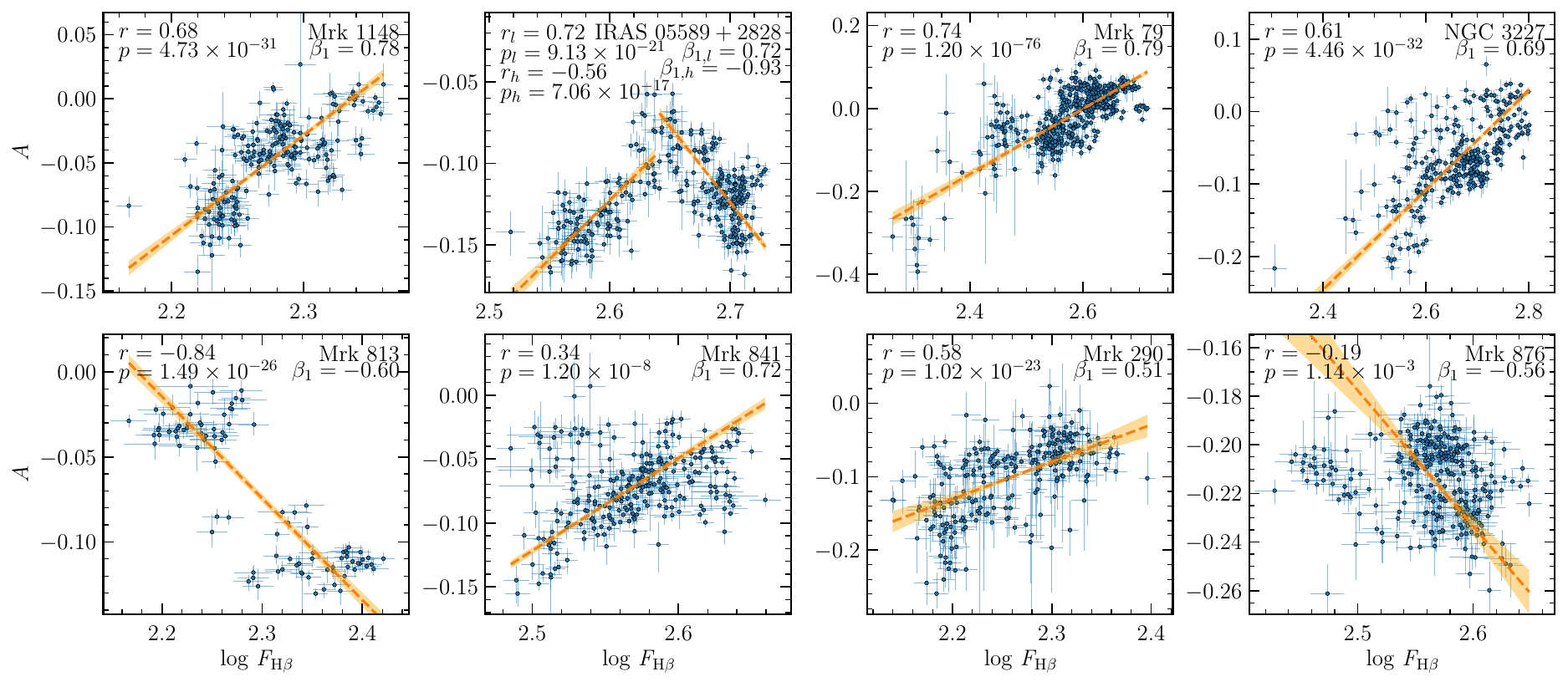}
  \caption{Correlations between H$\beta$ flux and asymmetry  $A$. The orange lines and regions represent the linear regressions and their uncertainties. The $F_{\rm H\beta}$ is in unit of $\rm 10^{-15}\,erg\,s^{-1}\,cm^{-2}$. In each panel, $r$ and $p$ are the Pearson's correlation coefficients and the corresponding null probabilities, and $\beta_1$ represents the slope.}
  \label{fig:ratio_A_all}
\end{figure*}
%%%%%%%%%%%%%%%%%%%%%%%%%%%
\begin{figure*}[ht]
    \centering
    \includegraphics[width=0.95\textwidth]{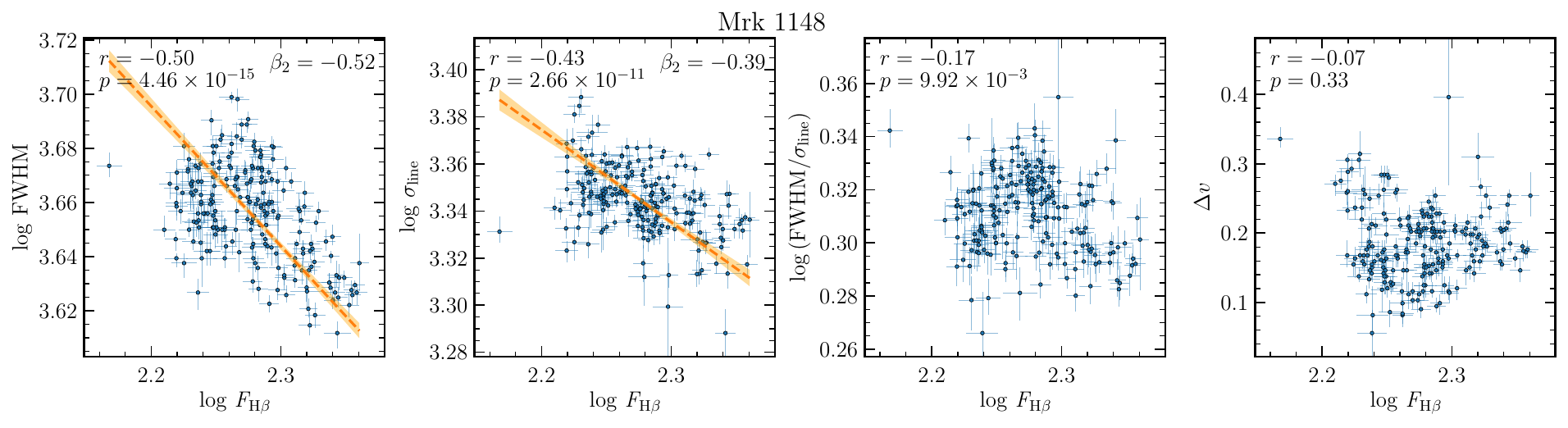}
    \includegraphics[width=0.95\textwidth]{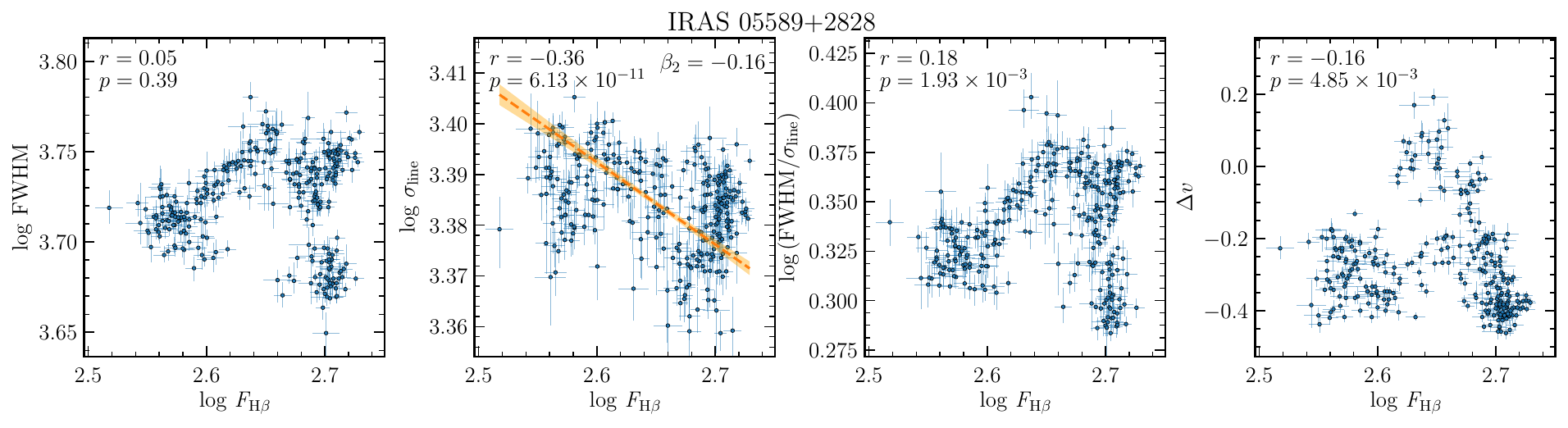}
    \includegraphics[width=0.95\textwidth]{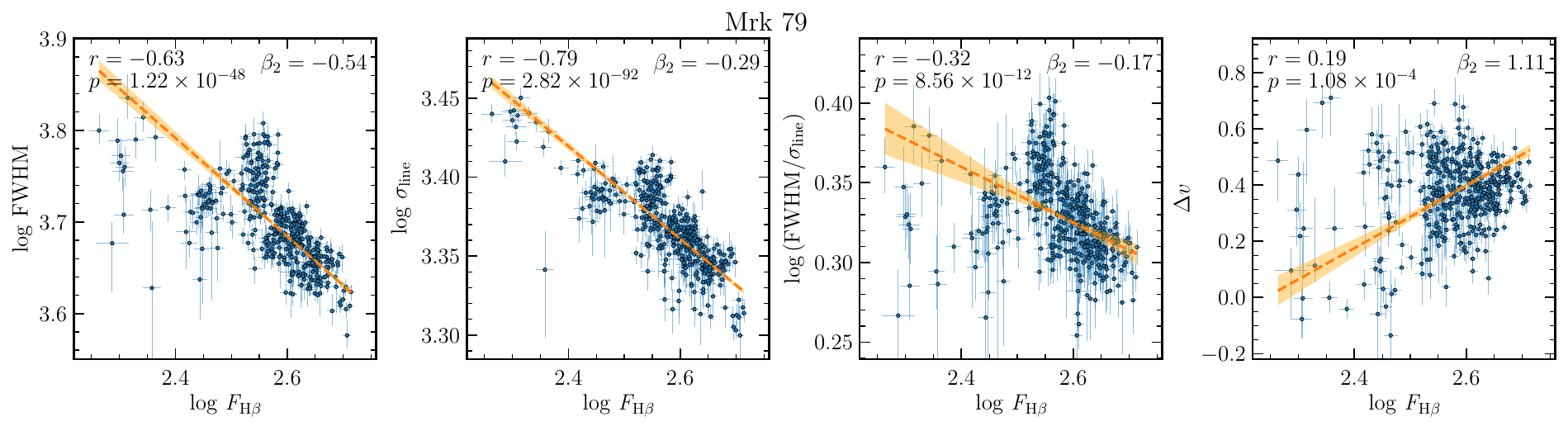}
    \includegraphics[width=0.95\textwidth]{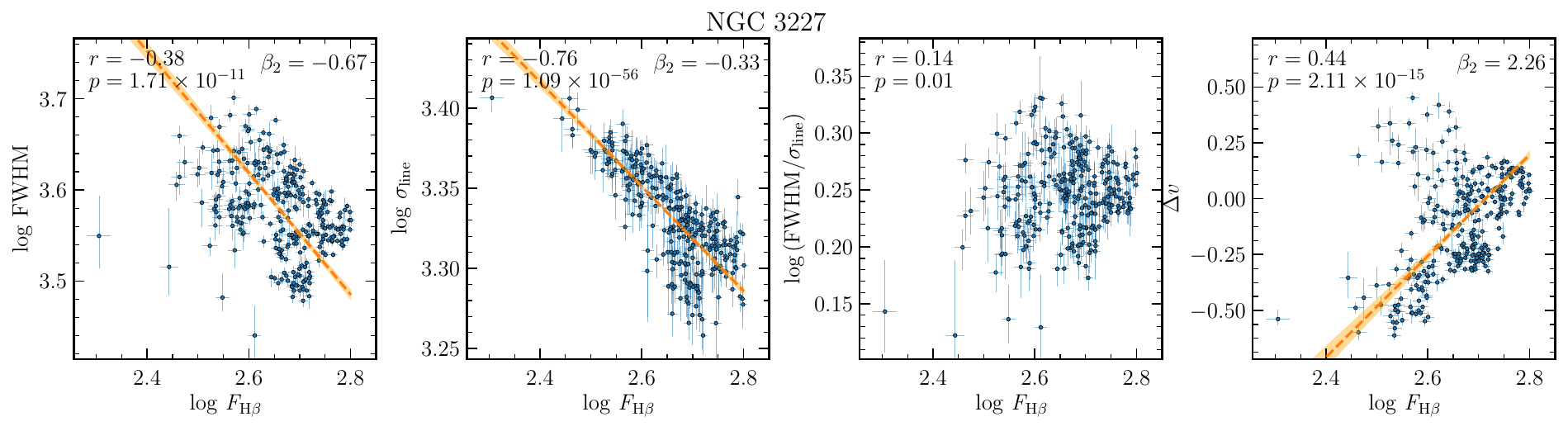}
    \caption{Correlations between H$\beta$ flux, and the FWHM, $\sigma_{\rm line}$, FWHM/$\sigma_{\rm line}$ and $\Delta v$. The orange lines and regions represent the linear regressions and their uncertainties. The $F_{\rm H\beta}$ is in unit of $\rm 10^{-15}\,erg\,s^{-1}\,cm^{-2}\,$. The FWHM and \sigmaline\ are in units of $\rm{km\,s^{-1}}$, and \deltav\ is in unit of $\rm{10^3\,km\,s^{-1}}$. In each panel, $r$ and $p$ are the Pearson's correlation coefficients and the null probabilities, and $\beta_2$ represents the slope. }
    \label{fig:ratio1}
\end{figure*}

\begin{figure*}[ht]
    \centering
    \includegraphics[width=0.95\textwidth]{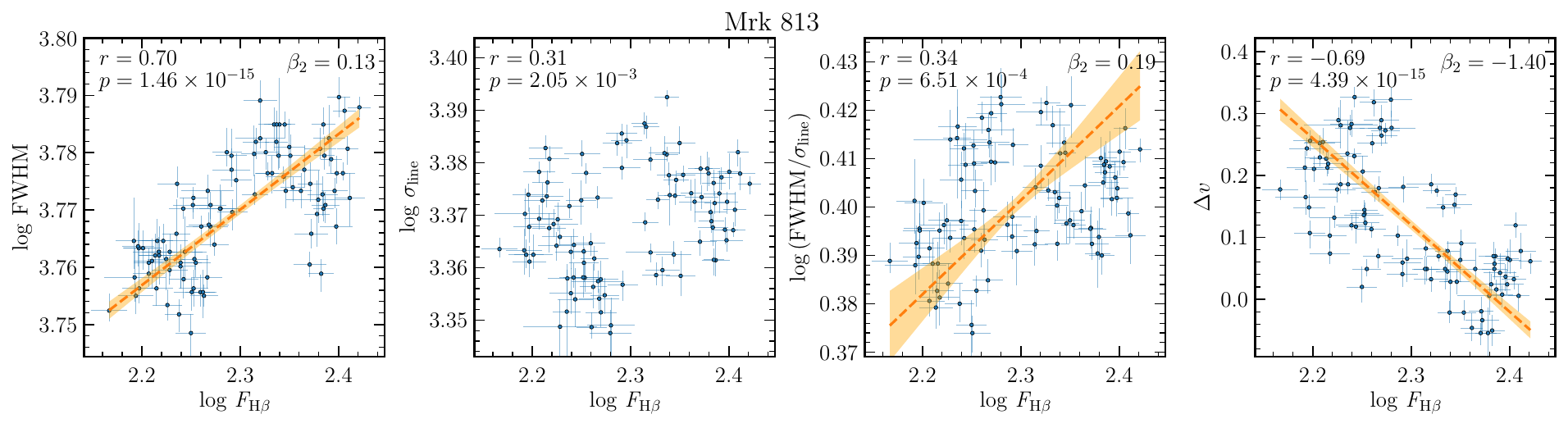}
    \includegraphics[width=0.95\textwidth]{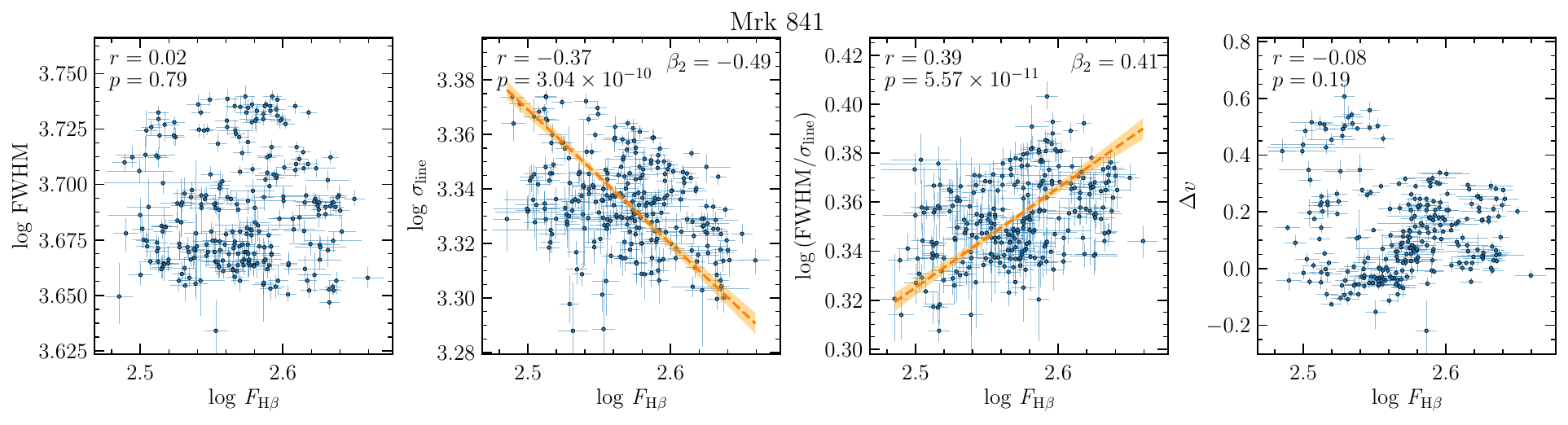}
    \includegraphics[width=0.95\textwidth]{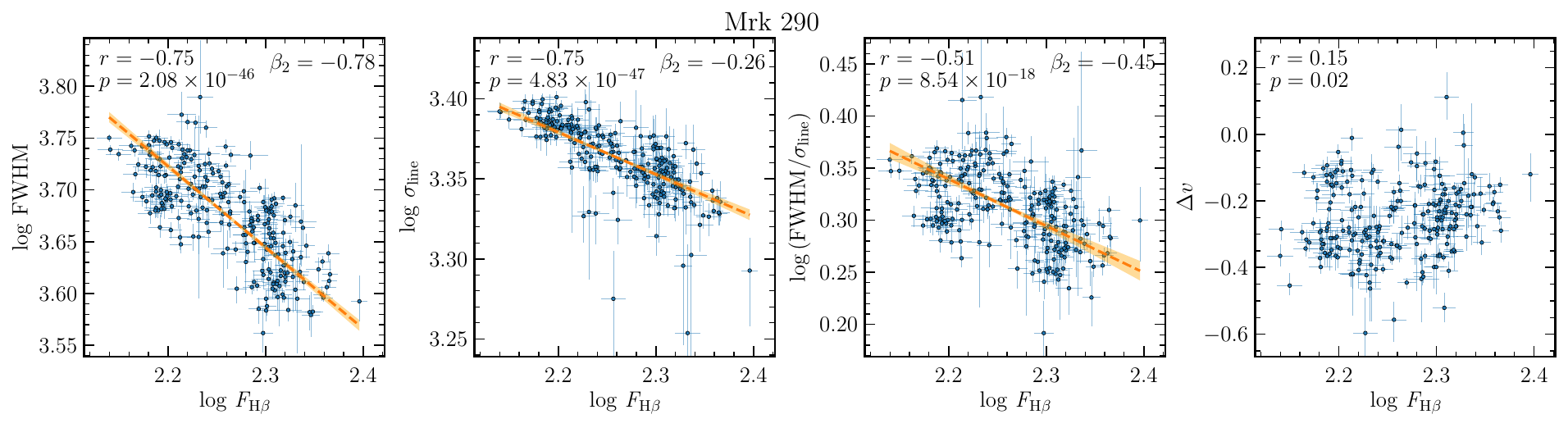}
    \includegraphics[width=0.95\textwidth]{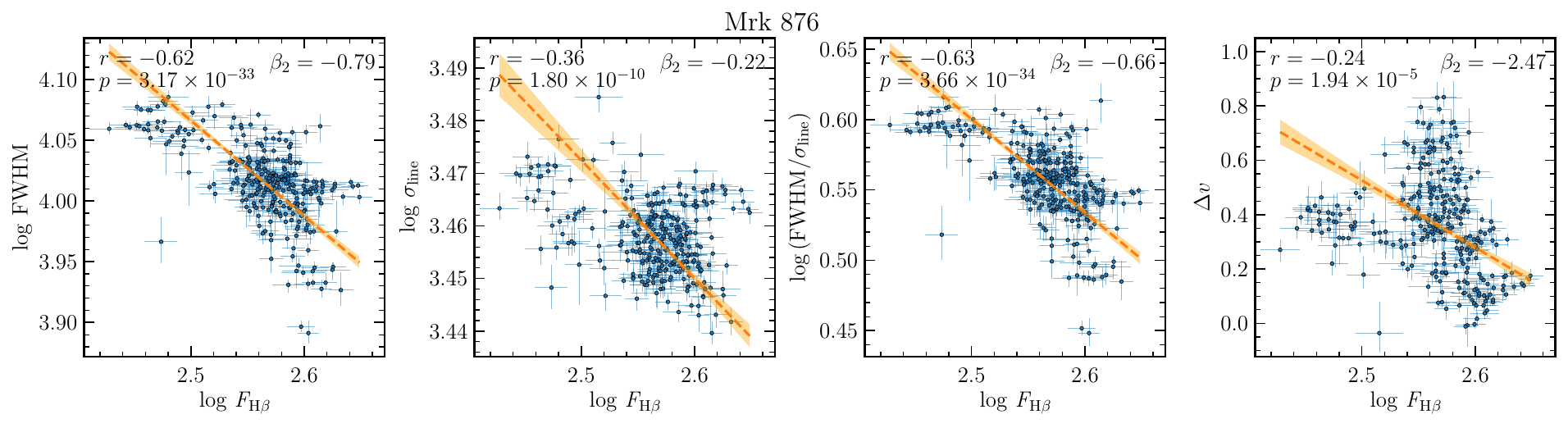}
    \addtocounter{figure}{-1}
    \caption{Continued.}
    \label{fig:ratio2}
\end{figure*}

%%%%%%%%%%%%%%%%%%%%%%%%%%%
\begin{figure*}[ht]
    \centering
    \includegraphics[width=1\textwidth]{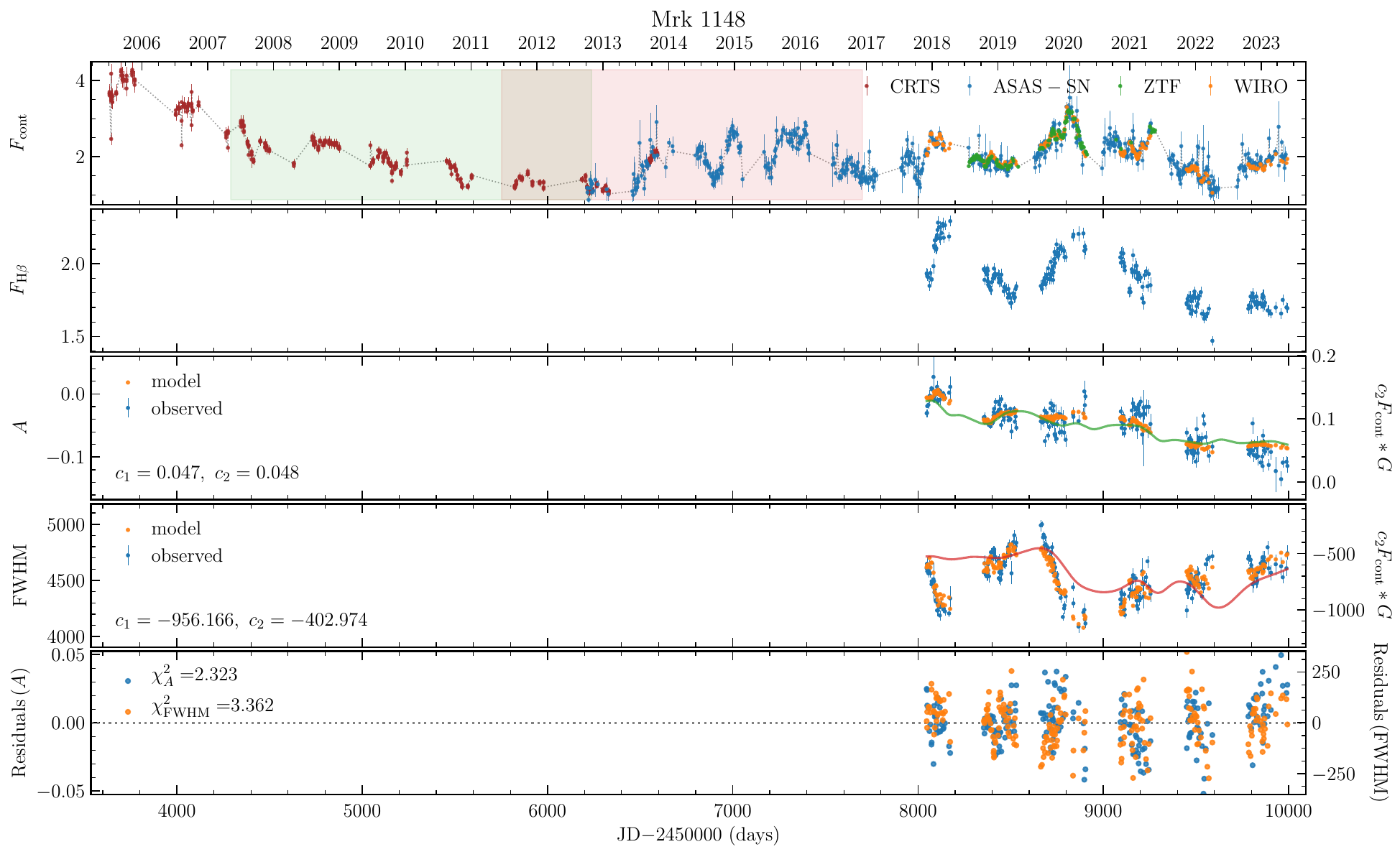}
    \includegraphics[width=0.45\textwidth]{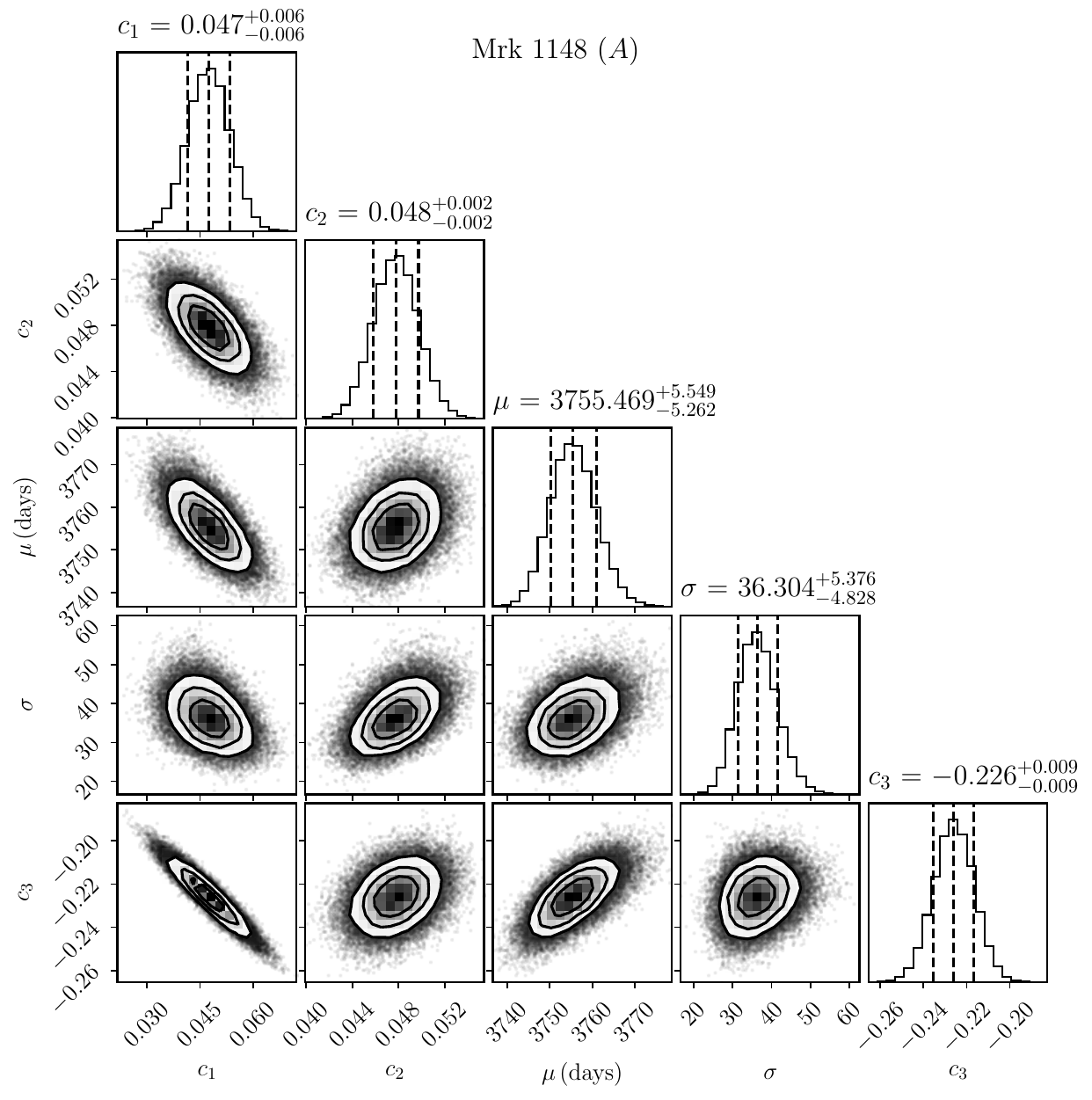}
    \includegraphics[width=0.45\textwidth]{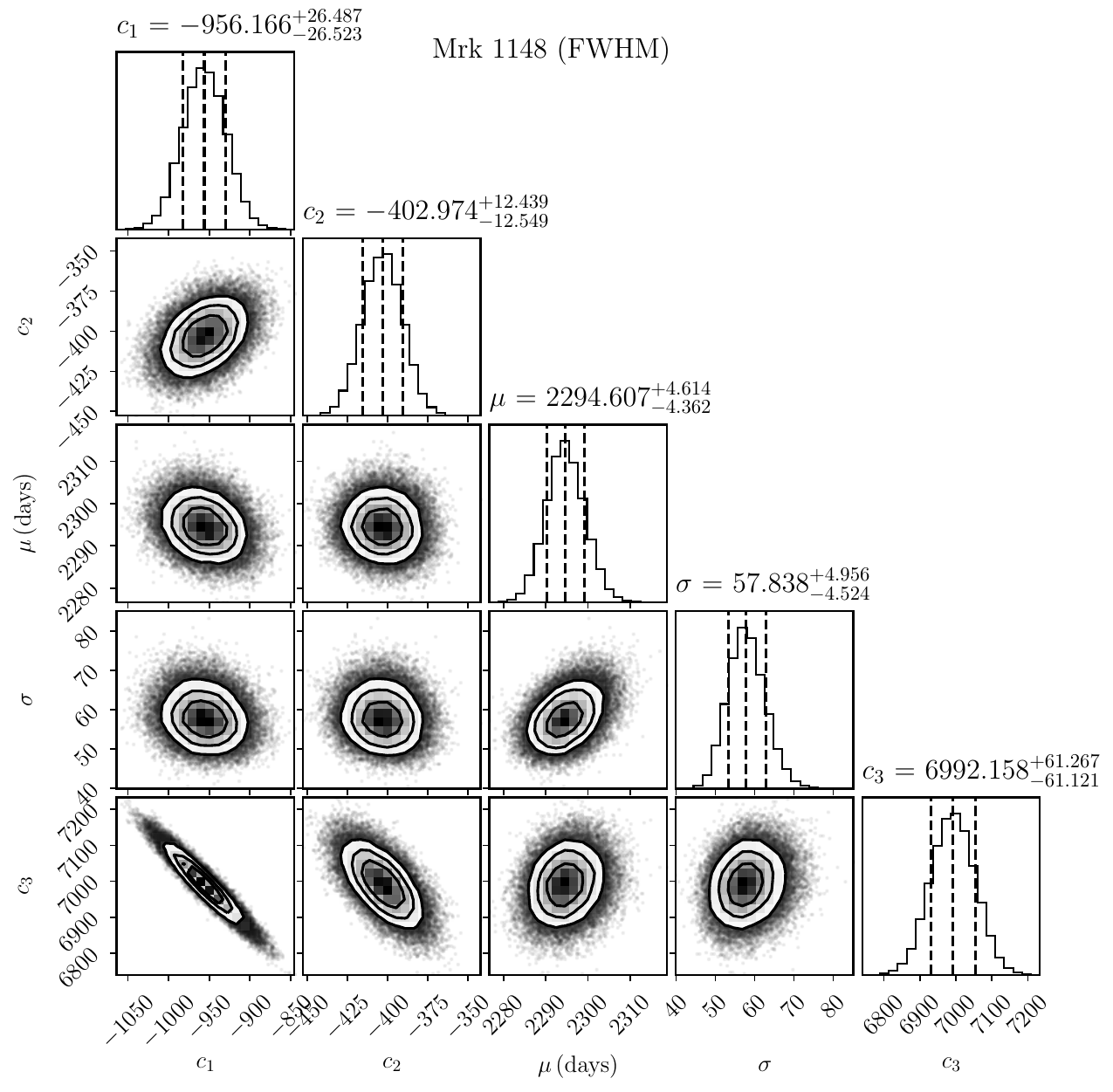}
    \caption{Decomposition of long-term trends. Panels on the top are the continuum (long-term) and H$\beta$ light curves in the units of $\rm 10^{-15}\,erg\,s^{-1}\,cm^{-2}\,$\AA$^{-1}$ and  $\rm 10^{-15}\,erg\,s^{-1}\,cm^{-2}$, and the temporal variations of the asymmetry $A$, FWHM, and the residuals of the decompositions. The orange points represent the best fit of the model described in Section \ref{sec:Long-term trend}. The green and red lines (in the third and fourth panels) are the long-term trends corresponding to the influence from the radiation pressure (the second term $c_2 F_{\mathrm{cont}}*G$ in Eqn \ref{equ:model for radiation pressure}), with the temporal spans of the corresponding continuum shown in the first panel. The bottom panels are the probability density distributions of the parameters, with the median values and $1\sigma$ errors marked on the top of the panels. The contours are at $1\sigma$, $1.5\sigma$ and $2\sigma$ confidence regions, and dashed lines in the 1D distributions are the 16\%, 50\%, and 84\% quantiles. The Python \texttt{corner} module developed by \cite{corner} is used for generating the plots. }
    \label{fig:longterm light curves}
\end{figure*}

\begin{figure*}[ht]
    \centering
    \includegraphics[width=1\textwidth]{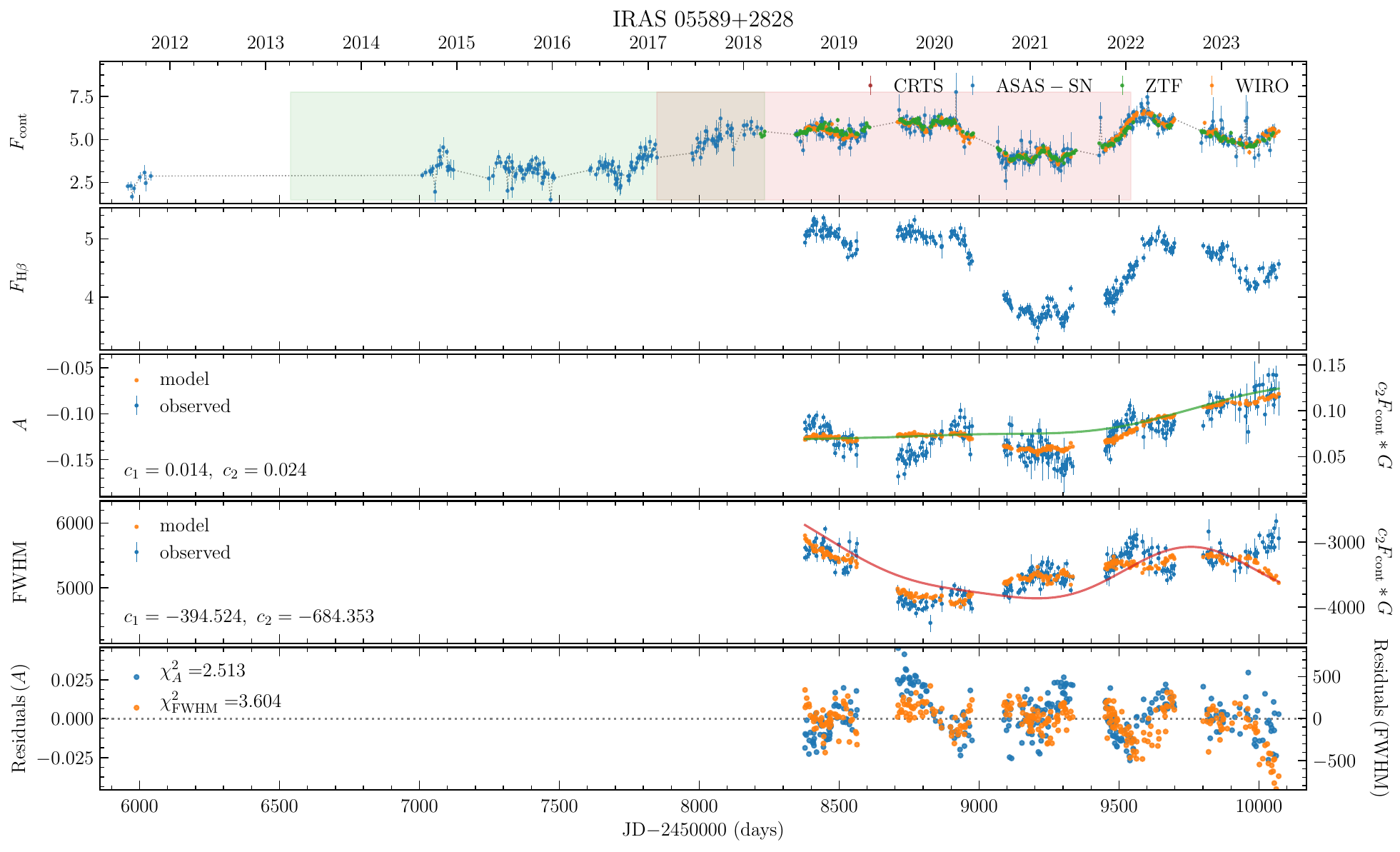}
    \includegraphics[width=0.45\textwidth]{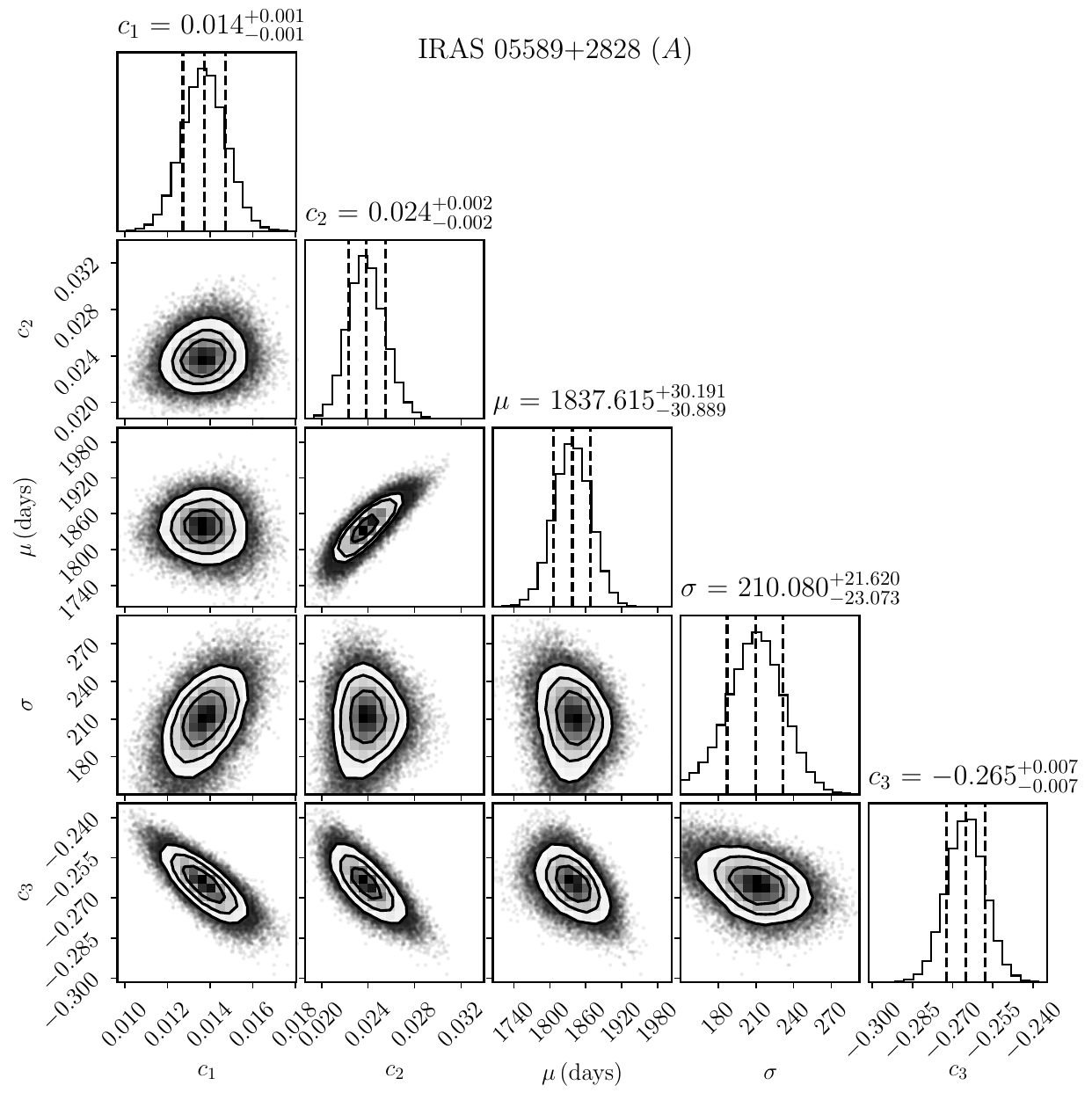}
    \includegraphics[width=0.45\textwidth]{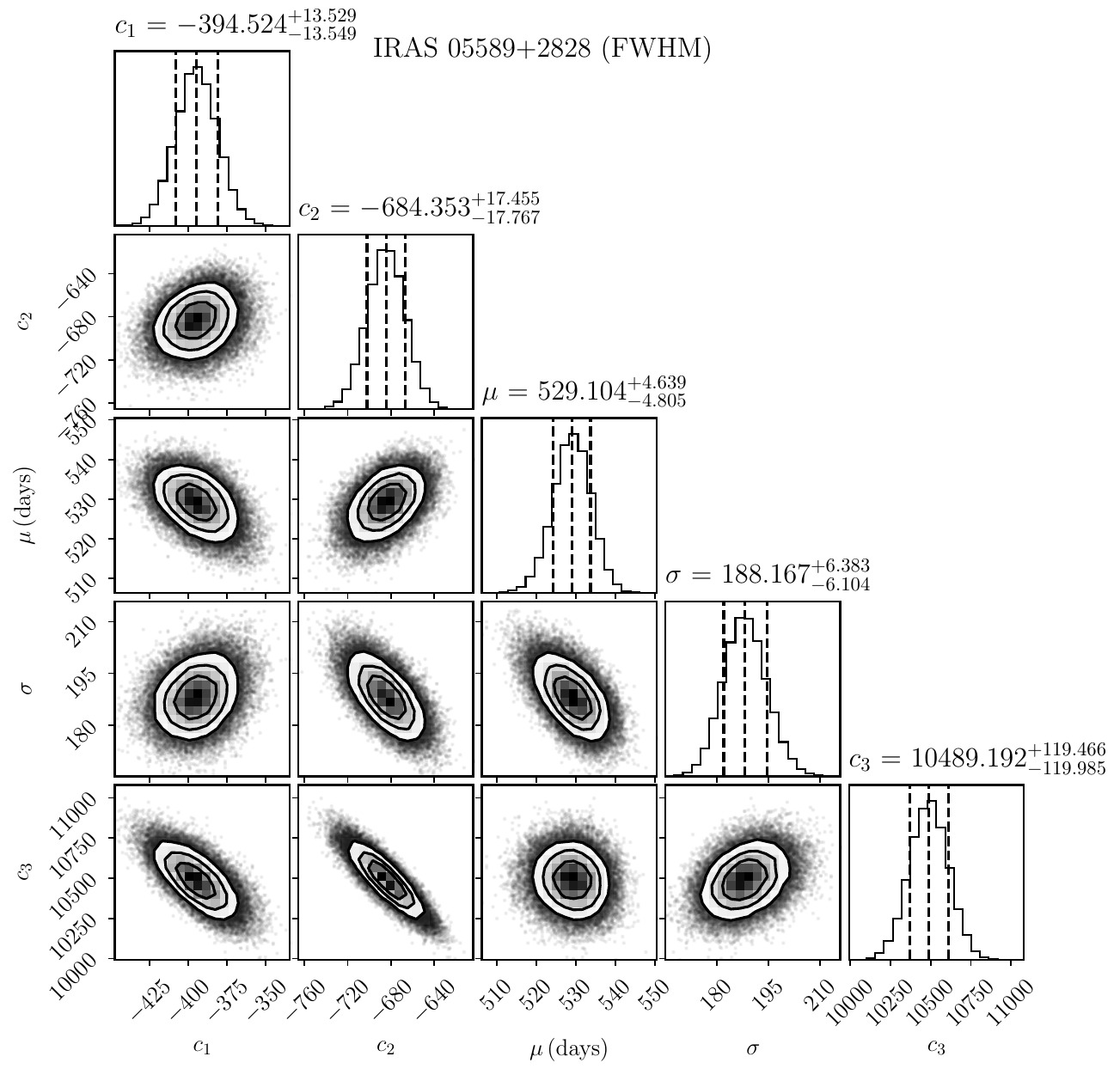}
    \addtocounter{figure}{-1}
    \caption{Continued.}
    \label{fig:longterm light curve2}
\end{figure*}

\begin{figure*}[ht]
    \centering
    \includegraphics[width=1\textwidth]{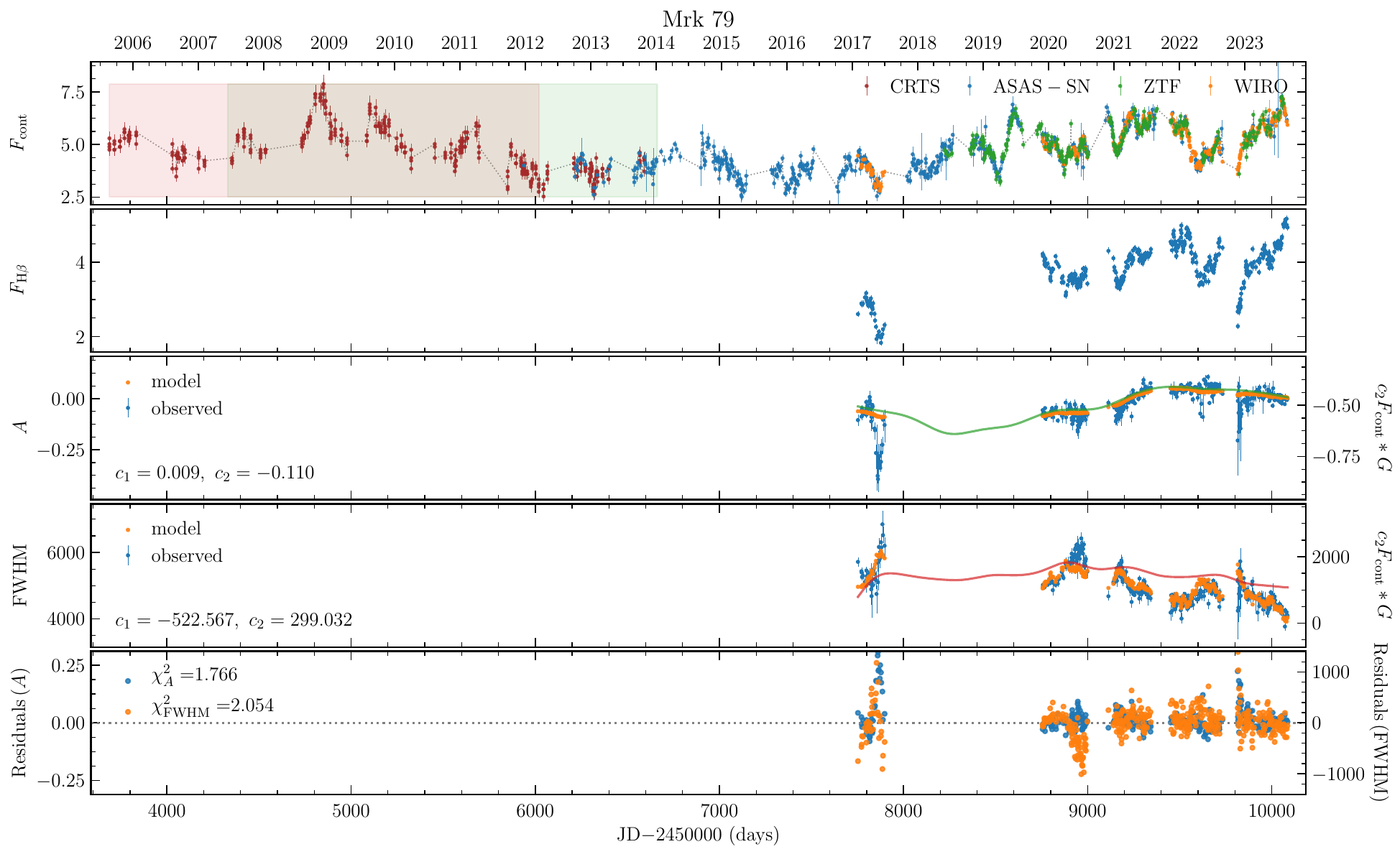}
    \includegraphics[width=0.45\textwidth]{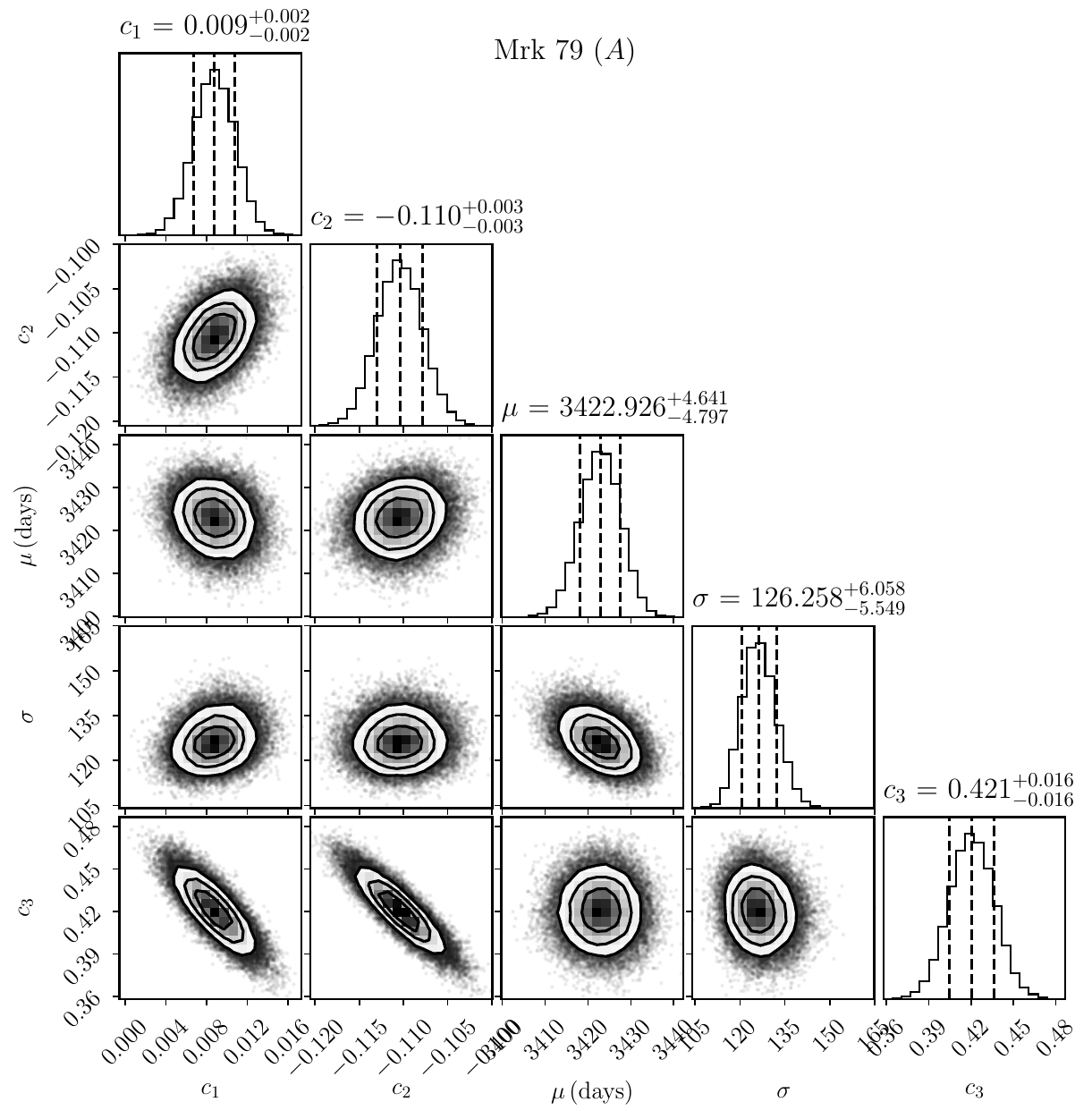}
    \includegraphics[width=0.45\textwidth]{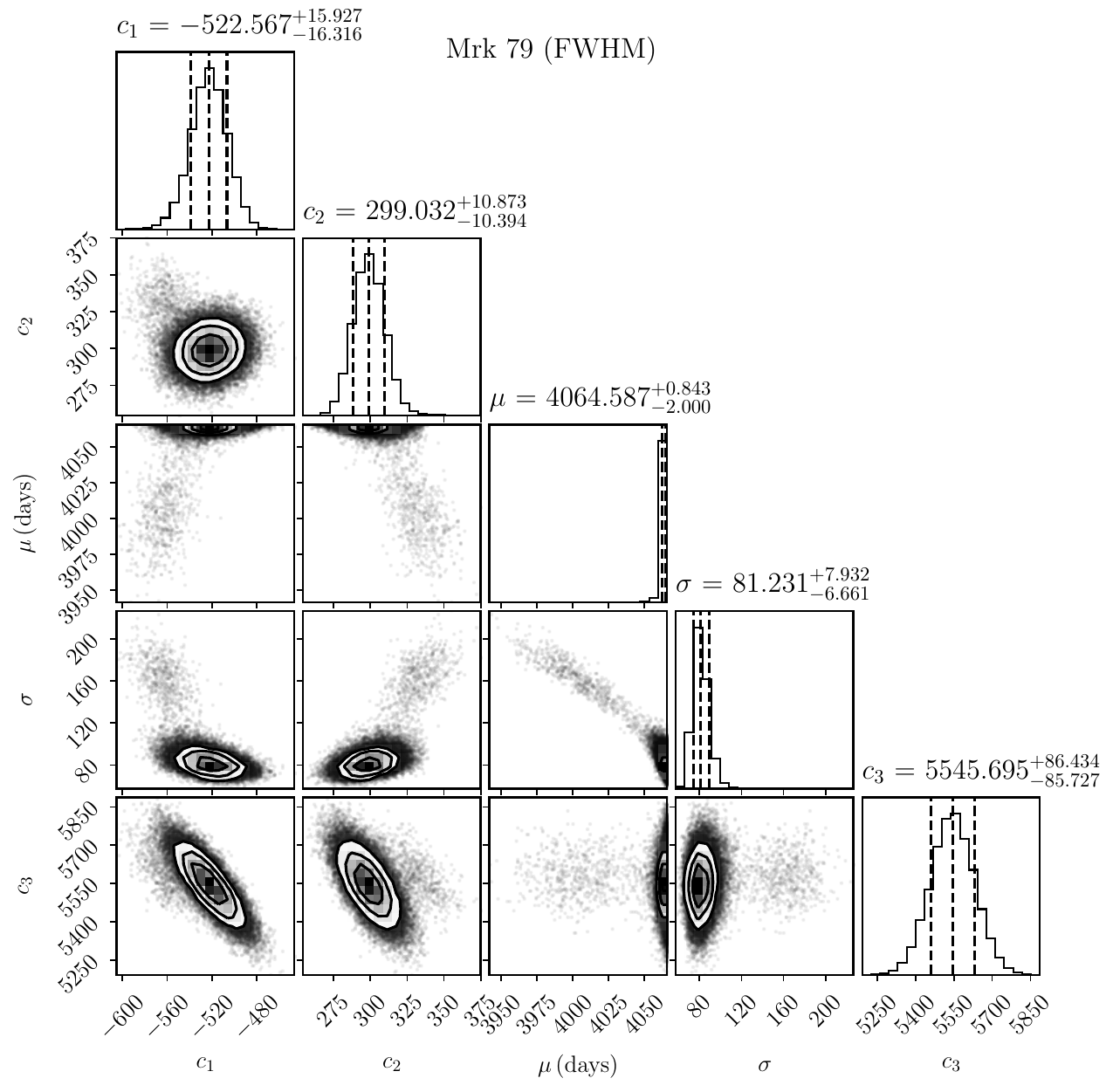}
    \addtocounter{figure}{-1}
    \caption{Continued.}
    \label{fig:longterm light curve3}
\end{figure*}

\begin{figure*}[ht]
    \centering
    \includegraphics[width=1\textwidth]{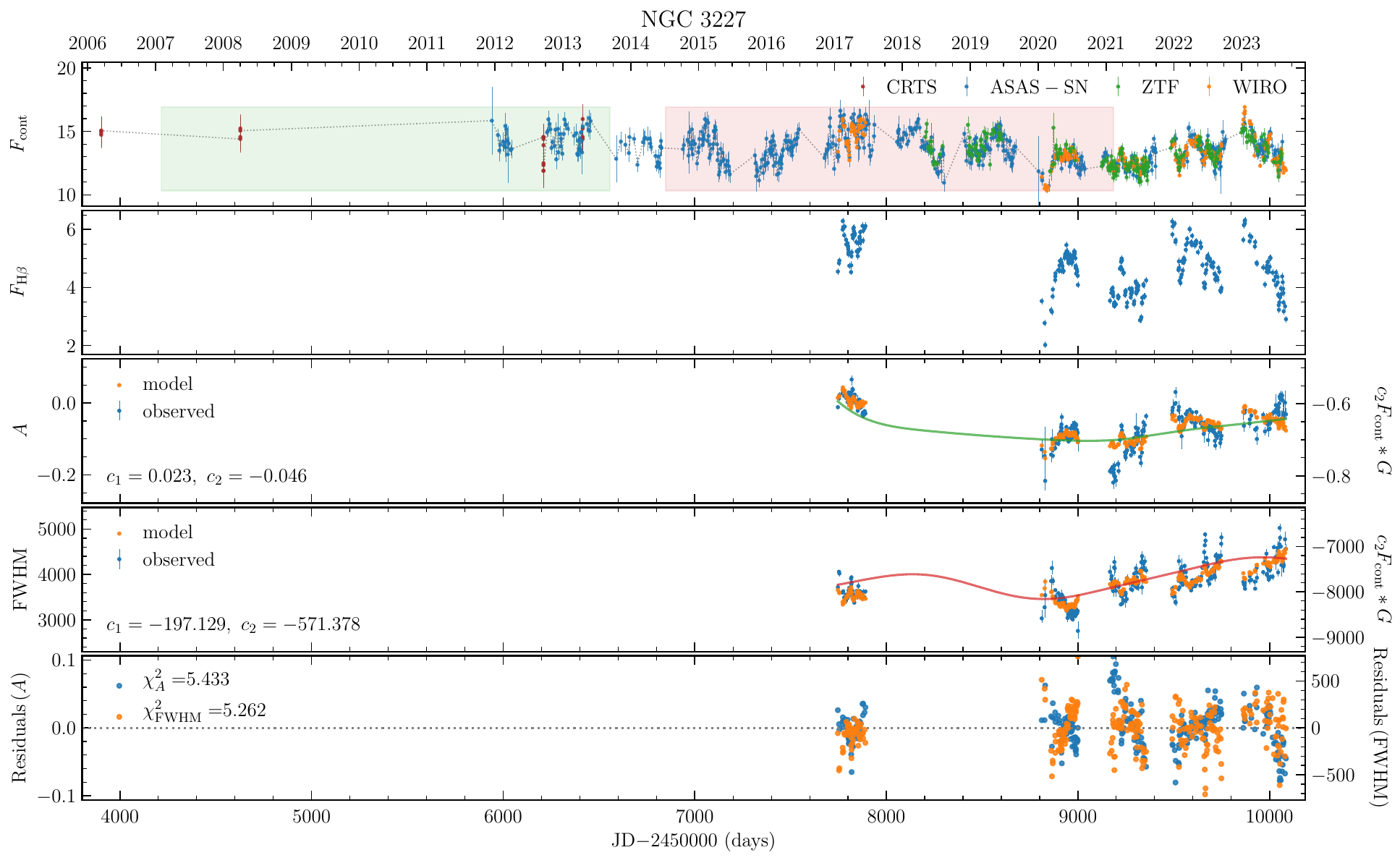}
    \includegraphics[width=0.45\textwidth]{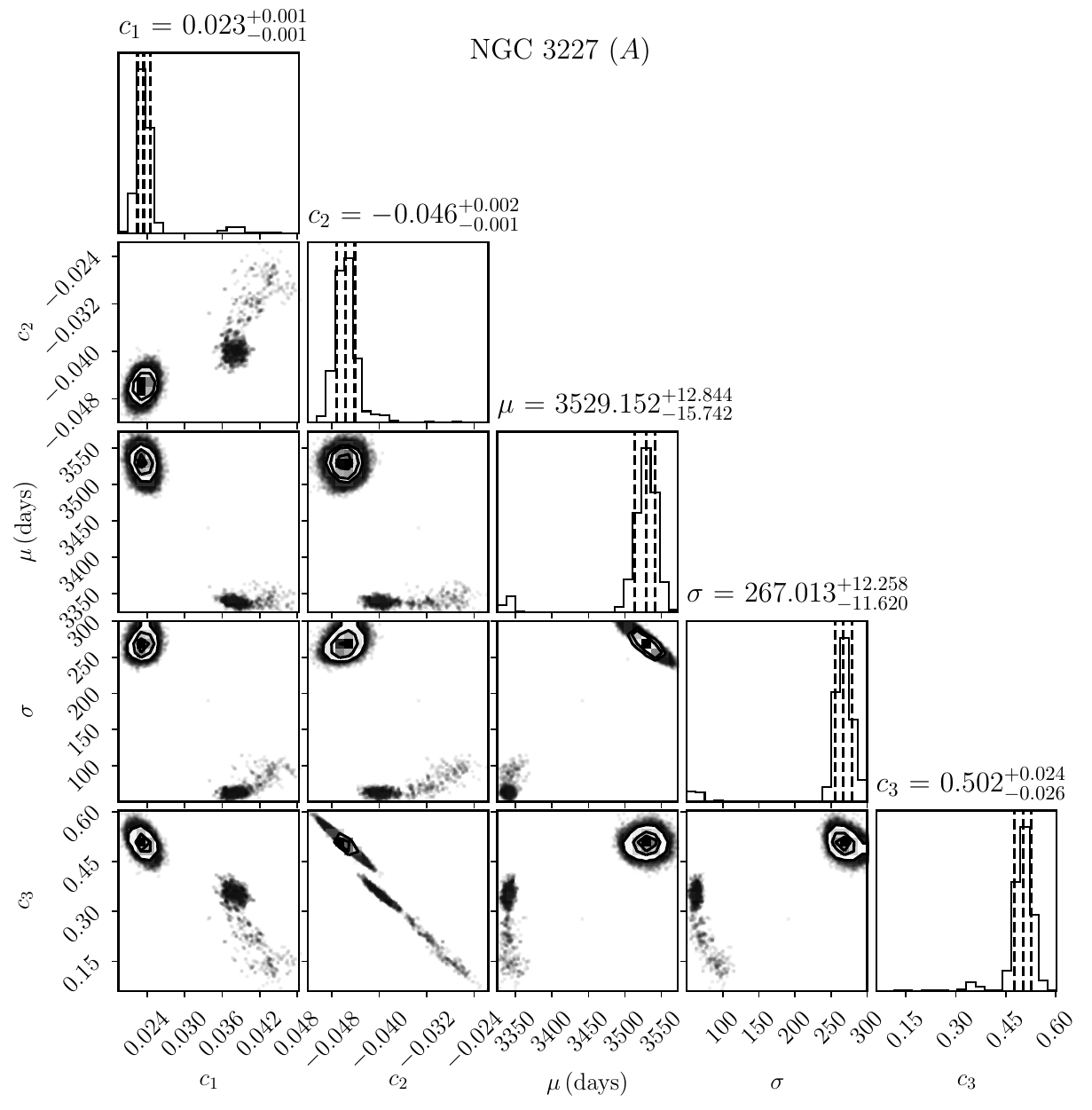}
    \includegraphics[width=0.45\textwidth]{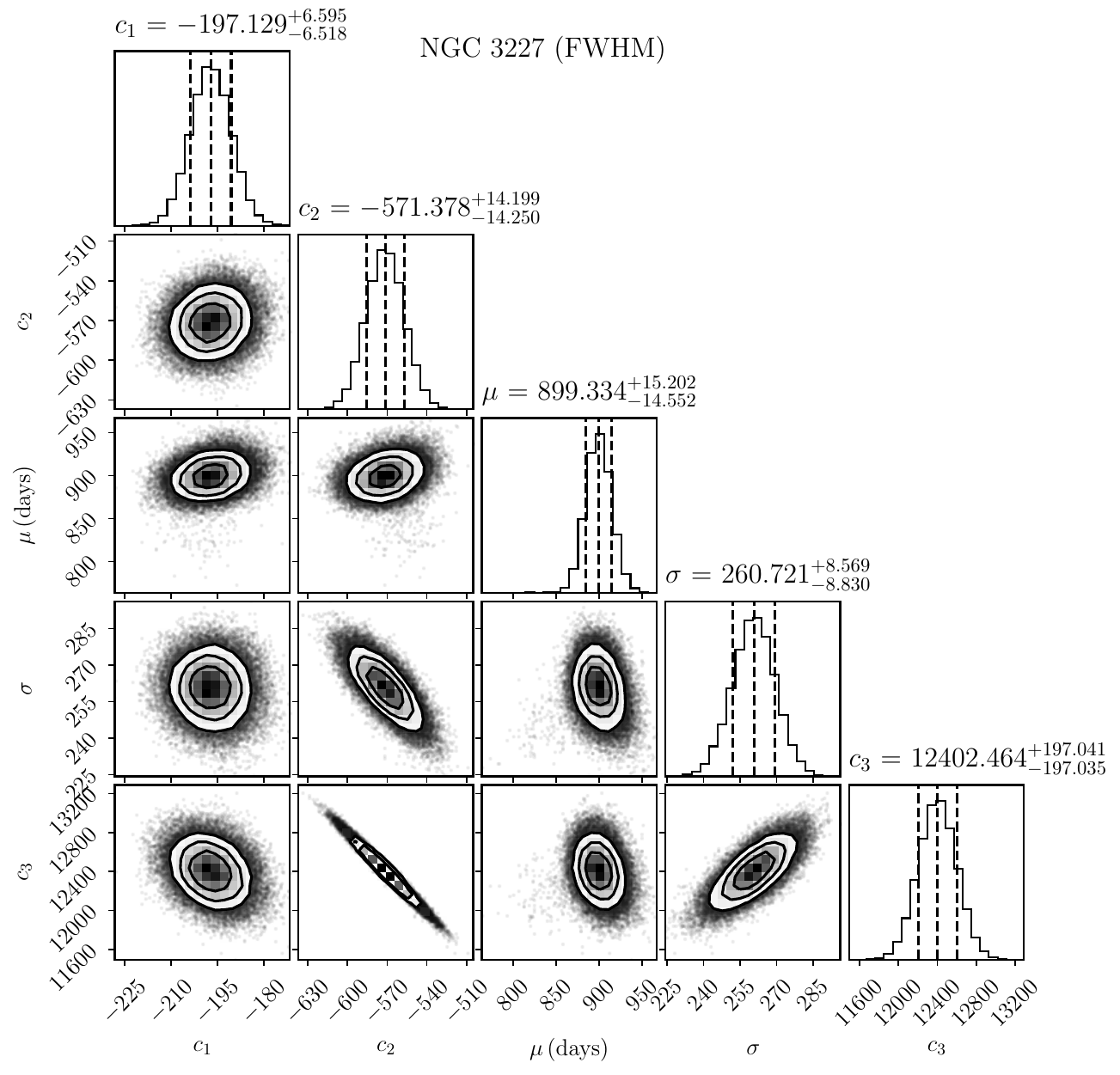}
    \addtocounter{figure}{-1}
    \caption{Continued.}
    \label{fig:longterm light curve4}
\end{figure*}

\begin{figure*}[ht]
    \centering
    \includegraphics[width=1\textwidth]{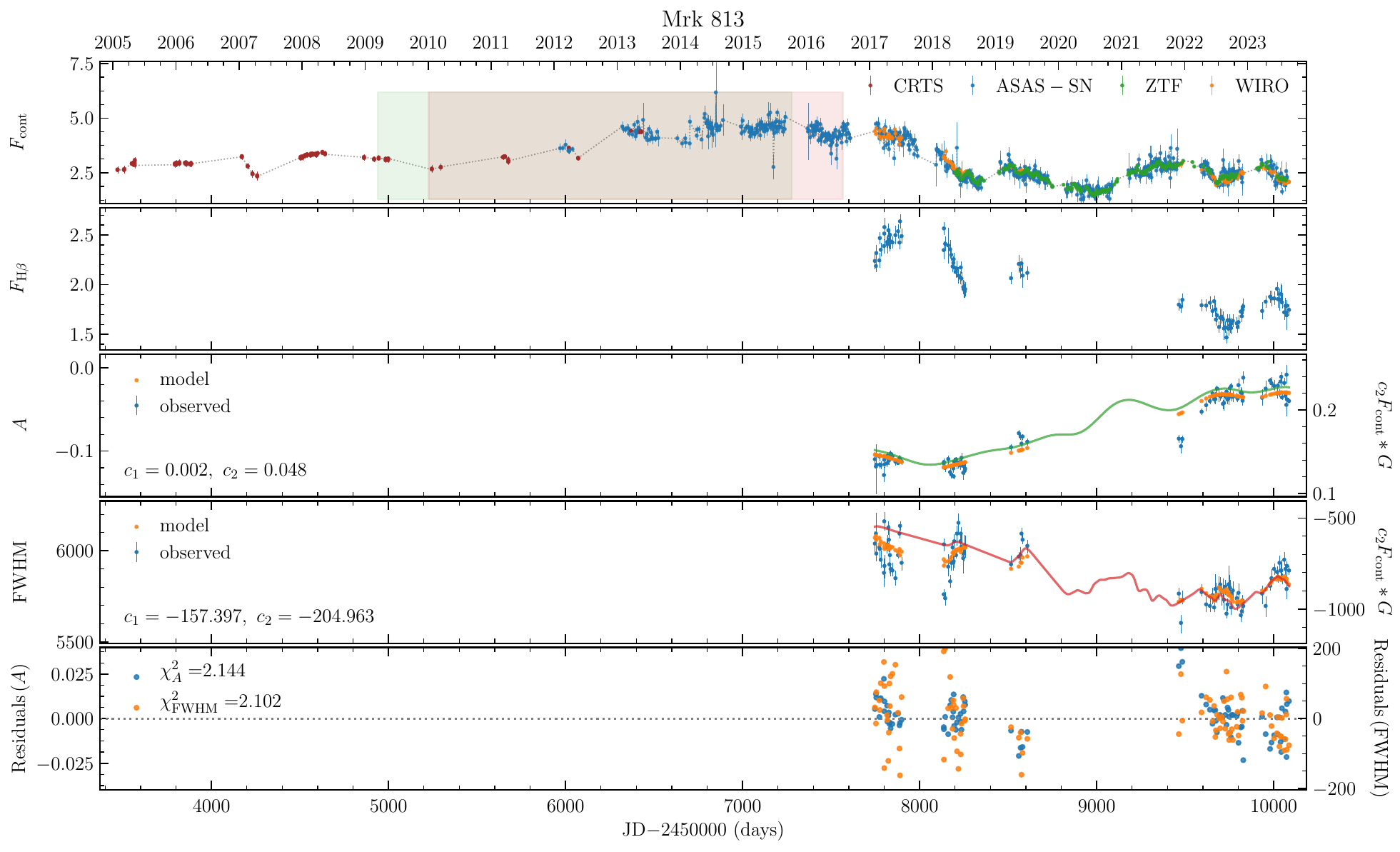}
    \includegraphics[width=0.45\textwidth]{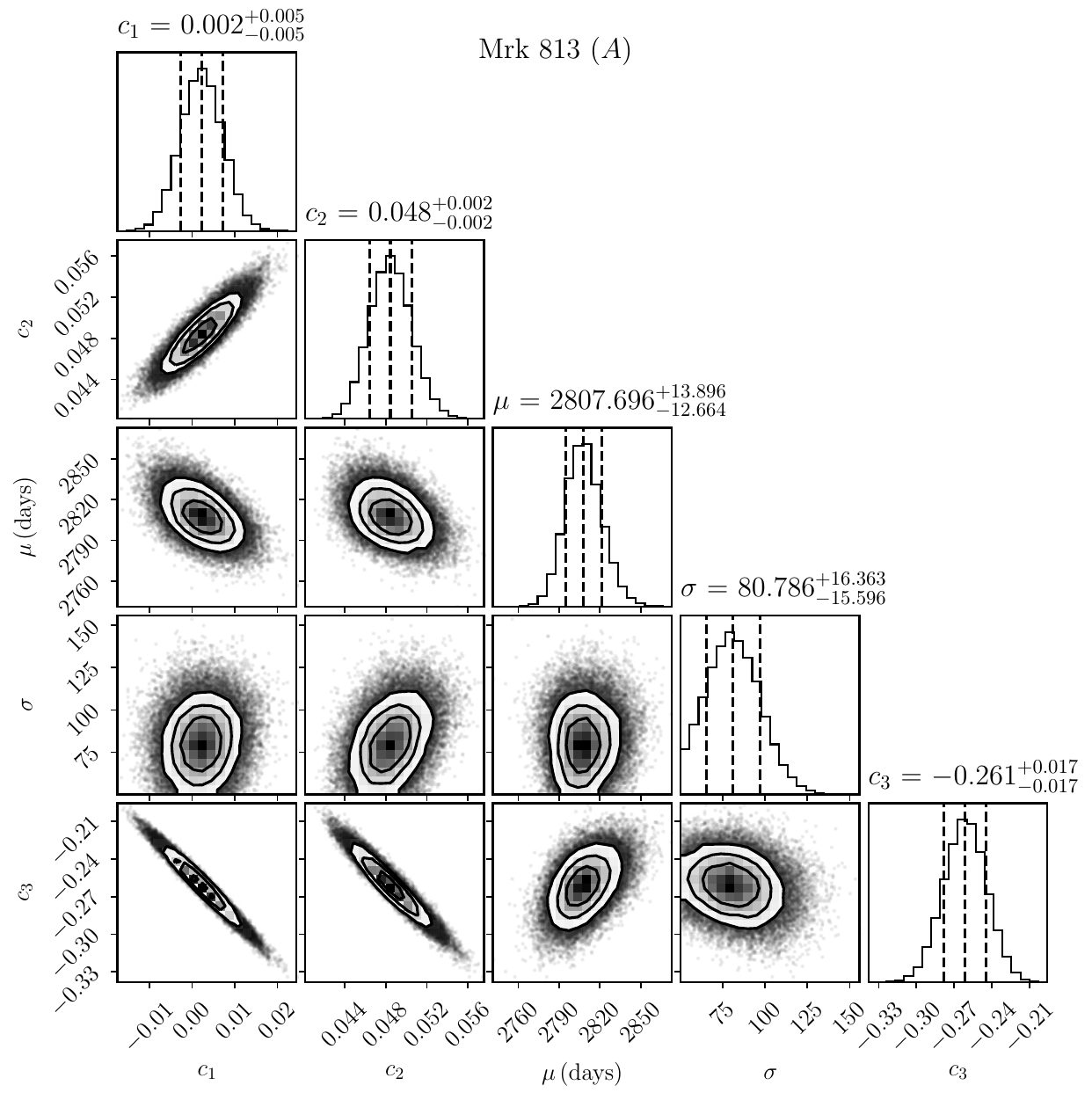}
    \includegraphics[width=0.45\textwidth]{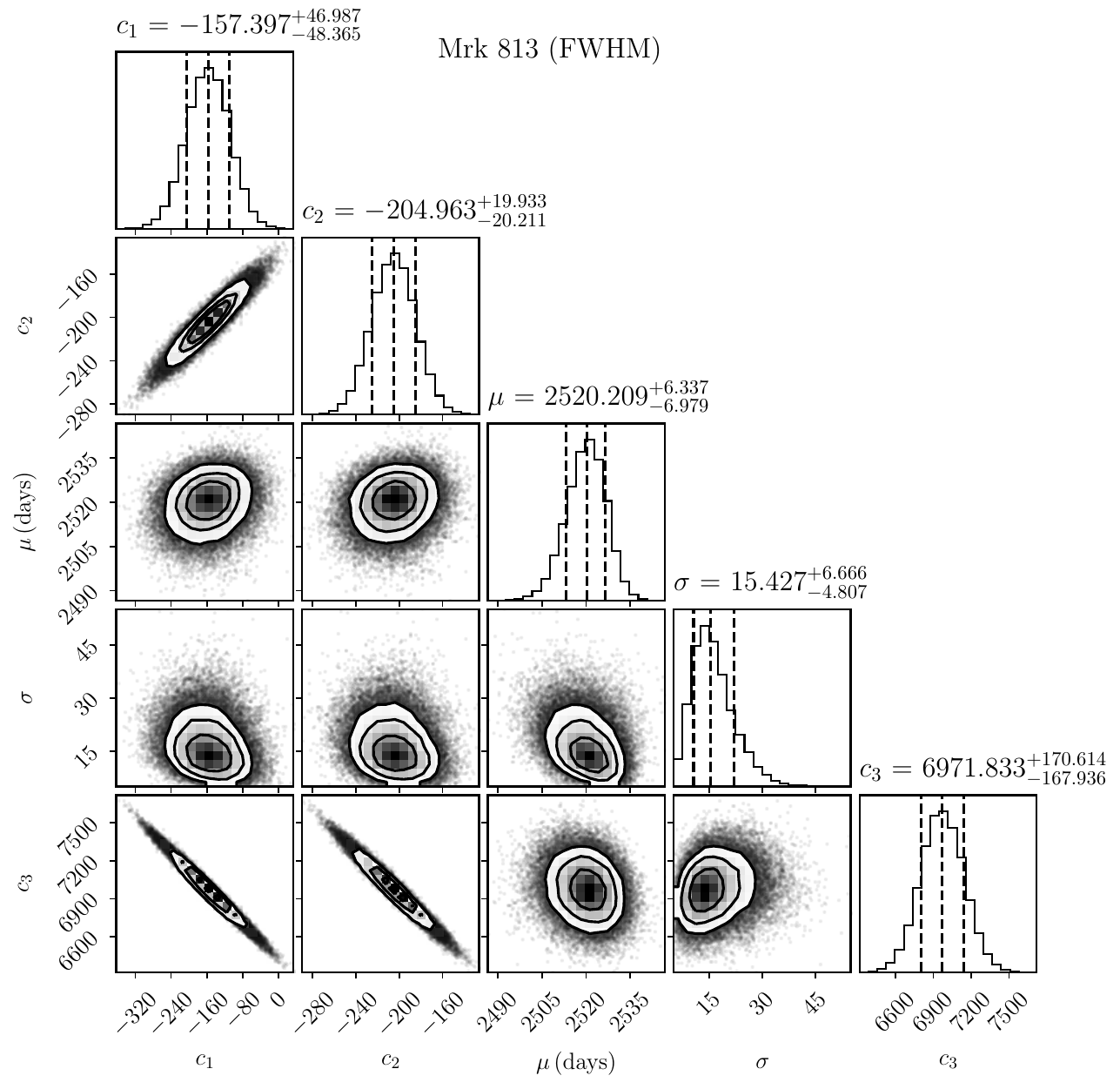}
    \addtocounter{figure}{-1}
    \caption{Continued.}
    \label{fig:longterm light curve5}
\end{figure*}

\begin{figure*}[ht]
    \centering
    \includegraphics[width=1\textwidth]{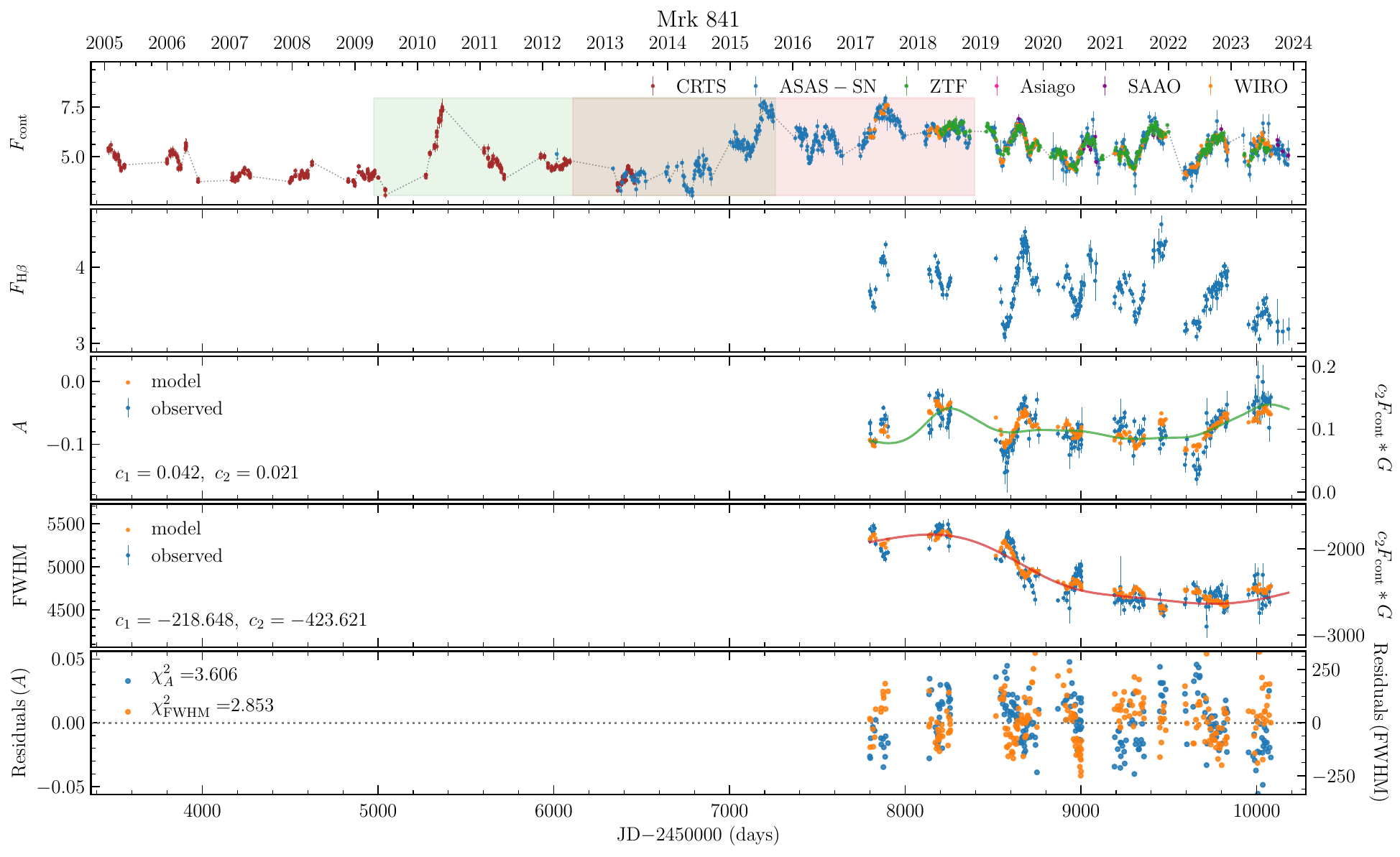}
    \includegraphics[width=0.45\textwidth]{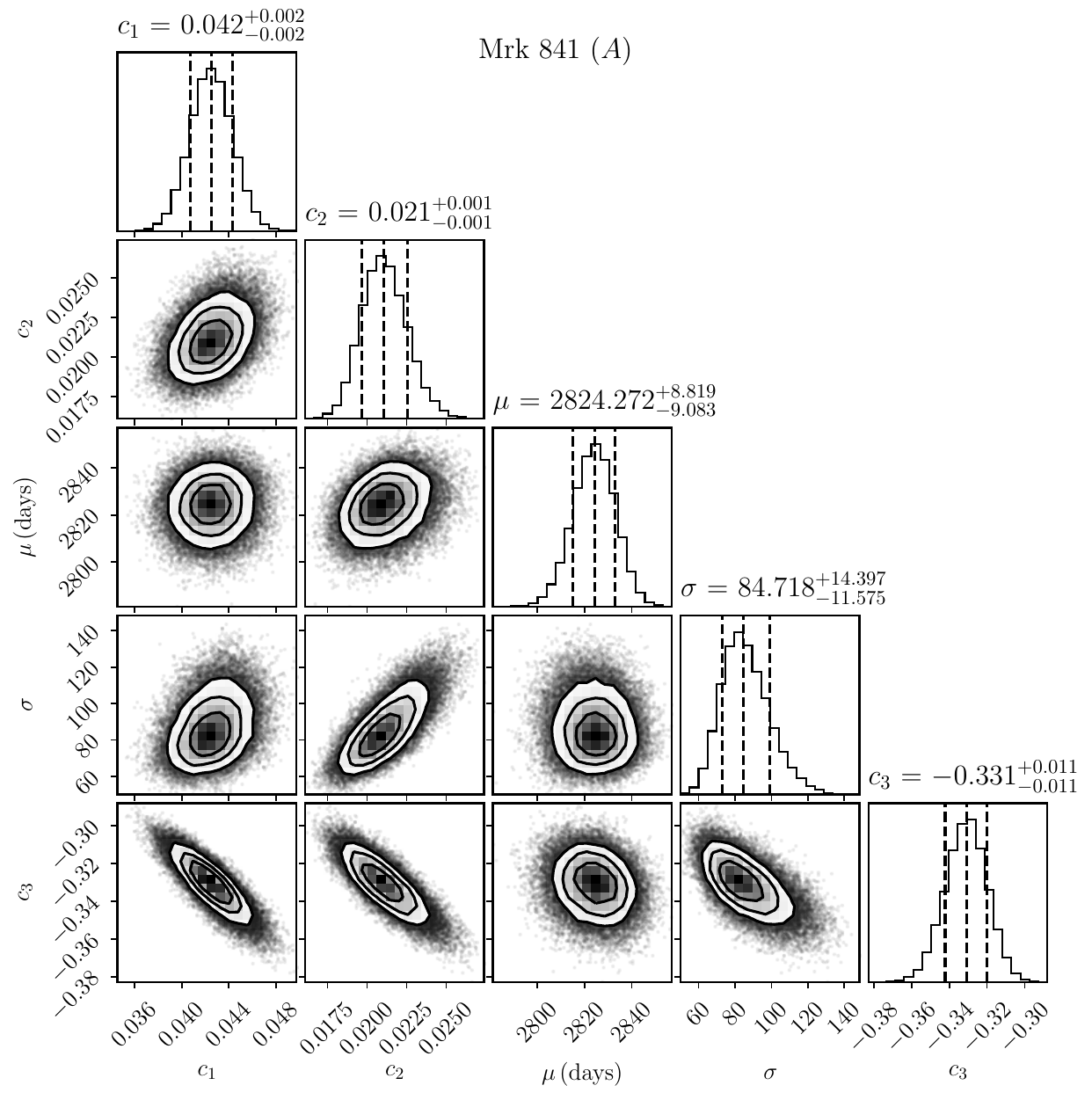}
    \includegraphics[width=0.45\textwidth]{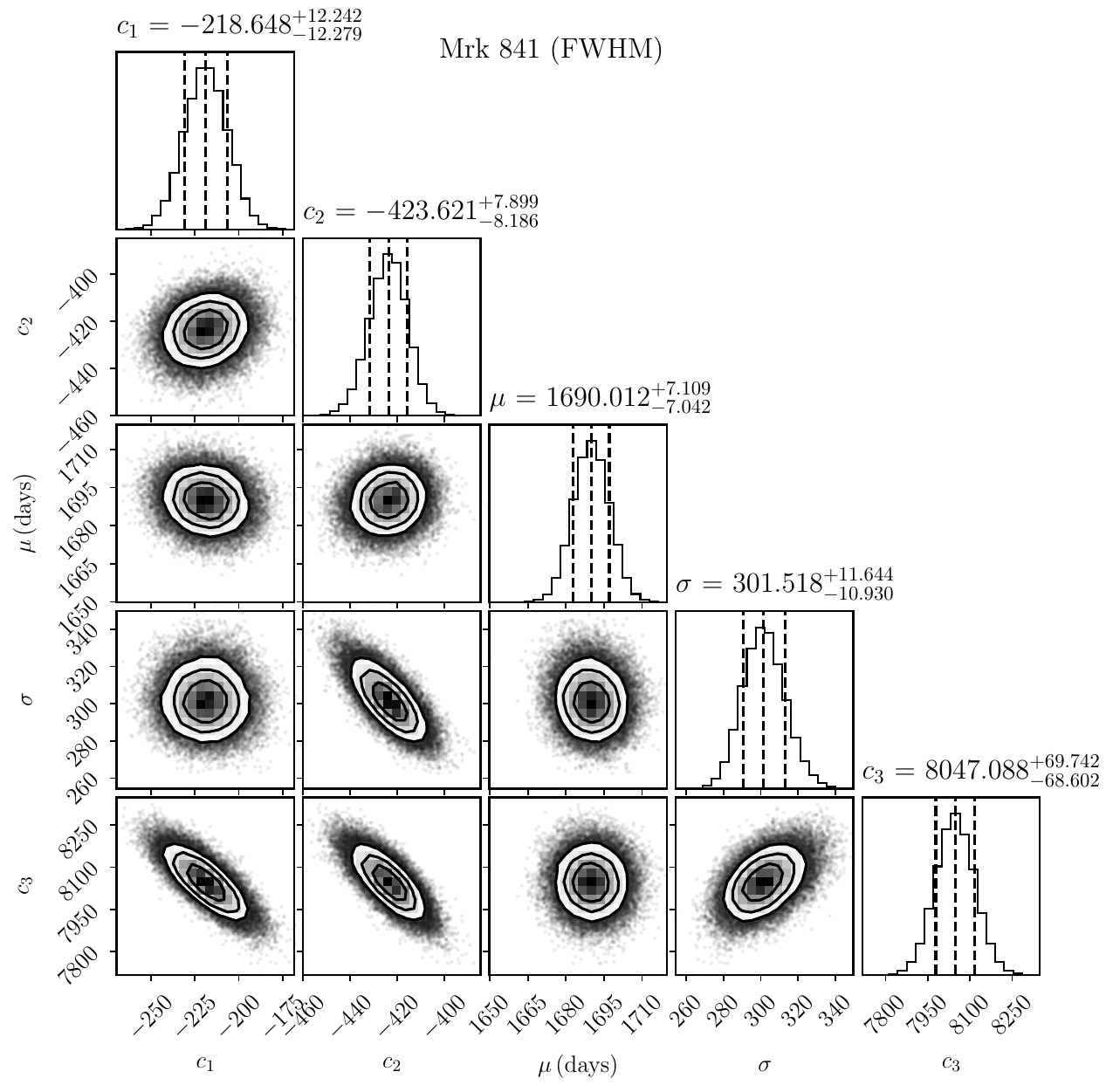}
    \addtocounter{figure}{-1}
    \caption{Continued.}
    \label{fig:longterm light curve6}
\end{figure*}

\begin{figure*}[ht]
    \centering
    \includegraphics[width=1\textwidth]{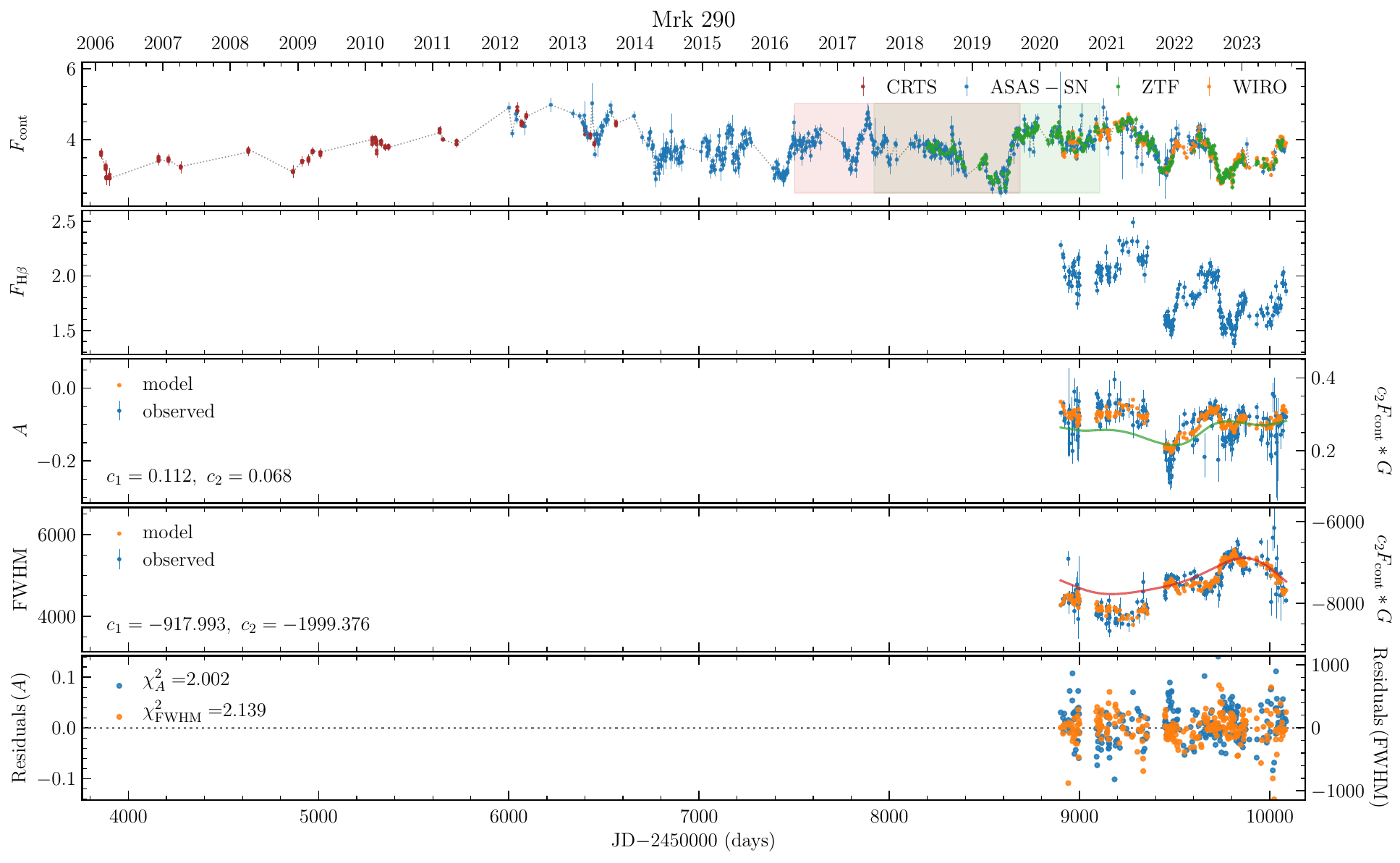}
    \includegraphics[width=0.45\textwidth]{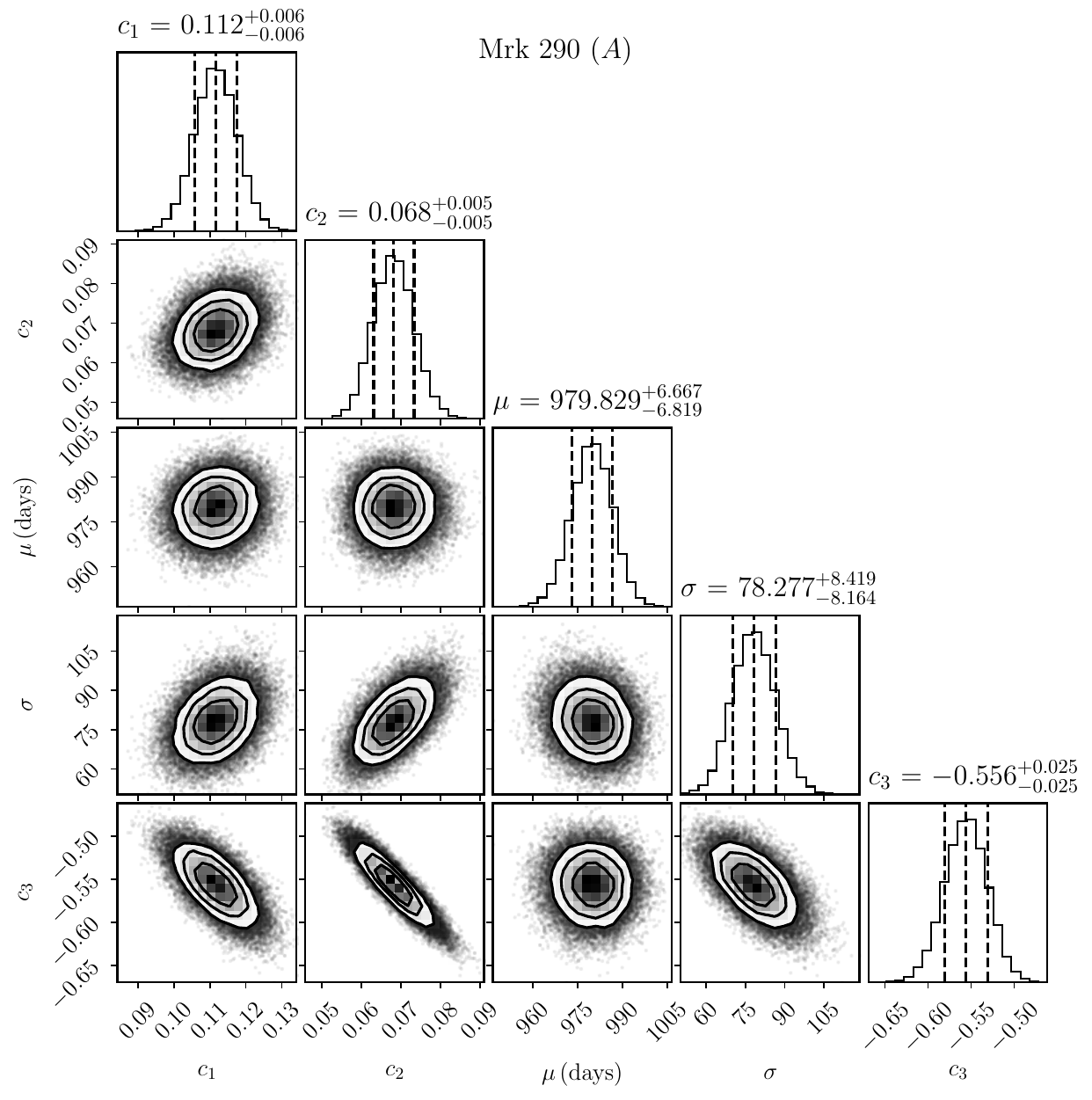}
    \includegraphics[width=0.45\textwidth]{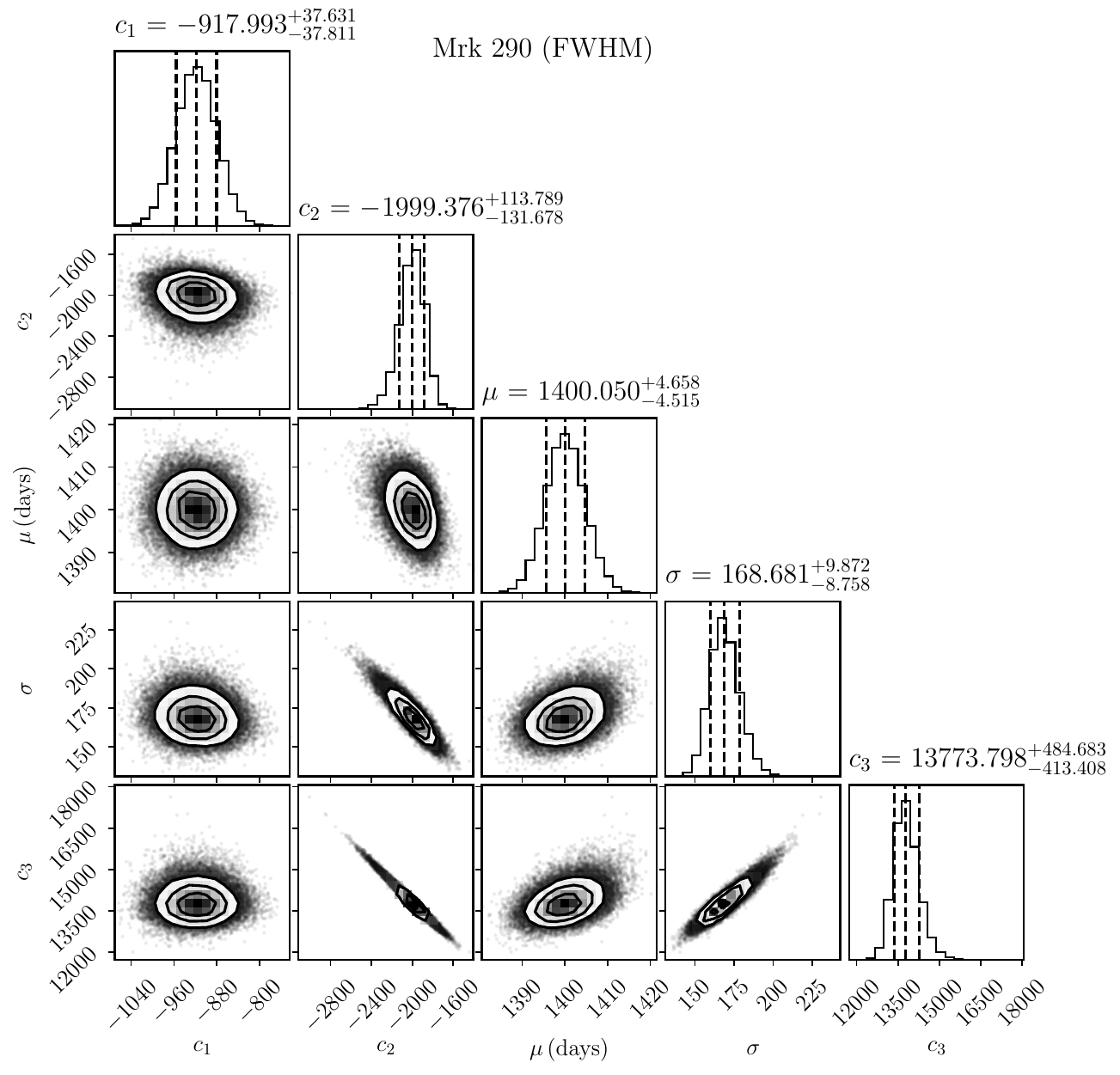}
    \addtocounter{figure}{-1}
    \caption{Continued.}
    \label{fig:longterm light curve7}
\end{figure*}

\begin{figure*}[ht]
    \centering
    \includegraphics[width=1\textwidth]{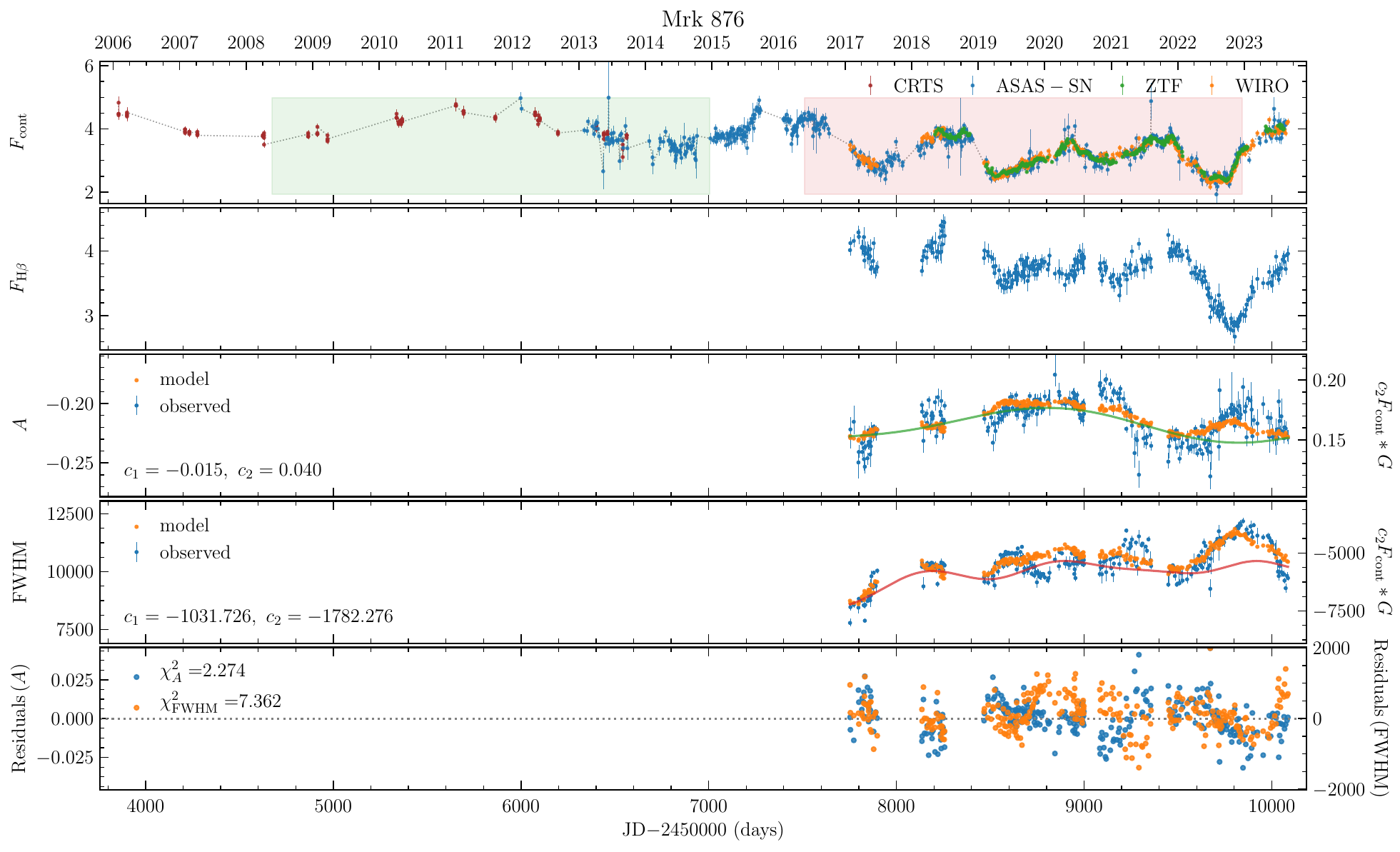}
    \includegraphics[width=0.45\textwidth]{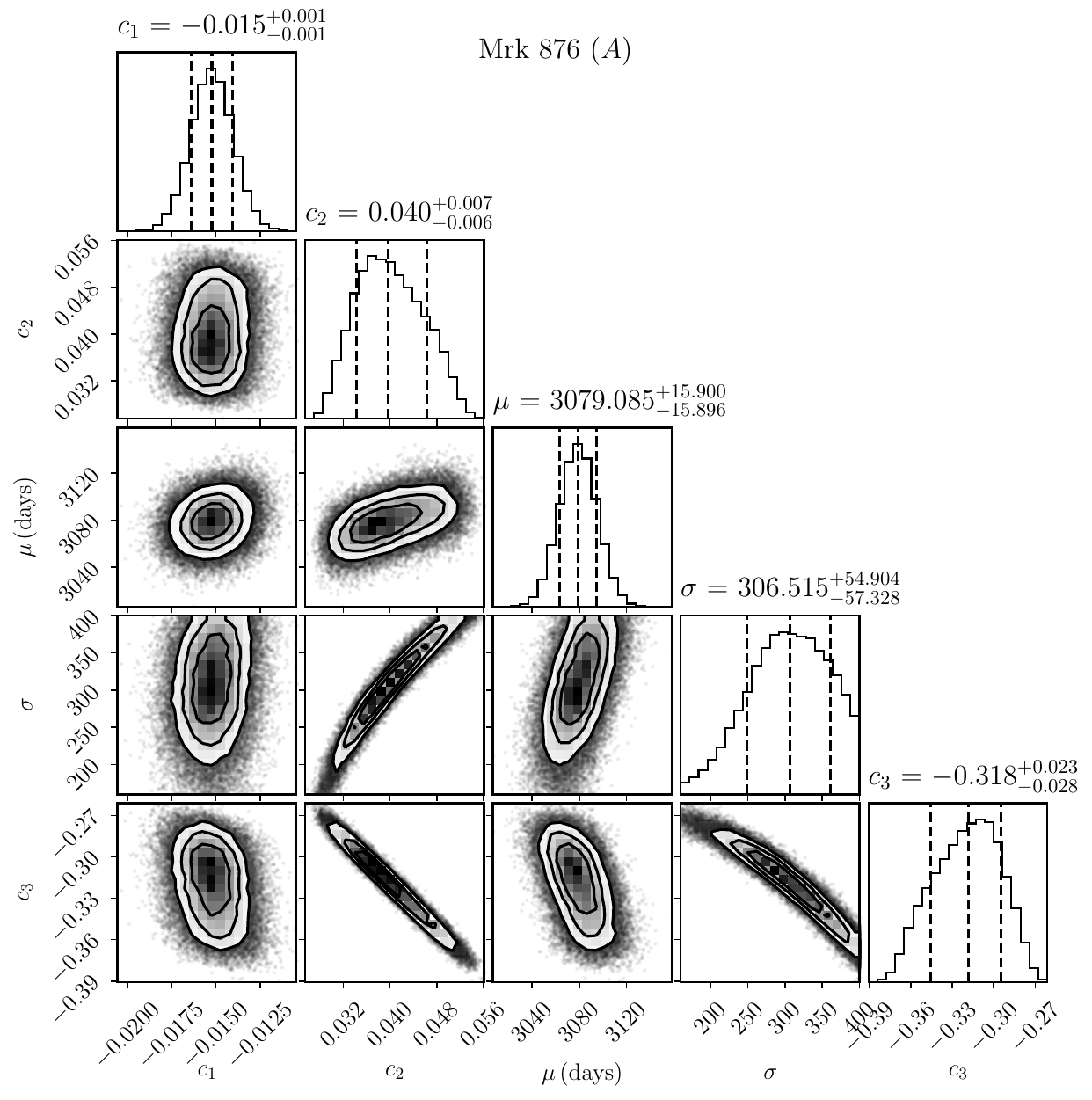}
    \includegraphics[width=0.45\textwidth]{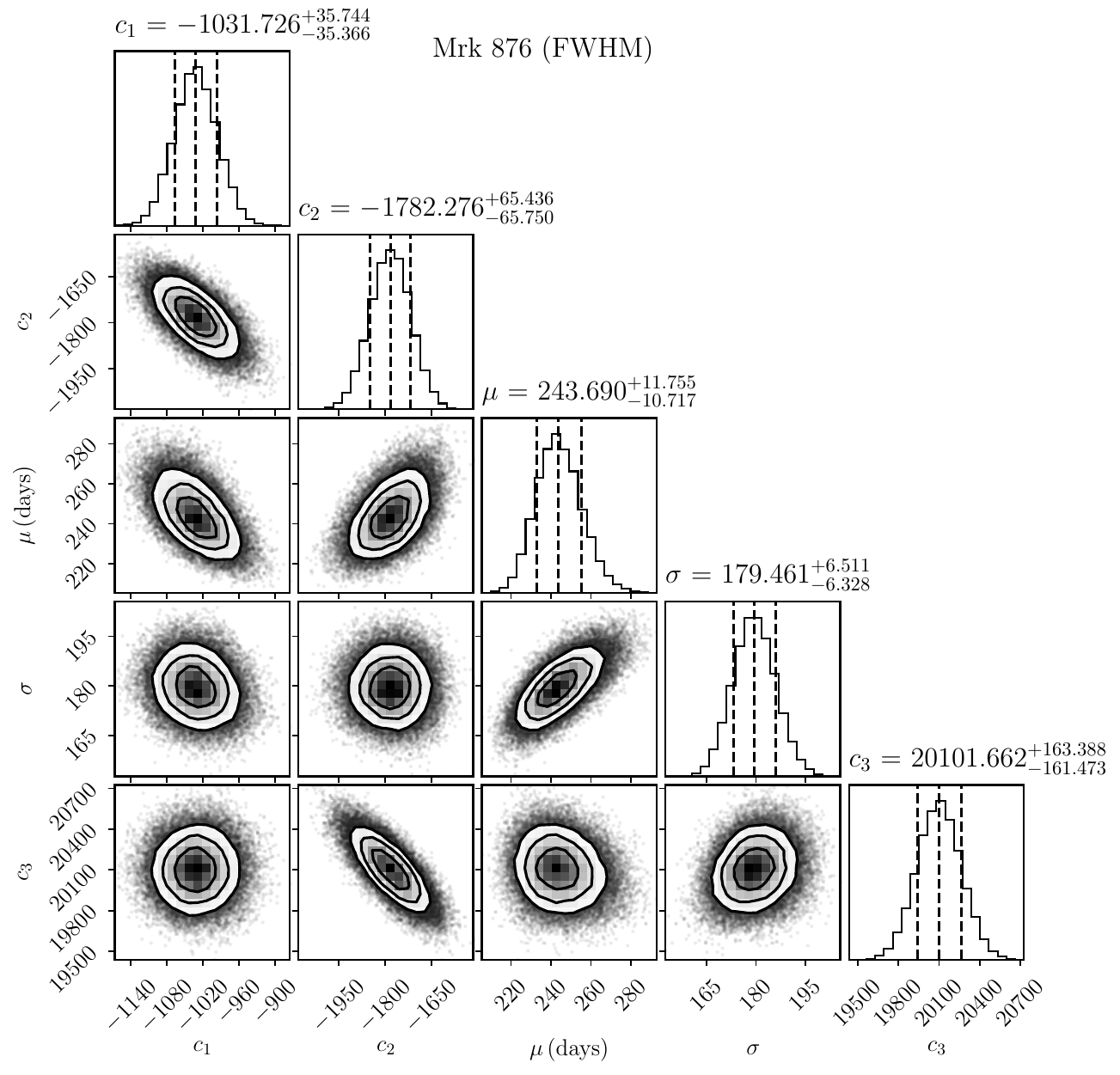}
    \addtocounter{figure}{-1}
    \caption{Continued.}
    \label{fig:longterm light curve8}
\end{figure*}
%%%%%%%%%%%%%%%%%%%%%%%%%%%
\begin{figure*}[ht]
    \centering
    \includegraphics[width=0.45\textwidth]{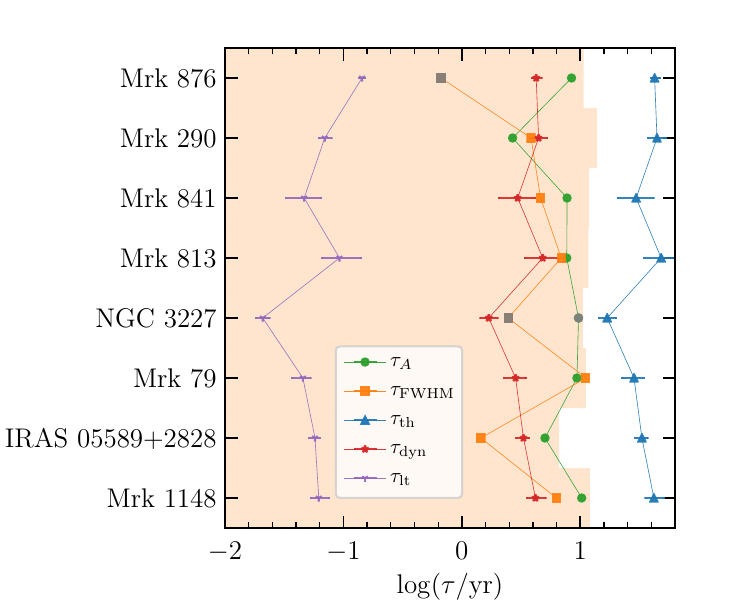}
    \caption{Comparisons between different timescales and the time delays of the long-term trends in $A$ (orange dots) and FWHM (green squares). The orange background exhibits the observational period covered by the campaign. The data points with reduced chi-square larger than 5.0 are marked in gray. }
    \label{fig:timescale}
\end{figure*}
%%%%%%%%%%%%%%%%%%%%%%%%%%%
\begin{figure*}[ht]
    \centering
    \includegraphics[width=1\textwidth]{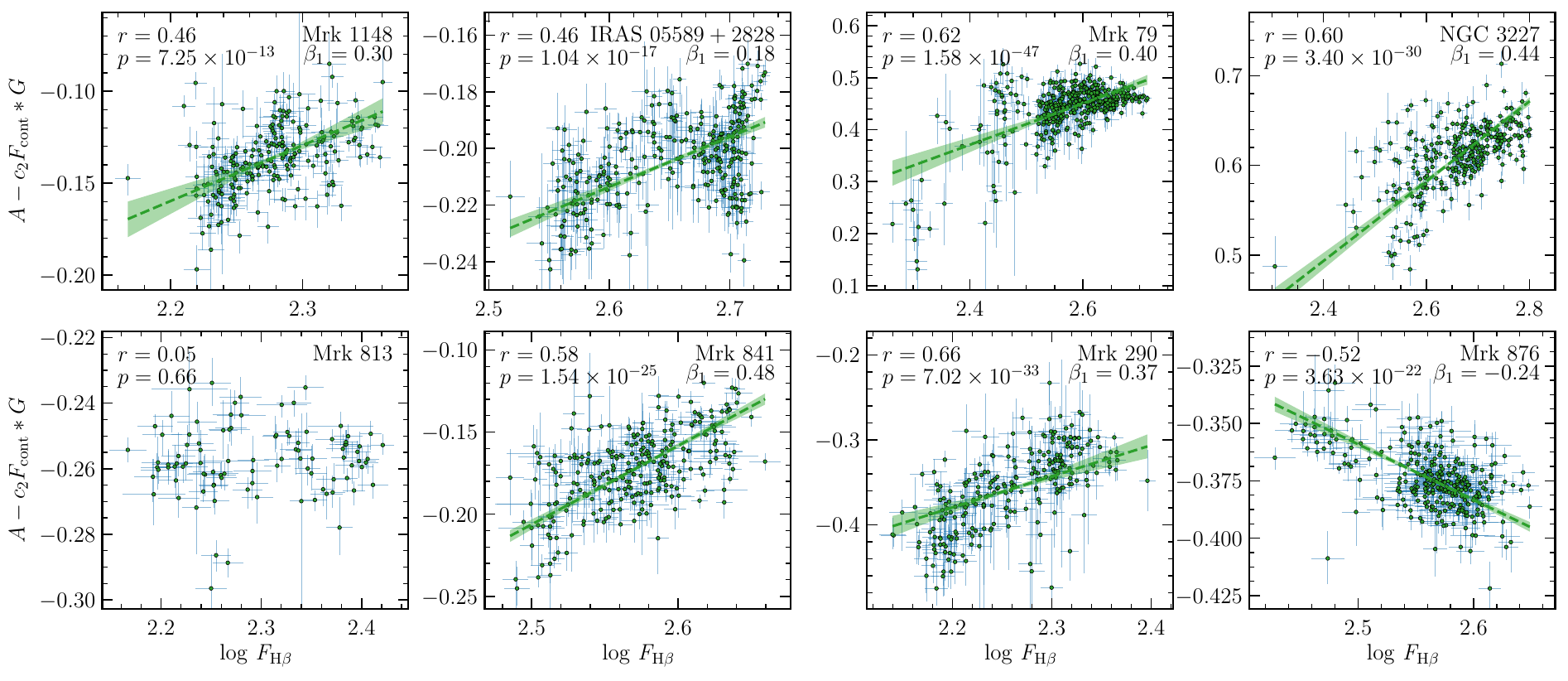}
    \caption{Correlations between H$\beta$ flux and asymmetry $A$ after subtracting the long-term contribution from the radiation pressure. }
    \label{fig:ratio_A_model}
\end{figure*}
%%%%%%%%%%%%%%%%%%%%%%%%%%%
\begin{figure*}[ht]
    \centering
    \includegraphics[width=1\textwidth]{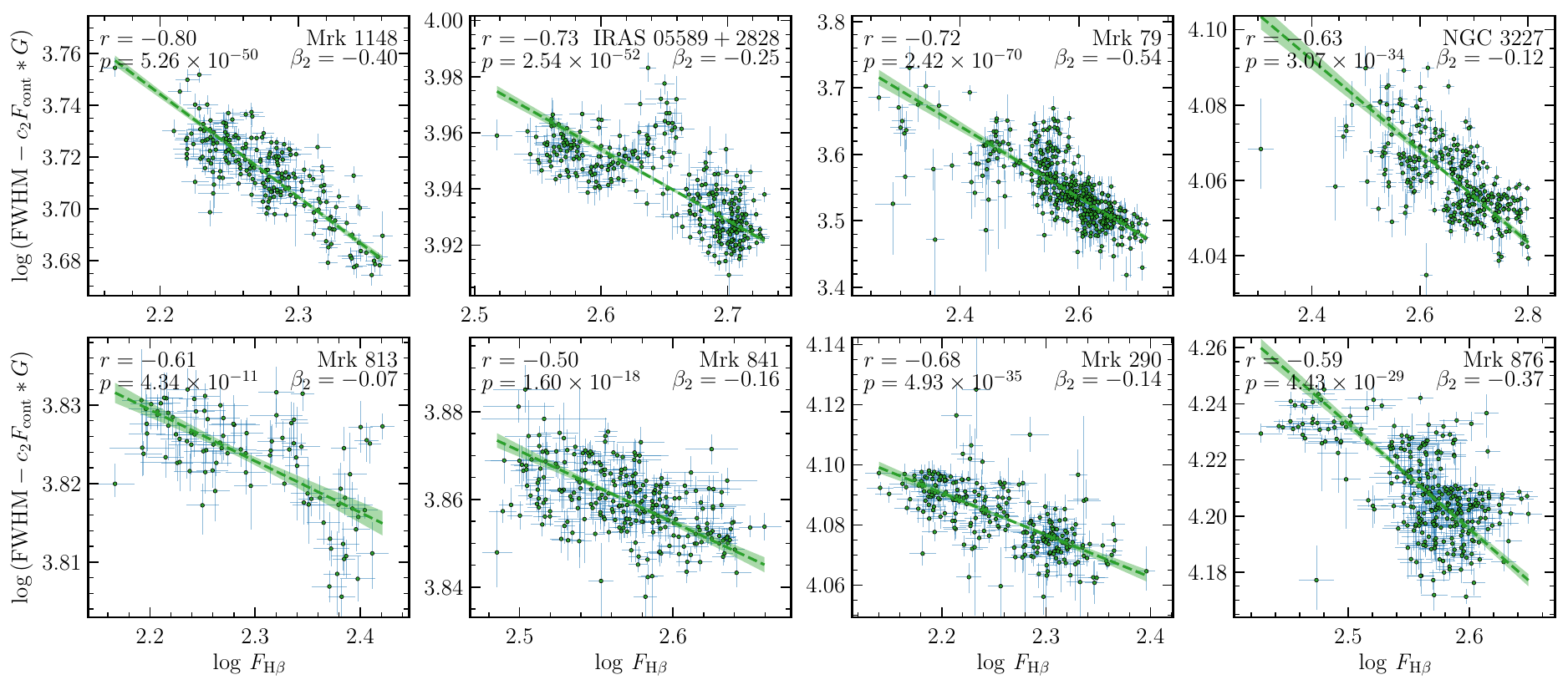}
    \caption{Correlations between H$\beta$ flux and FWHM after subtracting the long-term contribution from the radiation pressure. }
    \label{fig:ratio_FWHM_model}
\end{figure*}
%%%%%%%%%%%%%%%%%%%%%%%%%%%

\FloatBarrier

\end{document}